\documentclass[epj,nopacs,final]{svjour}
\usepackage{graphicx,color,amsmath,amssymb}
\usepackage{epsfig}
\usepackage{rotating}
\usepackage{dcolumn}
\usepackage{bm}
\usepackage{rotating,delarray,array,multirow}
\usepackage{makeidx,pifont,float}
\usepackage{lineno}

\usepackage{epstopdf}

\newcommand{\bc}{\begin{center}}
\newcommand{\ec}{\end{center}}
\newcommand{\be}{\begin{eqnarray}}
\newcommand{\ee}{\end{eqnarray}}
\newcommand{\bea}{\begin{eqnarray}}
\newcommand{\eea}{\end{eqnarray}}

\newcommand{\oo}{$^o$}
\newcommand{\er}{$\pm$}

\newcommand{\script}{\scriptsize}

\title{\boldmath $N^*$ resonances from $K\Lambda$ amplitudes in sliced bins in energy}
\titlerunning{$N^*$ resonances from  $K\Lambda$ amplitudes }
\authorrunning{A.V.~Anisovich \it et al.}
\author{A.V. Anisovich\inst{1,2}, V. Burkert\inst{3}, M.~Had\v{z}imehmedovi\'{c}\inst{6},
D.G.~Ireland\inst{5}, E.~Klempt\inst{1,3}, V.A. Nikonov\inst{1,2}, R.~Omerovi\'{c}\inst{6},
A.V. Sarantsev\inst{1,2}, J.~Stahov\inst{6}, A.~\v{S}varc\inst{4}, and U. Thoma\inst{1}\vspace{5mm}}
\institute{
\inst{1} Helmholtz-Institute f\"ur Strahlen- und Kernphysik der Universit\"at Bonn,
         Nussallee 14 - 16, 53115 Bonn, Germany \\
\inst{2} Particle and Nuclear Physics Institute, Orlova Rosha 1, 188300 Gatchina, Russia  \\
\inst{3} Thomas Jefferson Laboratory, 12000 Jefferson Avenue, Newport News, VA 23606, USA \\
\inst{4} Rudjer Boskovic Institute, Bijenicka cesta 54, P.O. Box 180, 10002 Zagreb, Croatia\\
\inst{5} SUPA, School of Physics and Astronomy, University of Glasgow, Glasgow G12 8QQ, United Kingdom \\
\inst{6} University of Tuzla, Faculty of Natural Sciences and Mathematics, Univerzitetska 4, 75000 Tuzla,
         Bosnia and Herzegovina
}

\date{Received: \today / Revised version: }

\abstract{
The two reactions $\gamma p\to K^+\Lambda$ and $\pi^-p\to K^0\Lambda$ are analyzed to determine
the leading photoproduction multipoles and the pion-induced partial wave amplitudes in slices
of the invariant mass. The multipoles and the partial-wave amplitudes are
simultaneously fitted in a multichannel Laurent+Pietarinen model (L+P model), which determines the poles in the complex energy plane on the second Riemann sheet close to the physical axes. The results from the L+P fit are compared with the results
of an energy-dependent fit based on the Bonn-Gatchina (BnGa) approach.
The study confirms the existence of several poles due to nucleon resonances in the region at about 1.9\,GeV
with quantum numbers $J^P = 1/2^+$, $3/2^+, 1/2^-, 3/2^-, 5/2^-$.
 }
\begin{document}
\maketitle

\section{Introduction\vspace{-1mm}}
The nucleon and its excited states are the simplest systems in which the non-abelian
character of strong interactions is manifest. Three quarks is the minimum quark content
of any baryon, and these three quarks carry the three fundamental colour charges of Quantum Chromodynamics (QCD), and combine to a colourless baryon.
At present it is, however, impossible to
calculate the spectrum of excited states from first principles, even though considerable
progress in lattice gauge calculations has been achieved~\cite{Edwards:2011jj}.
Models are therefore necessary when data are to be compared to predictions.

Quark models predict a rich excitation spectrum of the nucleon
\cite{Capstick:1986bm,Ferraris:1995ui,Glozman:1997ag,Loring:2001kx,Giannini:2015zia}.
In quark models, the resonances are classified in shells according to the energy levels of the
harmonic oscillator. The shell structure of the excitations is still seen in
the data and reproduced in lattice calculations \cite{Edwards:2011jj}. The first excitation shell
is predicted to house five $N^*$ and two $\Delta^*$ resonances with negative-parity; all of them
are firmly established. The second excitation shell contains {\it missing resonances}:
22 resonances (14 $N^*$'s and 8 $\Delta^*$'s) are predicted but 15 only are found in the
mass range below 2100\,MeV, and just 10 of them
(5 $N^*$'s and 5 $\Delta^*$'s) are considered as established, with three or four stars in the
notation of the Particle Data Group \cite{Olive:2016xmw}. Thus 9 $N^*$'s in the mass region
between 1700\,MeV and 2100\,MeV are predicted to exist which are unobserved or the evidence
for their existence is only fair or even poor. This deficit is known as the problem of
the {\it missing resonances} \cite{Koniuk:1979vw,Koniuk:1979vy}. The search
for missing resonances is one of the major aims of a number of experiments in which
the interaction of a photon beam in the GeV energy range with a hydrogen and deuterium target is
studied.

In $\pi N$ elastic and charge exchange scattering, the excited states may have isospin $I=1/2$ ($N^*$)
and $I=3/2$ ($\Delta^*$). A large amount of data was analyzed by the groups at
Karlsruhe-Helsinki (KH84)~\cite{Hohler:1984ux},
Carnegie-Mellon (CM) \cite{Cutkosky:1980rh} and at GWU \cite{Arndt:2006bf}. The
1850 - 2100\,MeV mass region -- where the missing resonances of the second excitation shell are
predicted in most constituent quark models (see, e.g.,~\cite{Capstick:1986bm,Ferraris:1995ui,%
Glozman:1997ag,Loring:2001kx,Giannini:2015zia,Capstick:1998uh,Capstick:2000qj}) -- is dominated by the production of $\Delta^*$ resonances with spin-parity $J^P=1/2^\pm,
3/2^\pm, 5/2^\pm$, $7/2^+$; nucleon resonances are difficult to establish in this mass range
due to the overwhelming background from $\Delta^*$ resonances.

The production of $\Lambda$ hyperons in pion and photo-in\-duced reactions, in contrast to
$\pi N$ elastic scattering, is ideally suited to search for new nucleon resonances and to confirm
resonances that are not yet well established~(see, e.g.,~\cite{Capstick:1998uh,Capstick:2000qj}
and references therein). Due to isospin conservation in strong interactions,
only $N^*$ resonances decay into $\Lambda K$ final states, there are no isospin $I=3/2$ contributions. Second,
the $\Lambda\to N\pi$ weak-interaction decay reveals the polarization $P$ of the $\Lambda$. Thus, the
recoil polarization is measurable. In $\pi N$ elastic scattering, the equivalent target polarization,
also called $P$,
requires the use of a polarized target. In photoproduction, a third advantage emerges: the process
is not suppressed
even when the $\pi N$ coupling constants of $N^*$ resonances in the second excitation
shell are small \cite{Capstick:1998uh,Capstick:2000qj}. Photoproduction may hence reveal
the existence of $N^*$ resonances coupling to $\pi N$ only weakly. Indeed, a number of
new resonances has been reported (or have been upgraded in the star rating) from a combined
analysis of a large number of pion and photo-produced reactions \cite{Anisovich:2011fc}.
Some of the ``new'' resonances had been observed before
\cite{Hohler:1984ux,Cutkosky:1980rh,Manley:1992yb,Penner:2002ma,Penner:2002md}
or were confirmed in later analyses \cite{Shrestha:2012va,Shrestha:2012ep,L+P2014}.
The evidence for the existence of the new states stemmed from energy-dependent fits to the data
using the BnGa approach \cite{Anisovich:2004zz,Anisovich:2006bc,Anisovich:2007zz,Denisenko:2016ugz}.
The reaction $\gamma p \to K^+\Lambda$ proved to be particularly useful \cite{Nikonov:2007br}.

The ultimate aim of experiments is to provide sufficient information that the data can be
decomposed into partial waves or multipoles of defined and unique spin-parity. It can either
be done through constructing an explicit theoretical model, or as we present here, through
the reconstruction of partial-wave amplitudes and of multipoles in a truncated partial wave
analysis. Limiting the partial wave series to low orbital angular momenta allows us to
overcome issues with the still relatively large errors in the measurements of observable
quantities.

The main goal of this paper is to test if $N^*$ resonances in the fourth resonance region
can be confirmed definitely from a fit to multipoles driving the excitation of partial
waves with defined spin-parity, and to extract their properties. This is done in two ways:
i.) In a standard way where a
theoretical model is constructed. Its free parameters are estimated by fitting to the experimental data
set base, and the partial waves of the final solution are analytically continued into the
complex energy plane to obtain poles. ii.) In a way which does not depend on detailed model
assumptions by using the Laurent+Pietarinen (L+P) method where a solution of the theoretical
model is replaced by a most general analytic function consisting of a number poles and
branch-cuts, which is embodied by a fast converging power series in a conformal variable.
This variable is generated by a conformal mapping of the complex energy plane onto a unit
circle. The first Riemann sheet is mapped to the outside of the unit circle, and the second
Riemann sheet -- where the poles are located -- into the inside of the unit circle.
In method ii.), poles are extracted by fitting to the single-energy
partial wave decomposition, as opposed to a direct global fit to the data.

Method i.), coupled-channel energy-dependent fits, exploits the full statistical potential
of the data. The effect of couplings to various other final states like
$N\pi$, $N\eta$, $\Sigma K$, $\Delta\pi$, etc. is taken into account exactly as well as all
correlations between the different amplitudes. However, all partial waves need to be
determined in one single fit, and it is difficult to verify the uniqueness of the results.
In method ii.) we use single channel L+P fits (SC L+P) where each channel is fitted
individually, and multi-channel L+P fits (MC L+P) where two or more channels are fitted
simultaneously. The main advantage of the model-independent approach is that we can fit
one partial wave at a time, and that we avoid any dependence on the quality of the model.
The drawback is that you first have to extract partial waves, and this procedure depends
on the choice of higher partial waves, introducing some model dependence.

\section{\label{Construction}\boldmath
Construction of $K\Lambda$ amplitudes in slices of their invariant mass}
\subsection{\boldmath The partial wave amplitudes for $\pi^-p\to K^0\Lambda$}
\subsubsection{Formalism}
The differential cross sections $d\sigma/d\Omega$ for the reaction \\
\mbox{$\pi^-p$ $\to K^0\Lambda$} receives contributions from a spin-non-flip and a
spin-flip amplitude, $f$ and $g$, according to the relation
\be
\frac{d\sigma}{d\Omega} = \frac{k}{q}(|f|^2 + |g|^2)\,,
\ee
where $q$ and $k$ are the initial and final meson momenta respectively in
the centre of mass frame~\cite{Hohler:1984ux}.
Both amplitudes depend on the invariant mass $W$ and $z=\cos\theta$,
with $\theta$ being the scattering angle. The two amplitudes can be expanded
into partial wave amplitudes~$A_l^\pm$
\be
\nonumber
 f(W,z)&=&\frac{1}{\sqrt{qk}}\sum\limits_{l=0}^{L} \big [(l\!+\!1)A_l^+(W)+ l A_l^-(W)\big ]
P_l(z) \;, \nonumber \\
 g(W,z)&=&\frac{1}{\sqrt{qk}} \sin \Theta \sum\limits_{l=1}^L \big [A_l^+(W)- A_l^-(W)\big ]
P'_l(z) \;,
\label{piN_expan}
\ee
where $P_l(z)$ are the Legendre polynomials. $J = |l \pm 1/2|$ is the total spin of the state.
The sign in the relation for $J$ defines the sign in $A_l^\pm$.

The $\Lambda\to N\pi$ decay can be used to determine the decay asymmetry with respect
to the scattering plane, called recoil asymmetry $P$. Assuming that
the target nucleon is fully polarized, $P$ can be defined as
\be
(1\pm P)\frac{d\sigma}{d\Omega} =  |f \pm i g|^2\,.
\ee
When the target proton is polarized longitudinally (along the pion beam line),
the spin transfer from proton to $\Lambda$ yields the spin rotation angle
$\beta$.
\be
\beta = arg\Big(\frac{f-i g}{f+i g}\Big) =tan^{-1}\Big(\frac{-2 \Re e(f^* g)}{|f|^2 - |g|^2}\Big)\,.
\ee
It is defined as $\beta =\arctan{(-R/A)}$, where
$A$ and $R$ are the polarization components in direction of the $\Lambda$ and its
orthogonal component in the scattering plane. $R$ and $A$ are
given by
\be
R = \frac{2 \Re e(f^* g)}{|f|^2 + |g|^2}\;,\;\;\; A = \frac{|f|^2 -
|g|^2}{|f|^2 + |g|^2}\,.
\ee
The polarization variables are constrained by the relation
\be
\label{norm}
P^2+A^2+R^2 = 1.
\ee
\subsubsection{Fits to the data}
Data on the reaction $\pi^-p\to K^0\Lambda$ were taken in
Chicago \cite{Knasel:1975rr} and at
the NIMROD accelerator at the Rutherford Laboratory \cite{Baker:1978qm,Saxon:1979xu,Bell:1983dm}.
From these data, the partial wave amplitudes $A_l^\pm$ defined in eqn. (\ref{piN_expan})
should be derived.

\begin{figure*}
\begin{center}
\begin{tabular}{ccccc}
\hspace{-3mm}\includegraphics[width=0.195\textwidth,height=0.162\textwidth]{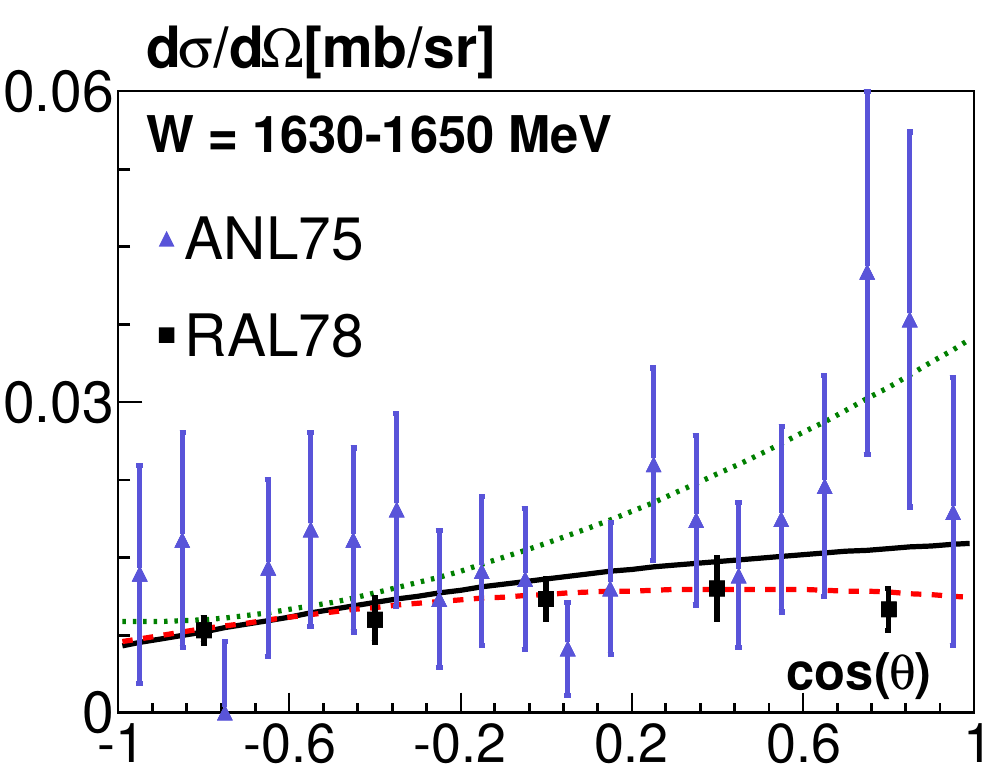}&
\hspace{-4mm}\includegraphics[width=0.195\textwidth,height=0.162\textwidth]{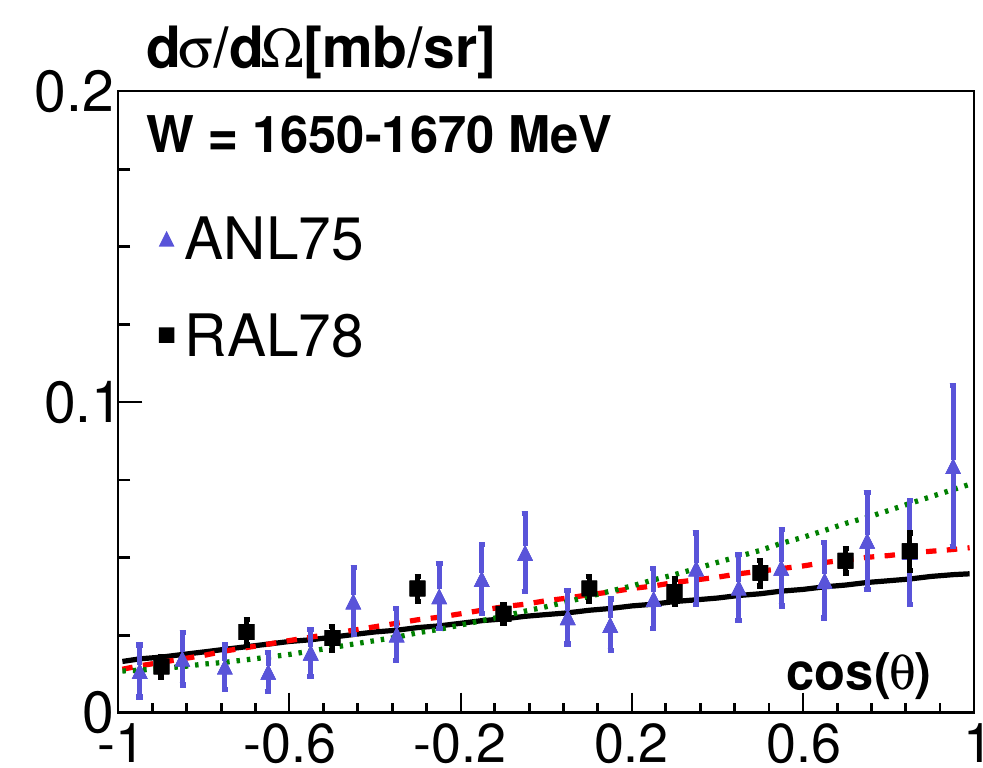}&
\hspace{-4mm}\includegraphics[width=0.195\textwidth,height=0.162\textwidth]{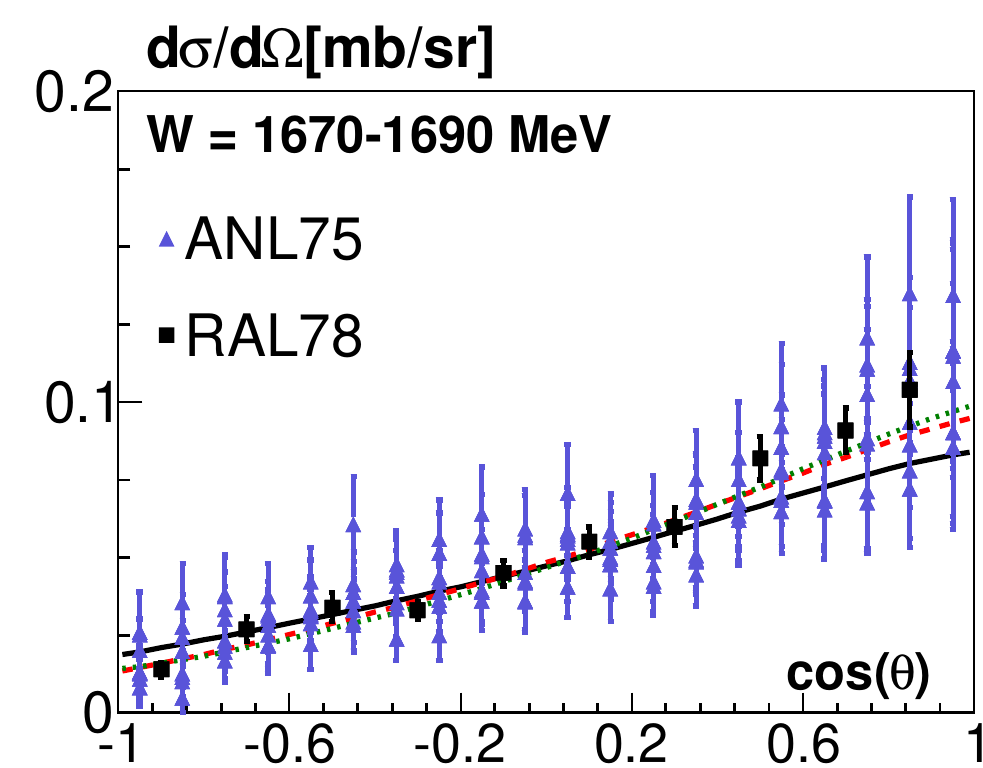}&
\hspace{-4mm}\includegraphics[width=0.195\textwidth,height=0.162\textwidth]{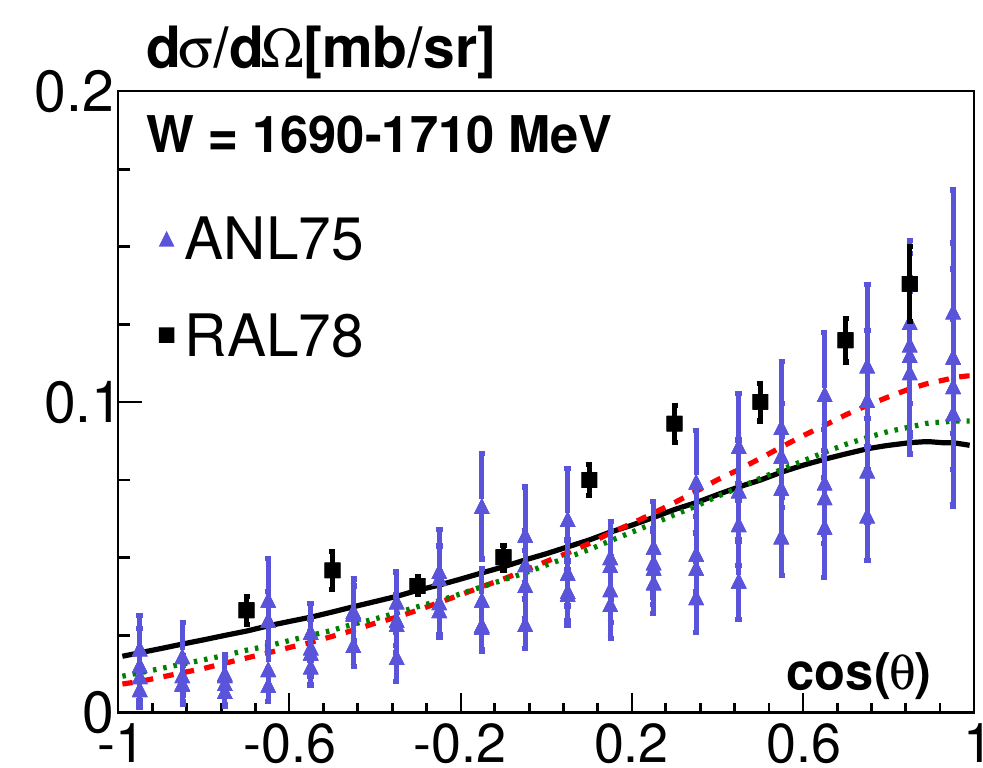}&
\hspace{-4mm}\includegraphics[width=0.195\textwidth,height=0.162\textwidth]{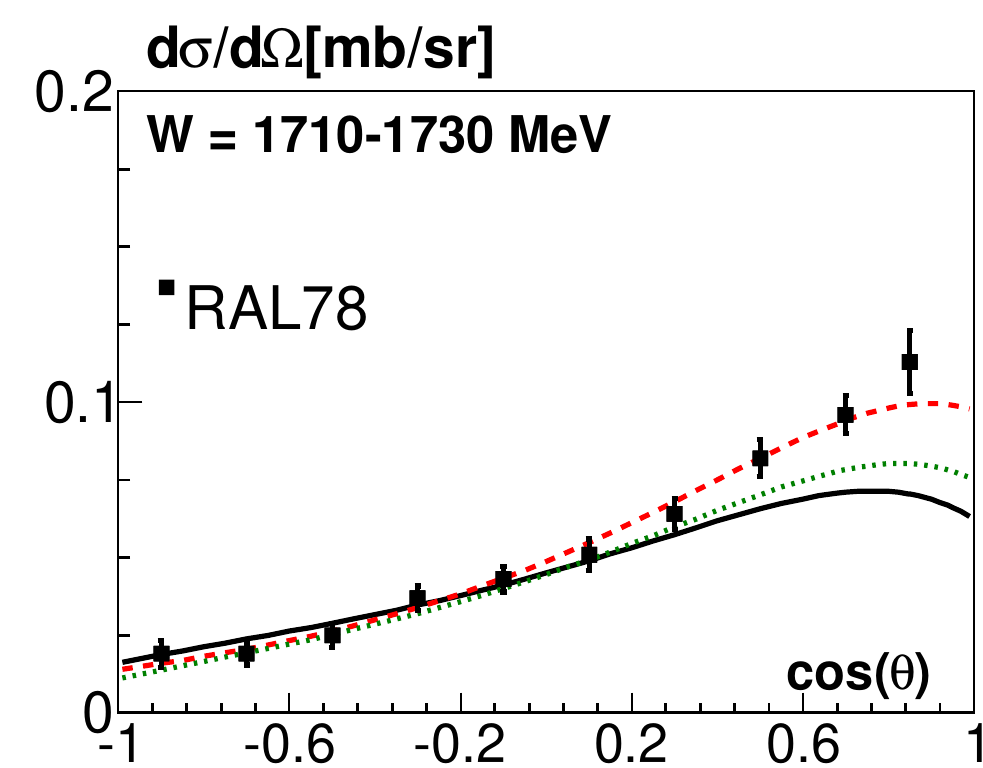}\\
\hspace{-3mm}\includegraphics[width=0.195\textwidth,height=0.162\textwidth]{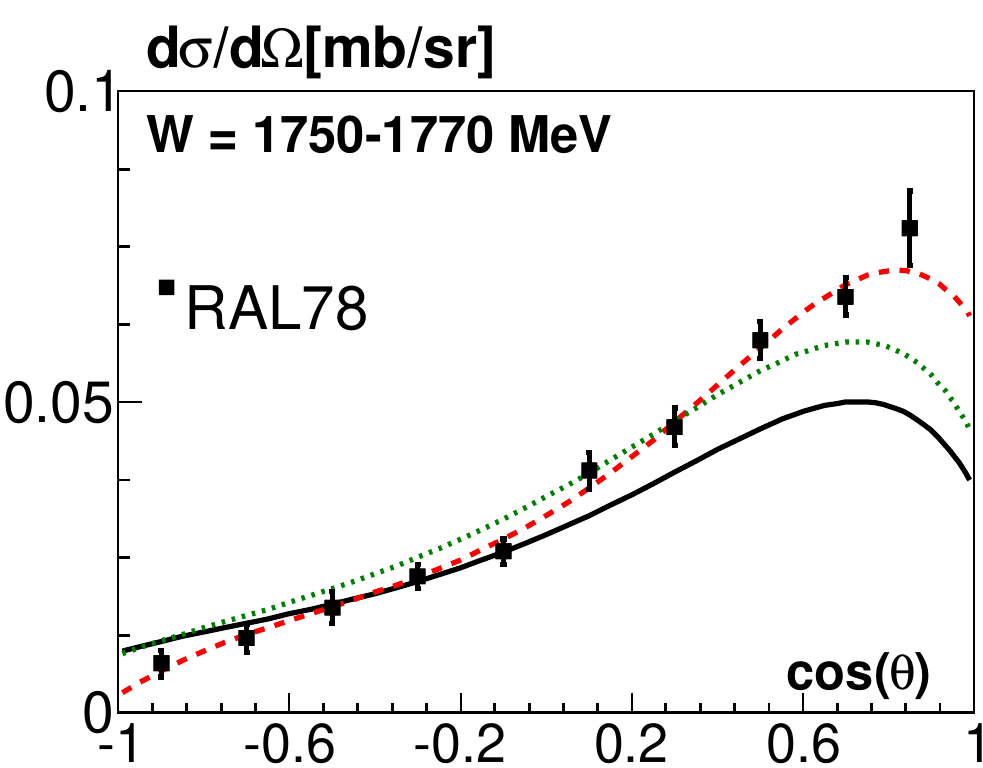}&
\hspace{-4mm}\includegraphics[width=0.195\textwidth,height=0.162\textwidth]{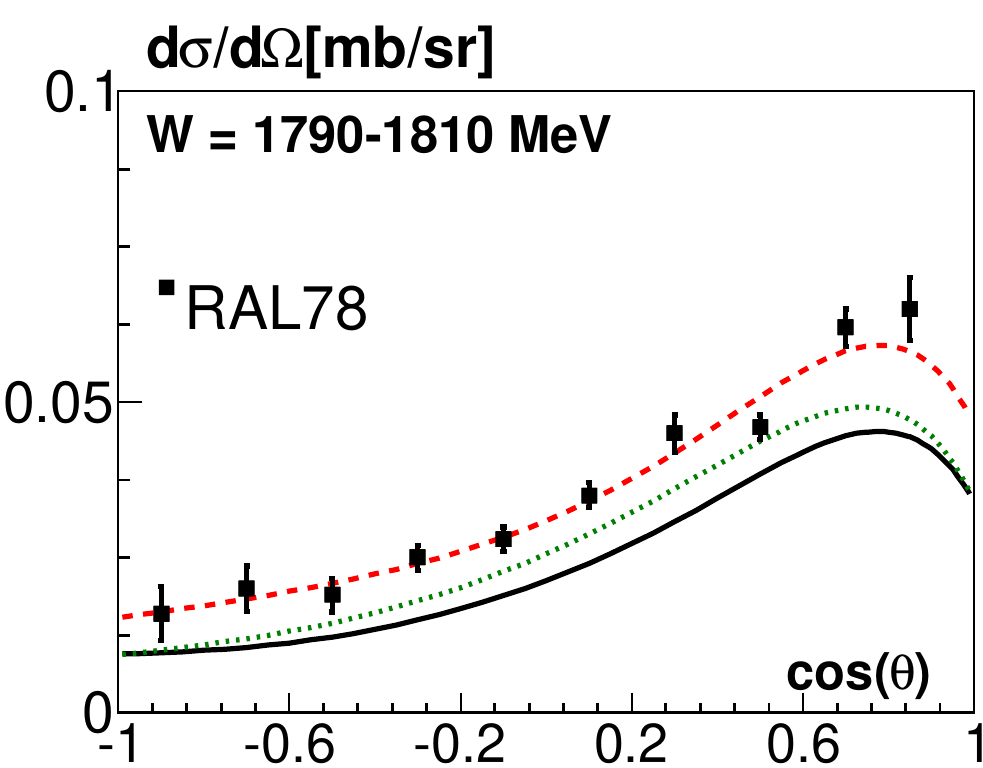}&
\hspace{-4mm}\includegraphics[width=0.195\textwidth,height=0.162\textwidth]{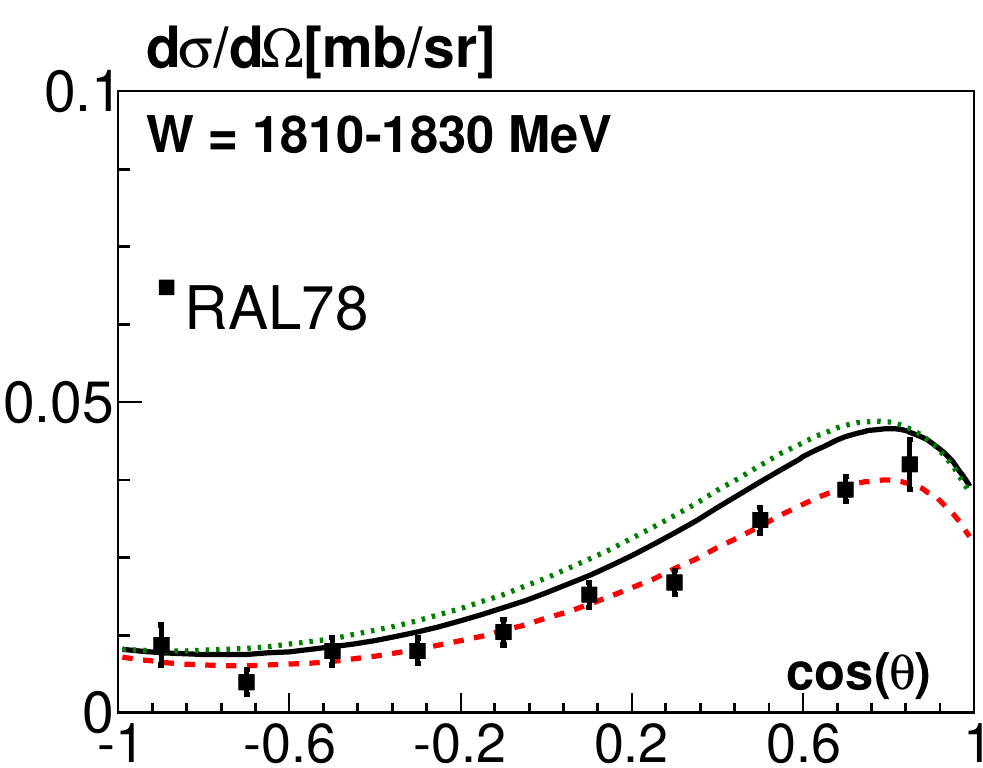}&
\hspace{-4mm}\includegraphics[width=0.195\textwidth,height=0.162\textwidth]{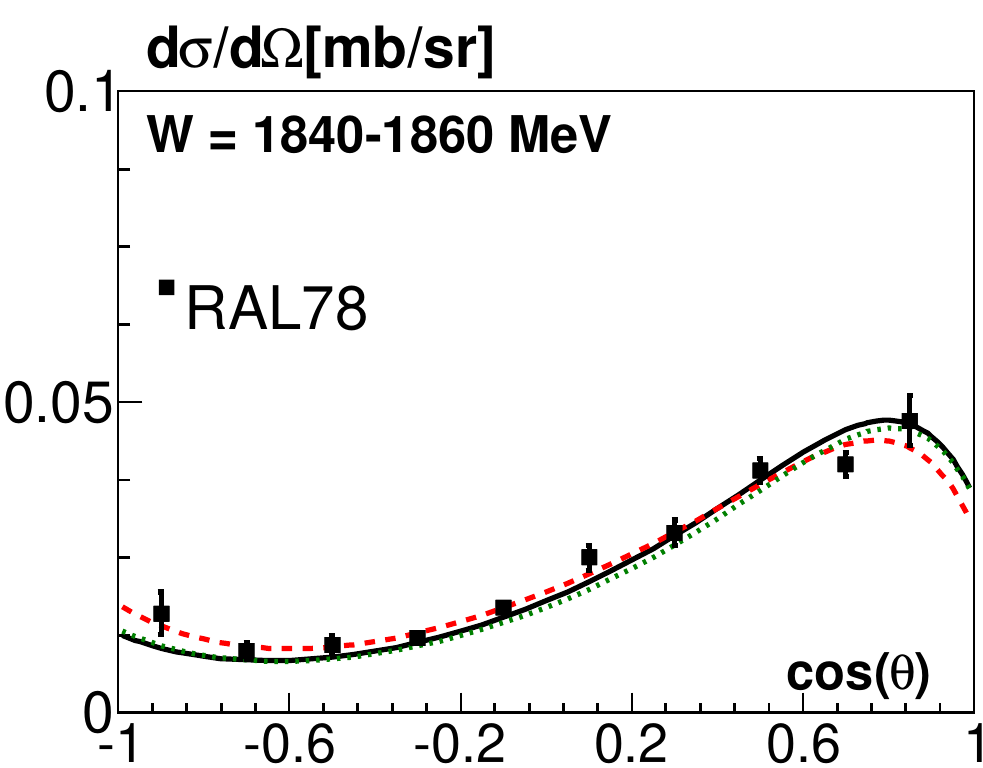}&
\hspace{-4mm}\includegraphics[width=0.195\textwidth,height=0.162\textwidth]{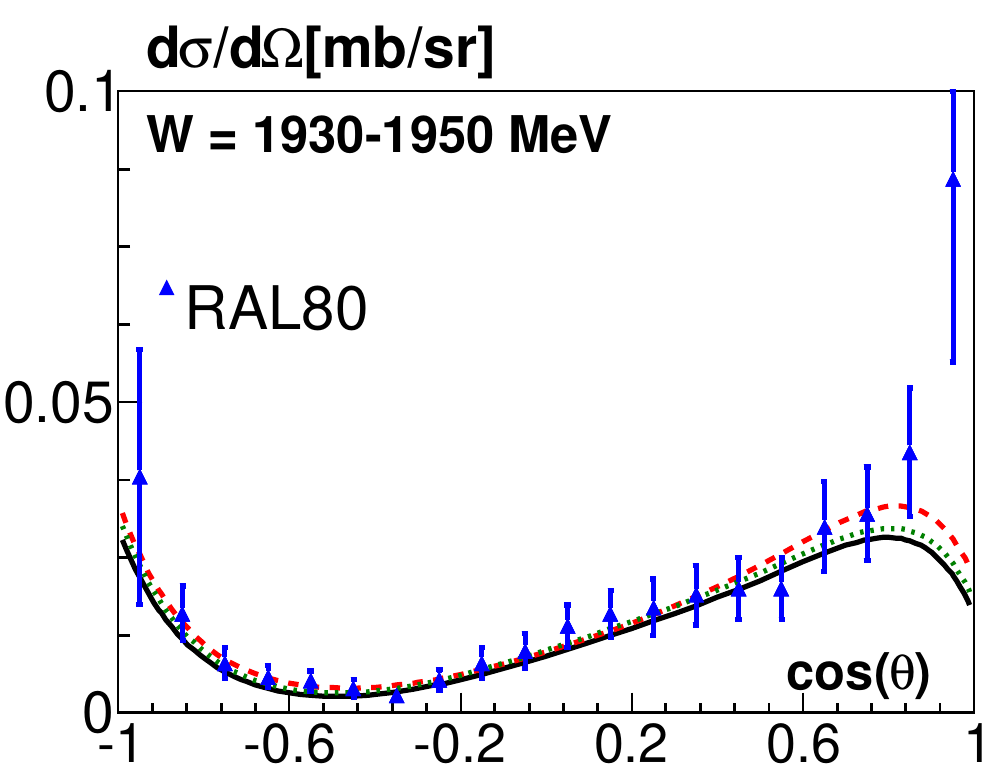}\\
\hspace{-4mm}\includegraphics[width=0.195\textwidth,height=0.162\textwidth]{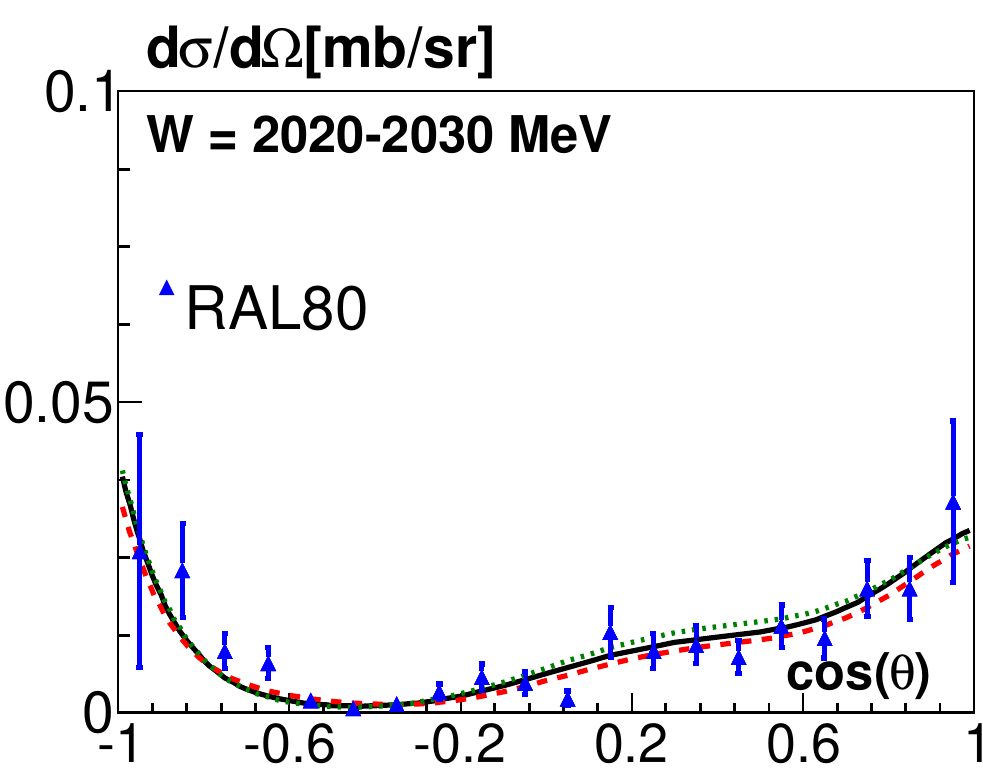}&
\hspace{-4mm}\includegraphics[width=0.195\textwidth,height=0.162\textwidth]{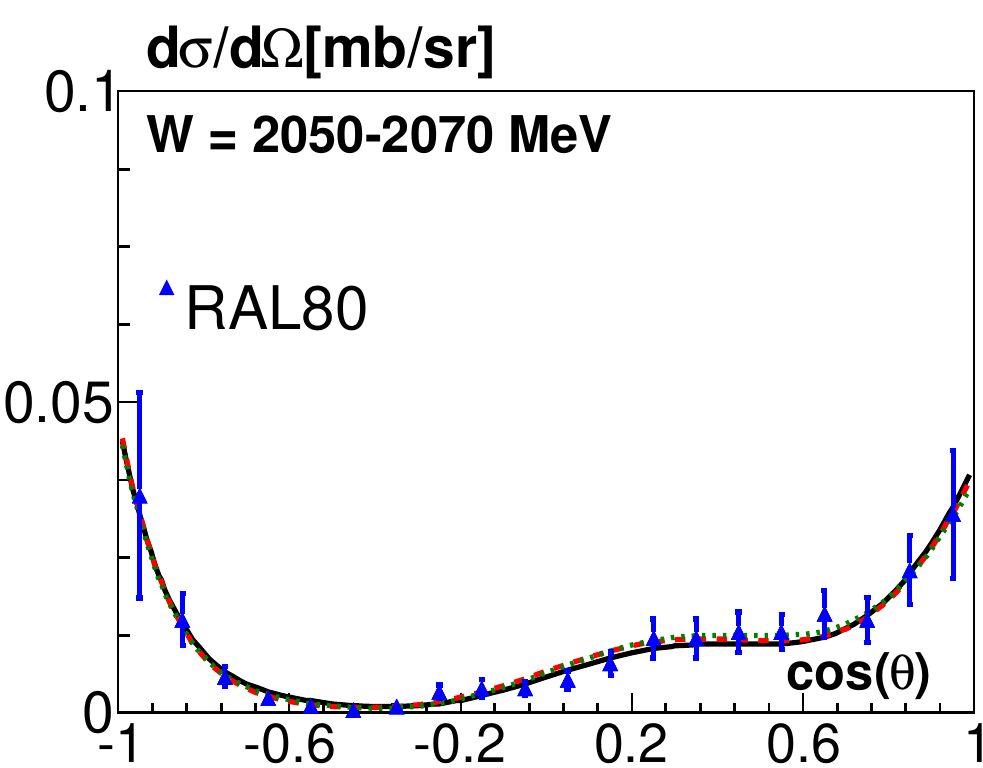}&
\hspace{-4mm}\includegraphics[width=0.195\textwidth,height=0.162\textwidth]{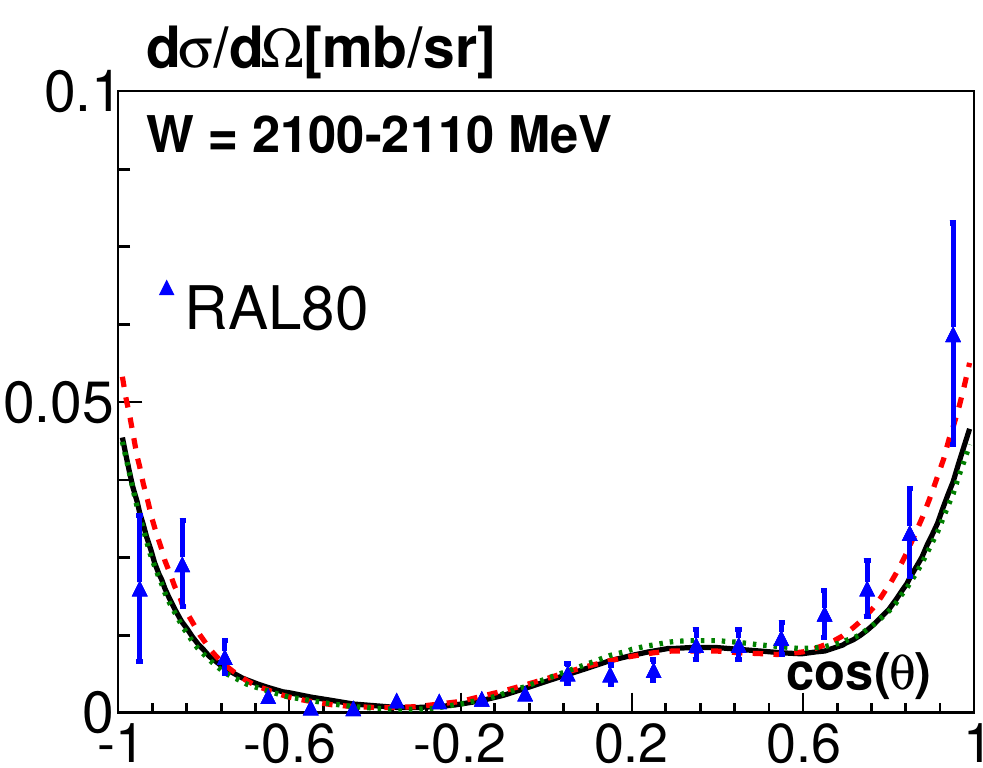}&
\hspace{-4mm}\includegraphics[width=0.195\textwidth,height=0.162\textwidth]{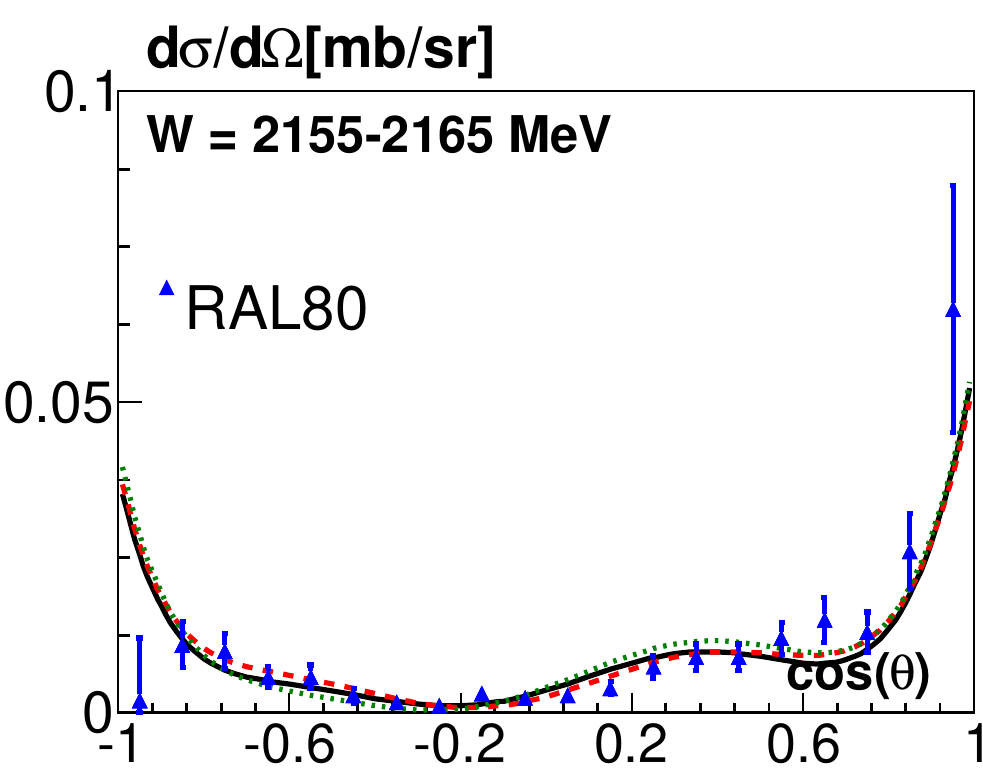}&
\hspace{-4mm}\includegraphics[width=0.195\textwidth,height=0.162\textwidth]{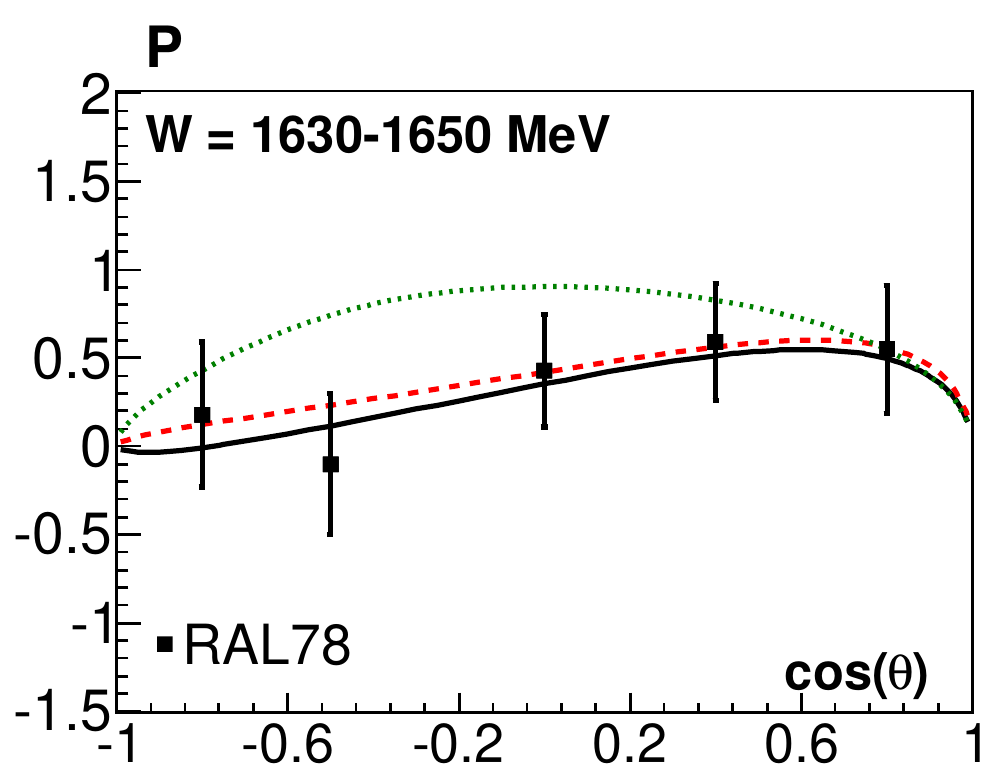}\\
\hspace{-4mm}\includegraphics[width=0.195\textwidth,height=0.162\textwidth]{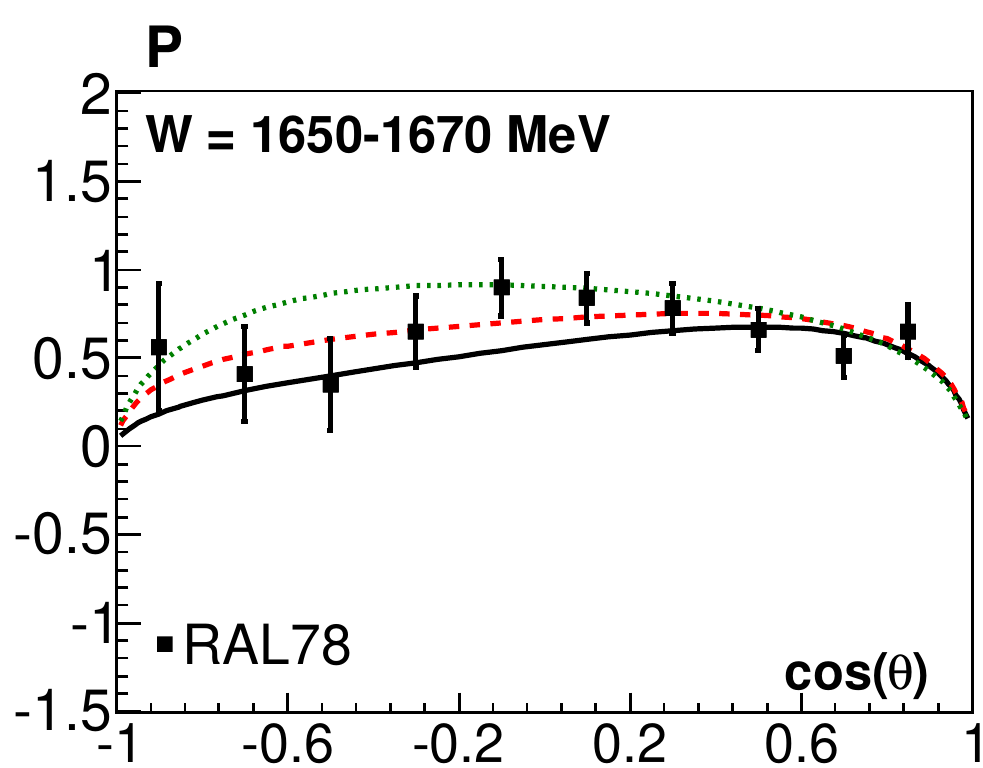}&
\hspace{-4mm}\includegraphics[width=0.195\textwidth,height=0.162\textwidth]{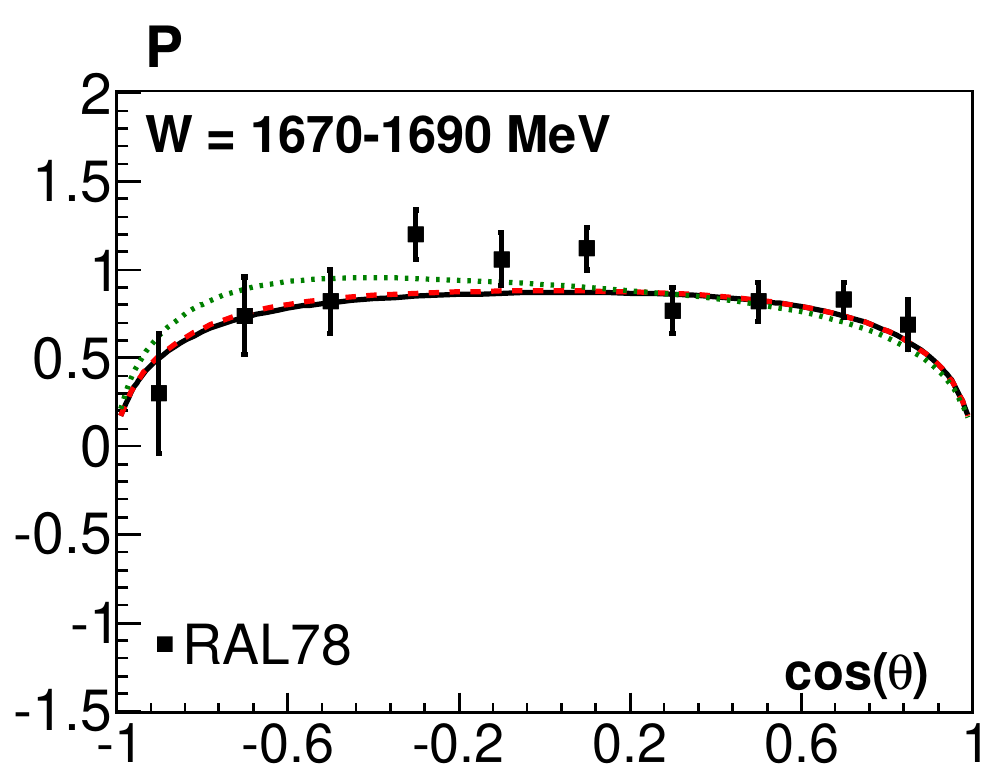}&
\hspace{-4mm}\includegraphics[width=0.195\textwidth,height=0.162\textwidth]{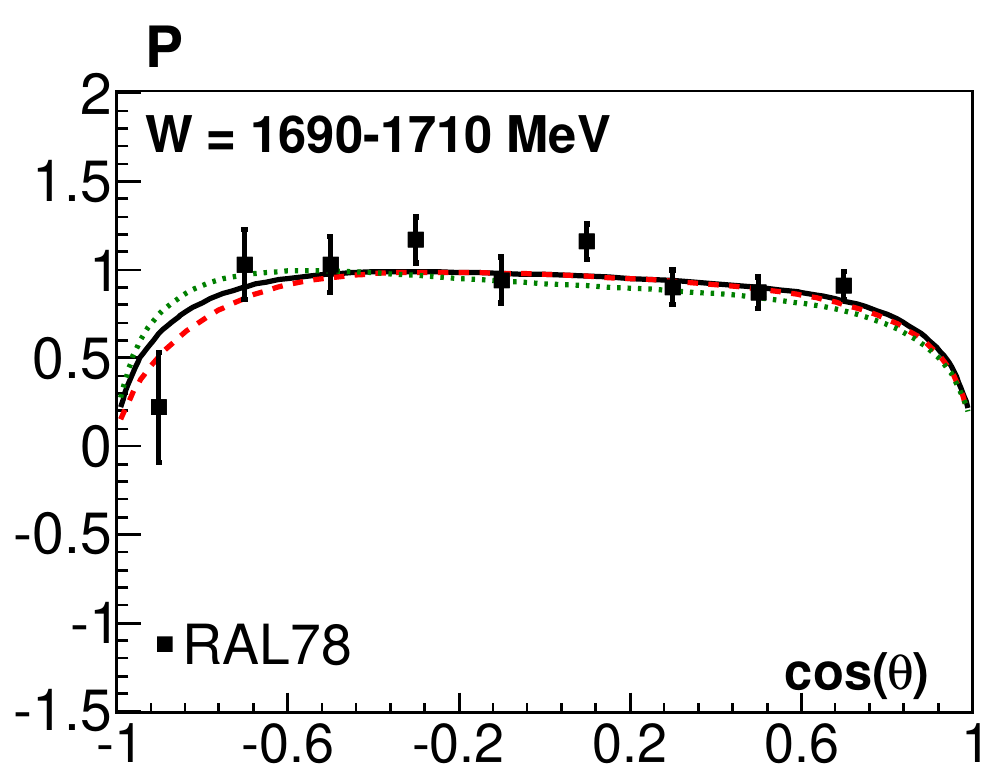}&
\hspace{-4mm}\includegraphics[width=0.195\textwidth,height=0.162\textwidth]{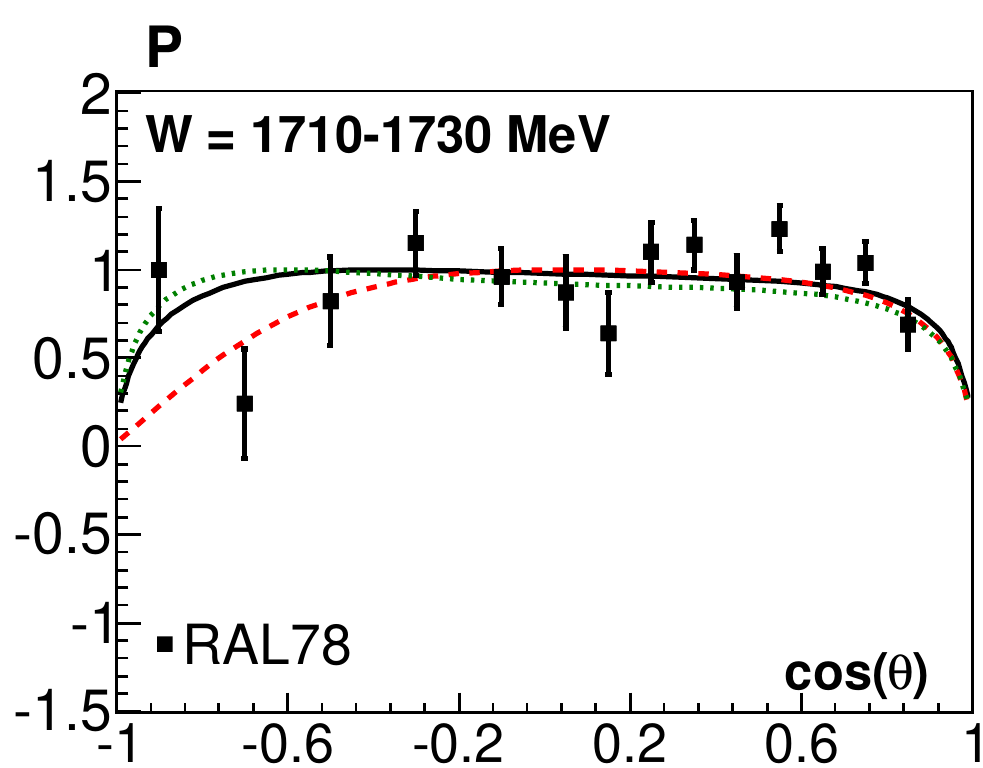}&
\hspace{-4mm}\includegraphics[width=0.195\textwidth,height=0.162\textwidth]{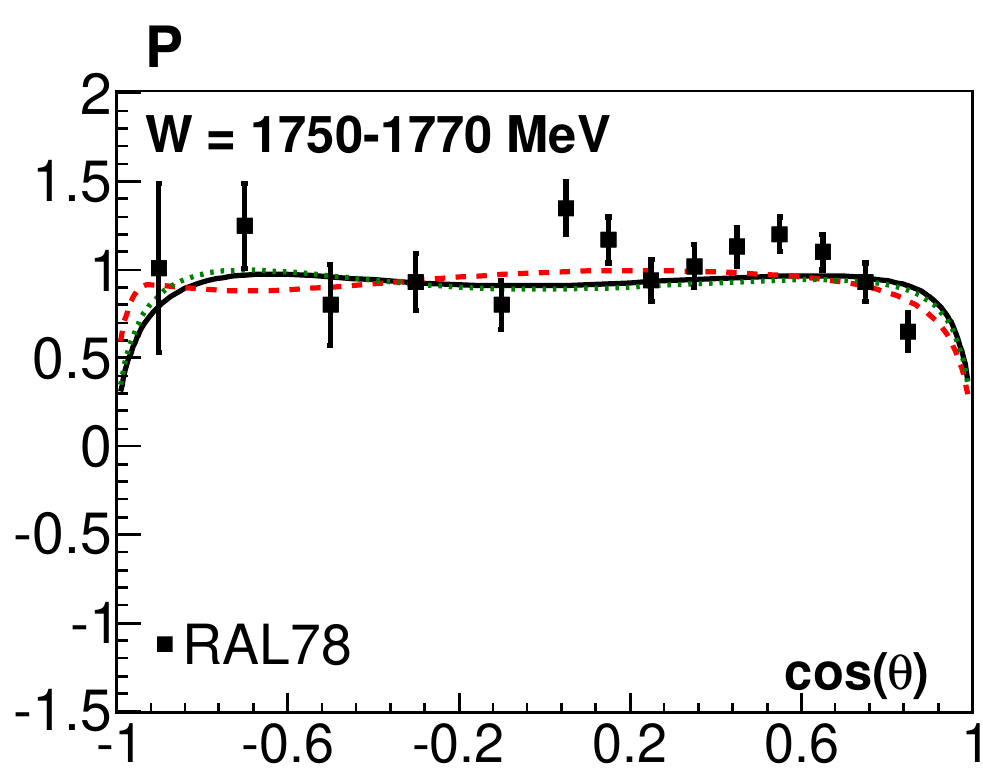}\\
\hspace{-4mm}\includegraphics[width=0.195\textwidth,height=0.162\textwidth]{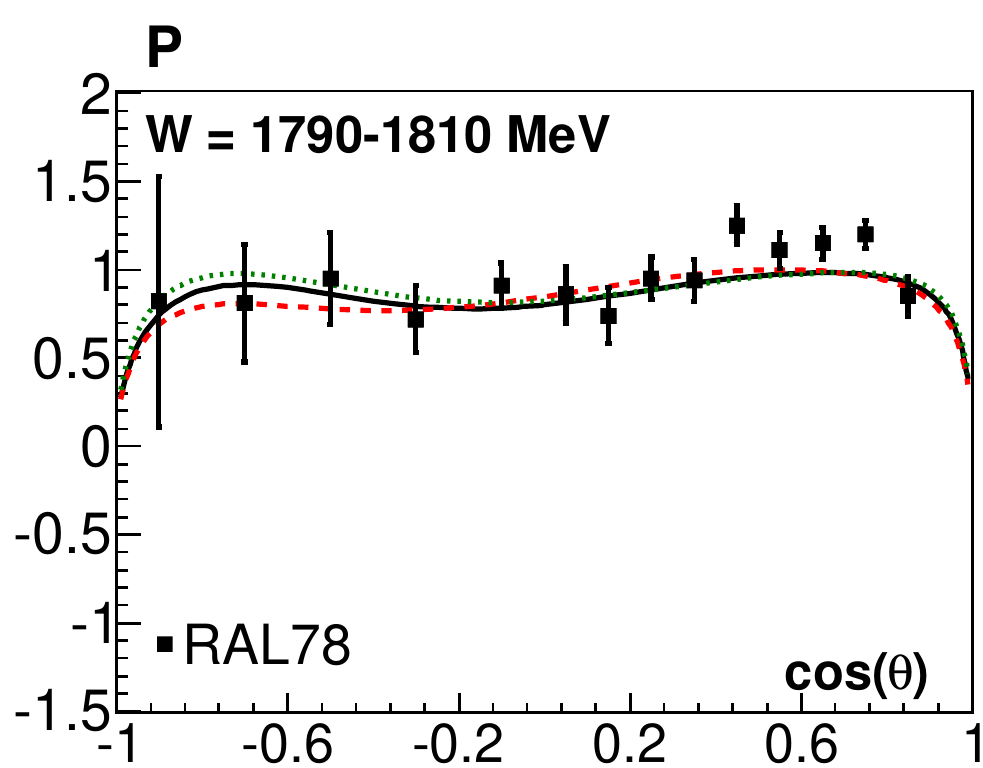}&
\hspace{-4mm}\includegraphics[width=0.195\textwidth,height=0.162\textwidth]{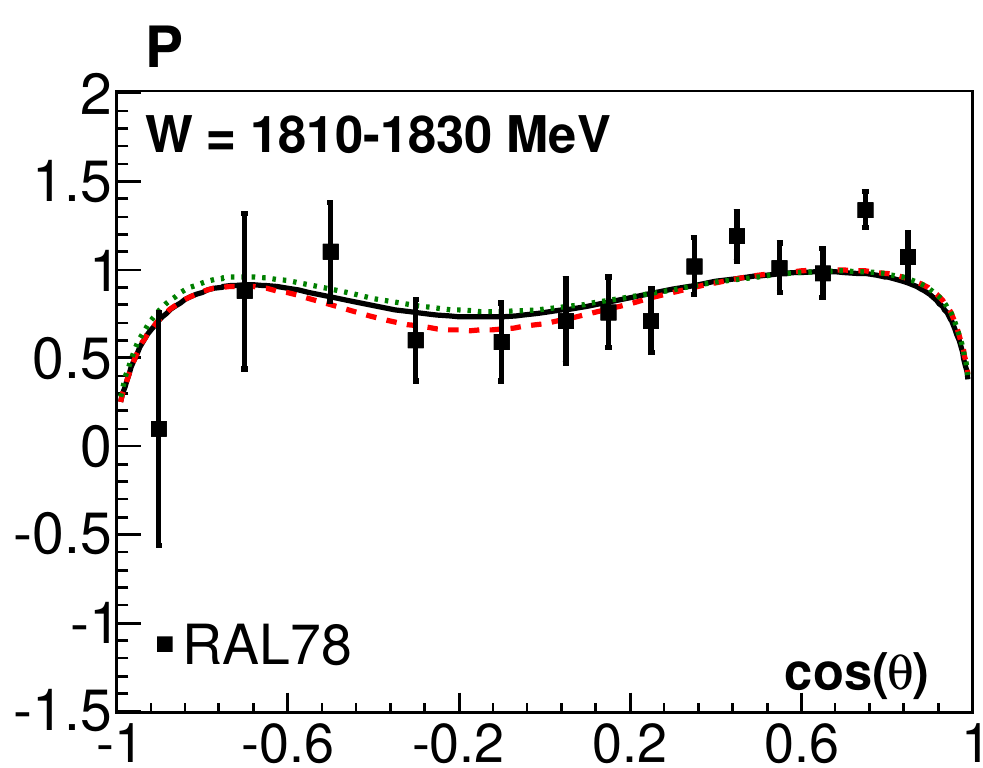}&
\hspace{-4mm}\includegraphics[width=0.195\textwidth,height=0.162\textwidth]{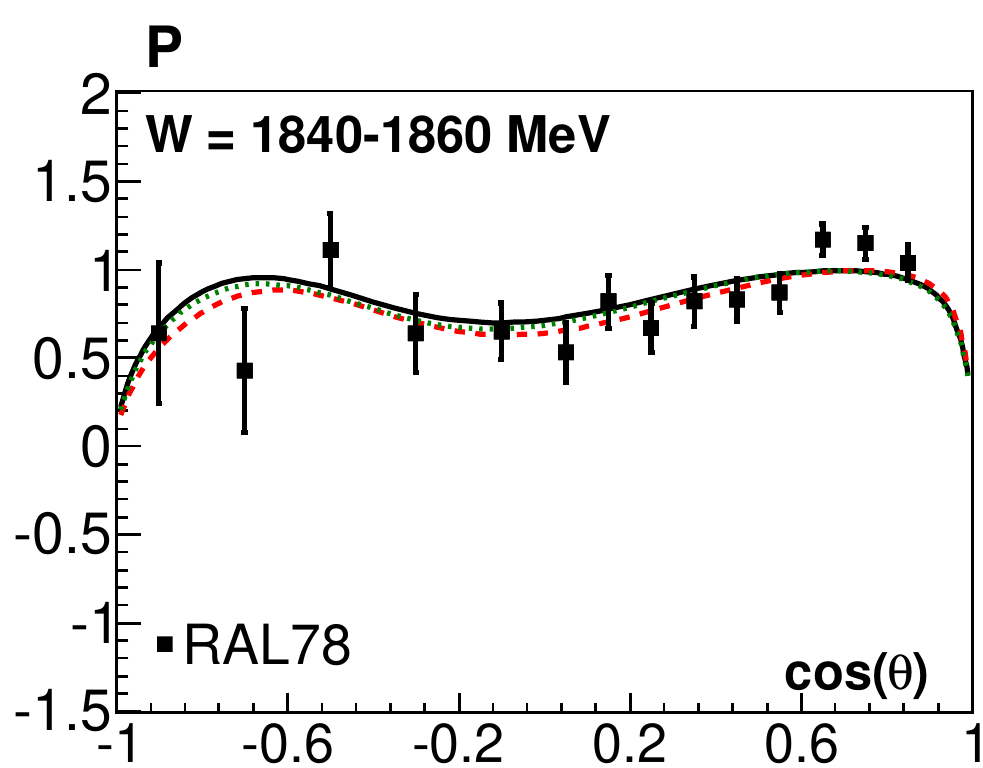}&
\hspace{-4mm}\includegraphics[width=0.195\textwidth,height=0.162\textwidth]{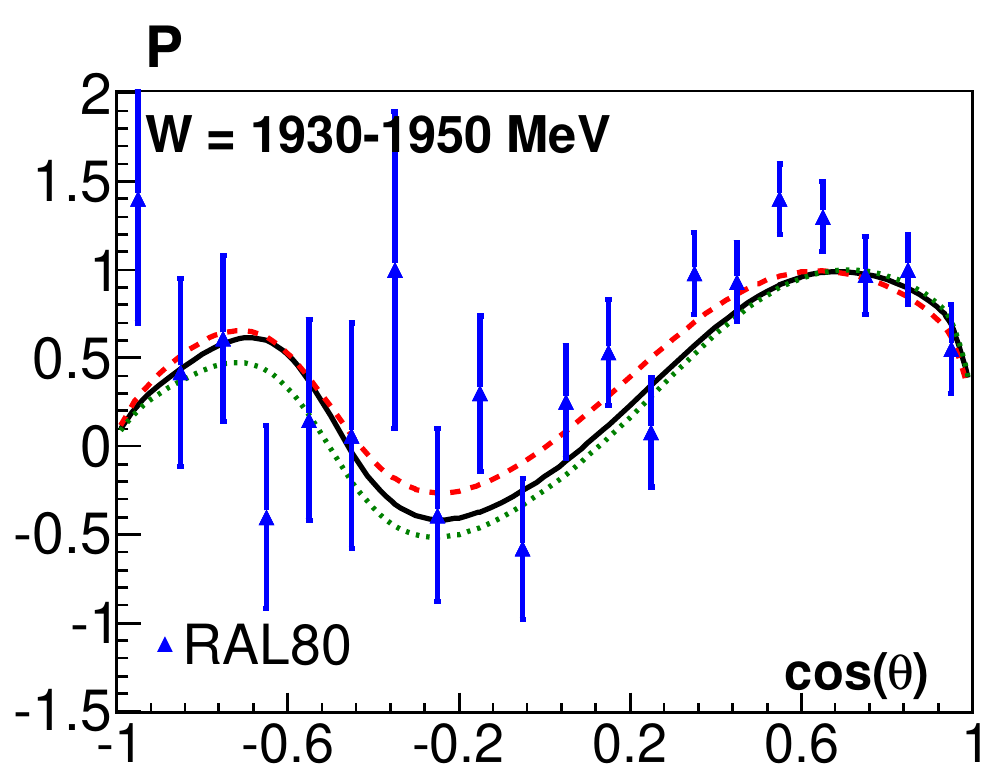}&
\hspace{-4mm}\includegraphics[width=0.195\textwidth,height=0.162\textwidth]{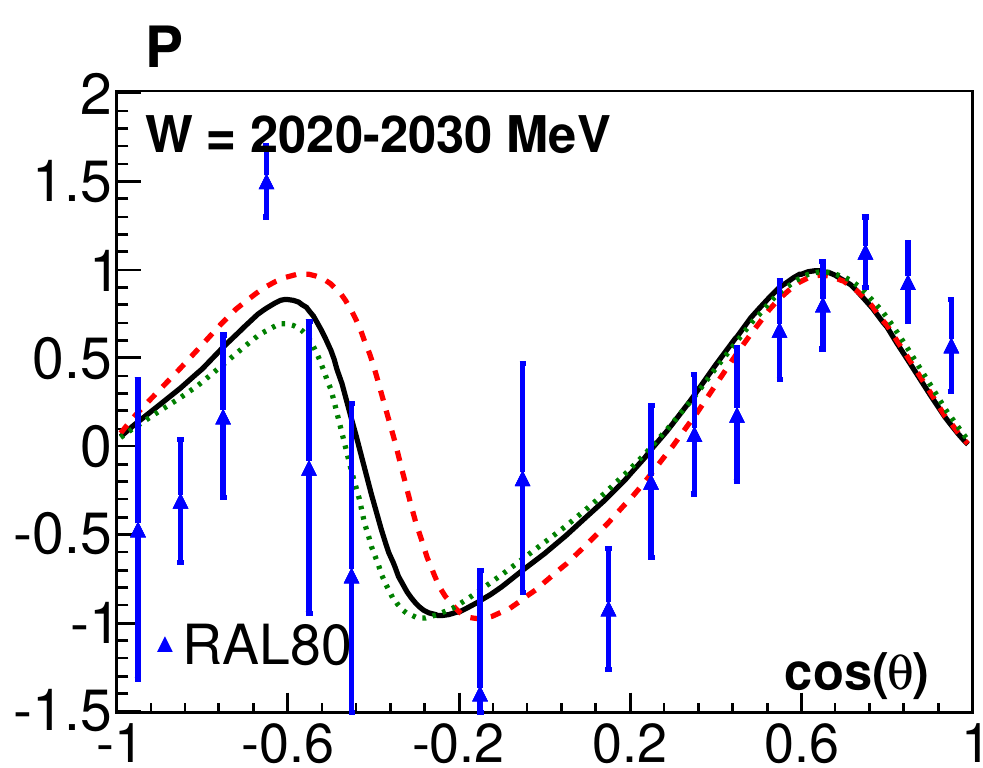}\\
\hspace{-4mm}\includegraphics[width=0.195\textwidth,height=0.162\textwidth]{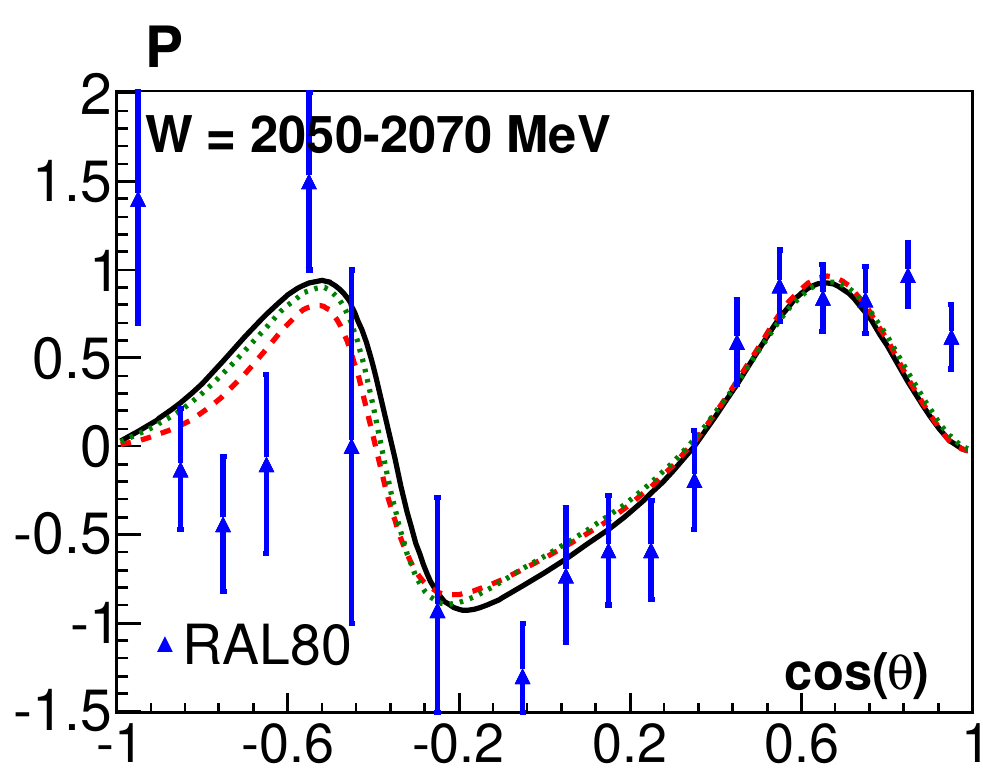}&
\hspace{-4mm}\includegraphics[width=0.195\textwidth,height=0.162\textwidth]{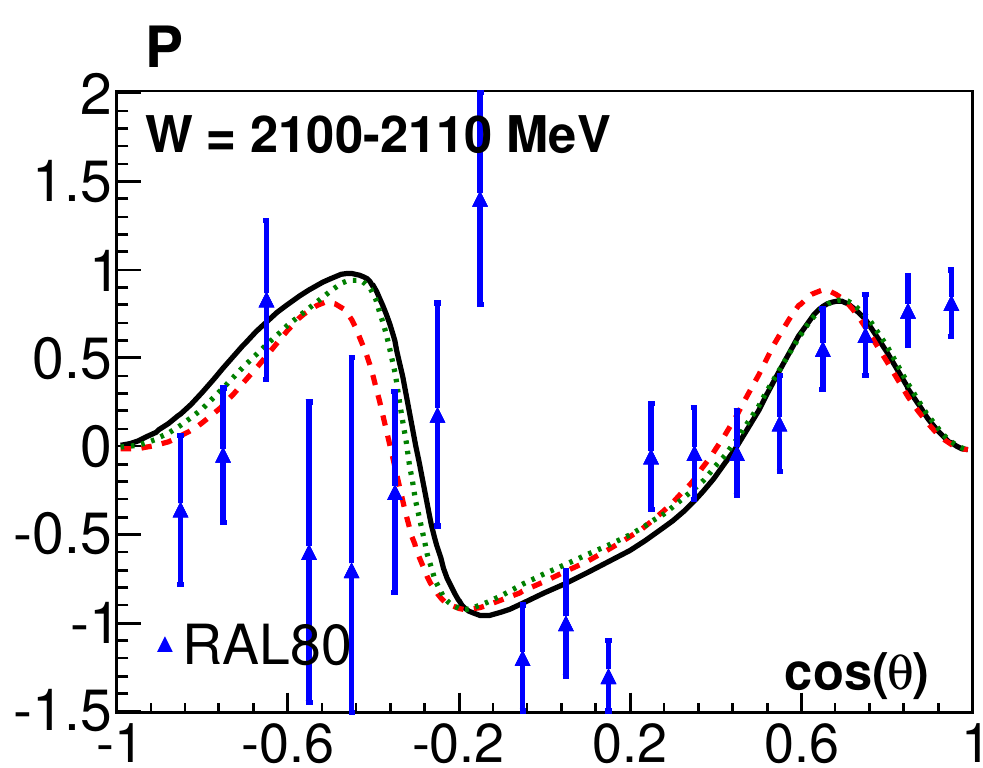}&
\hspace{-4mm}\includegraphics[width=0.195\textwidth,height=0.162\textwidth]{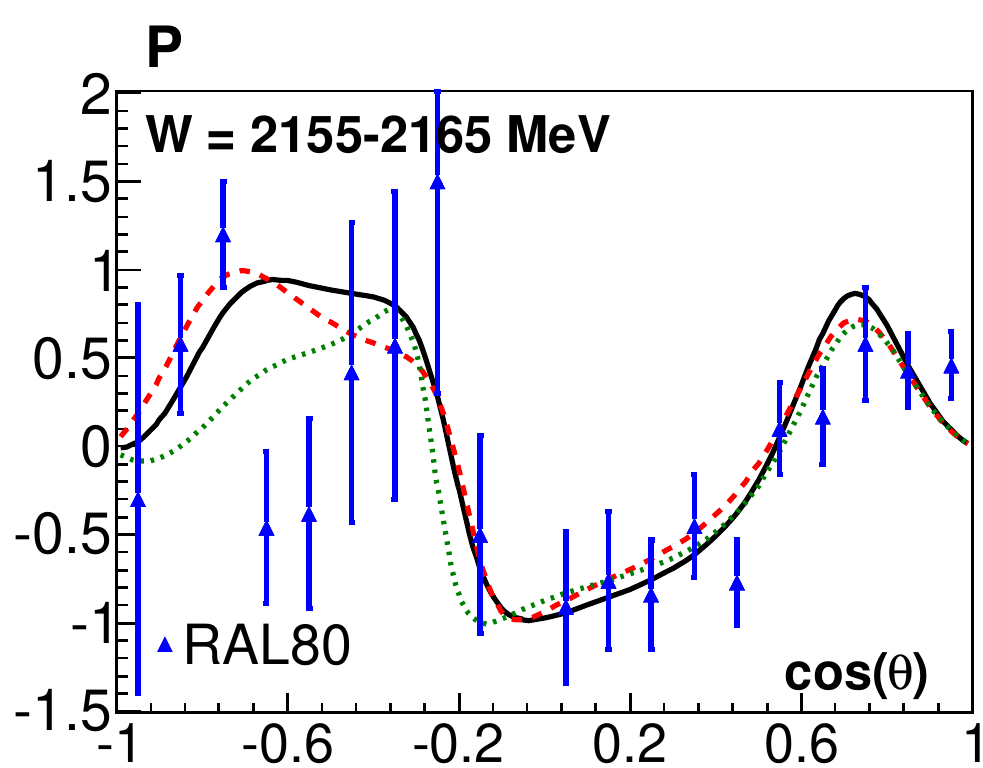}&
\hspace{-4mm}\includegraphics[width=0.195\textwidth,height=0.162\textwidth]{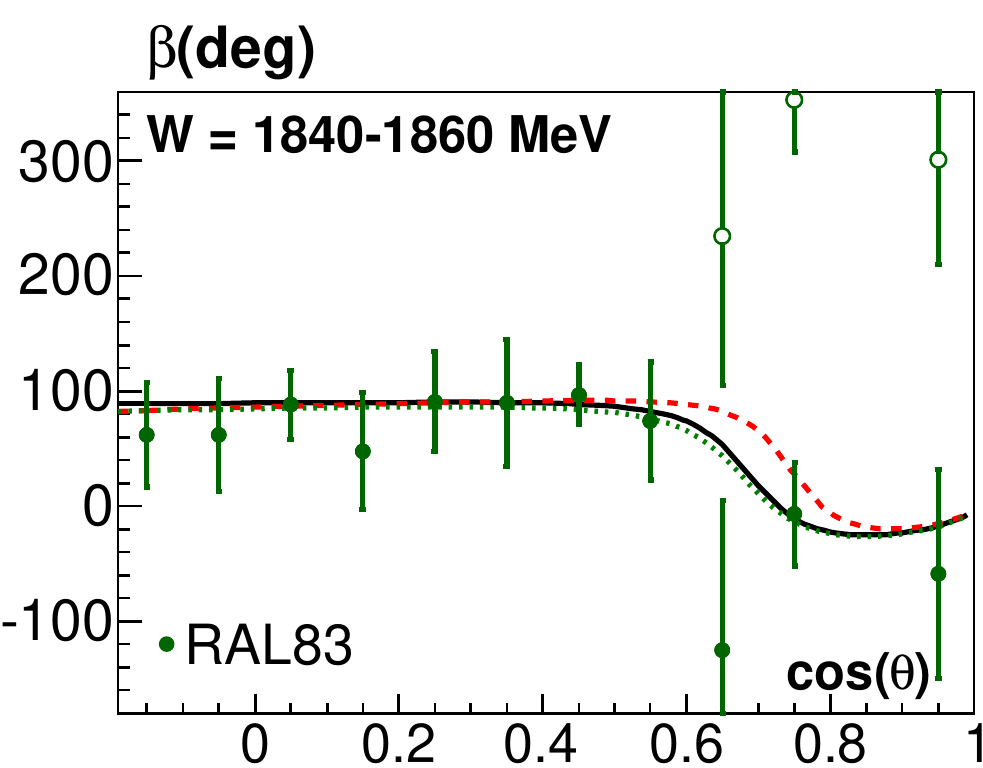}&
\hspace{-4mm}\includegraphics[width=0.195\textwidth,height=0.162\textwidth]{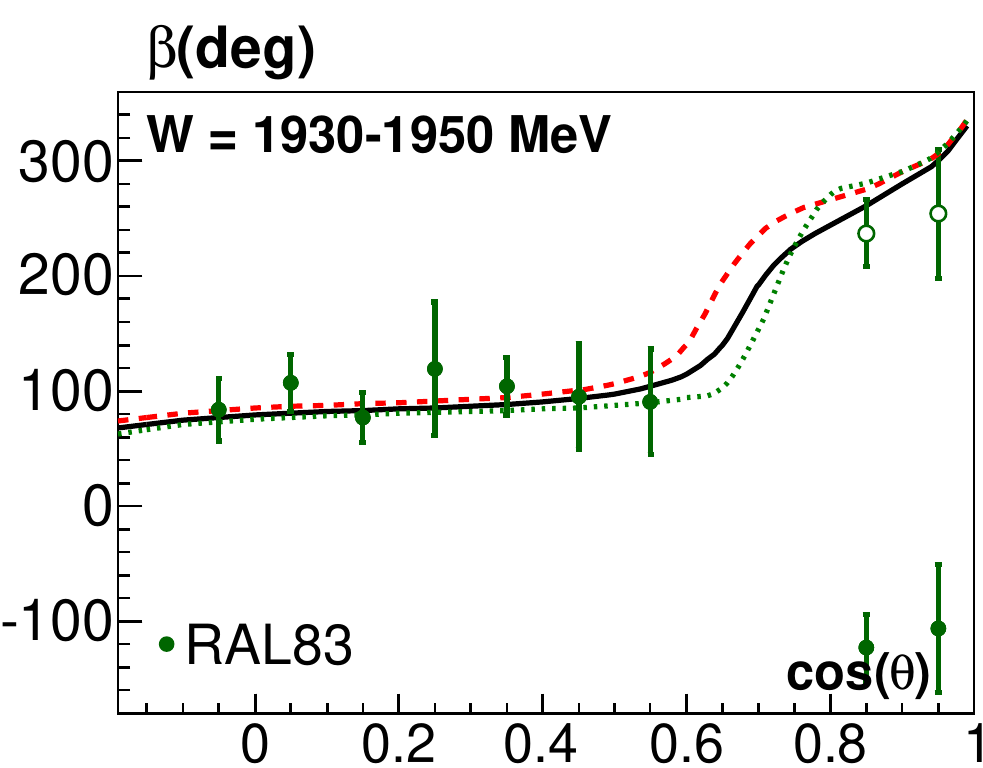}\\
\hspace{-4mm}\includegraphics[width=0.195\textwidth,height=0.162\textwidth]{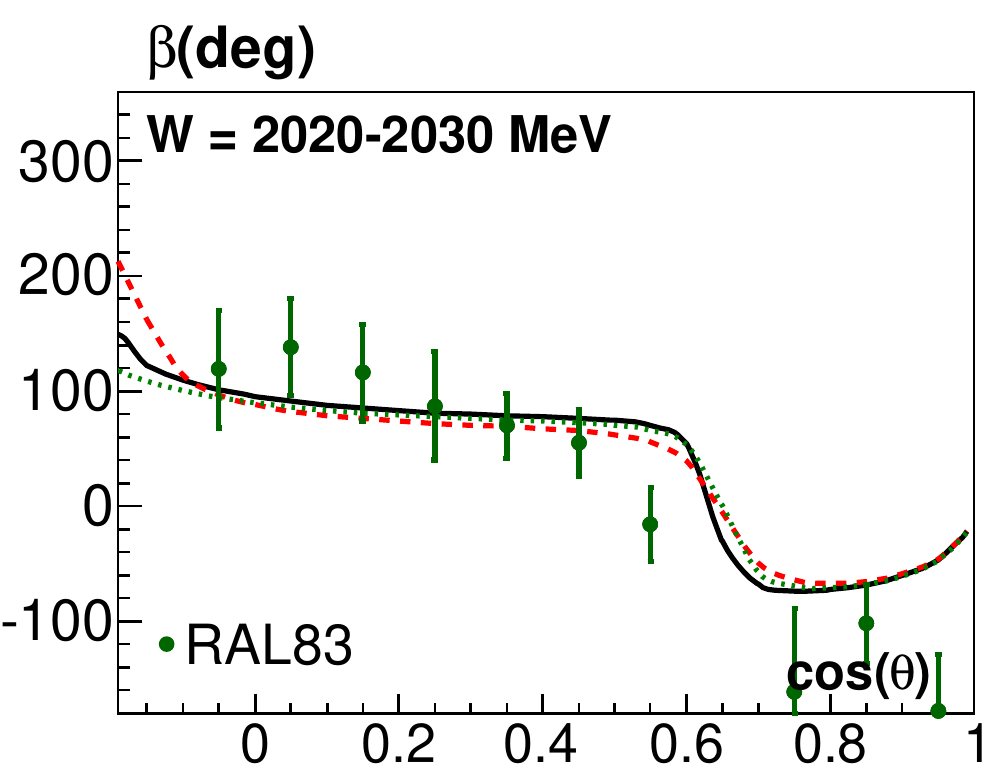}&
\hspace{-4mm}\includegraphics[width=0.195\textwidth,height=0.162\textwidth]{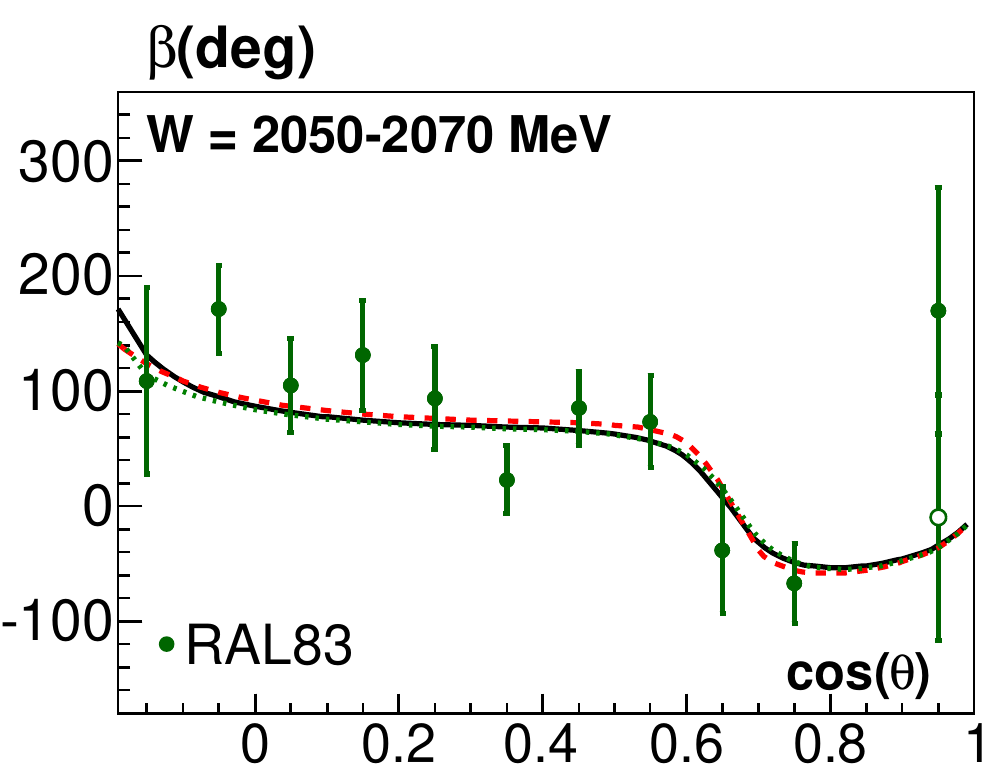}&
\hspace{-4mm}\includegraphics[width=0.195\textwidth,height=0.162\textwidth]{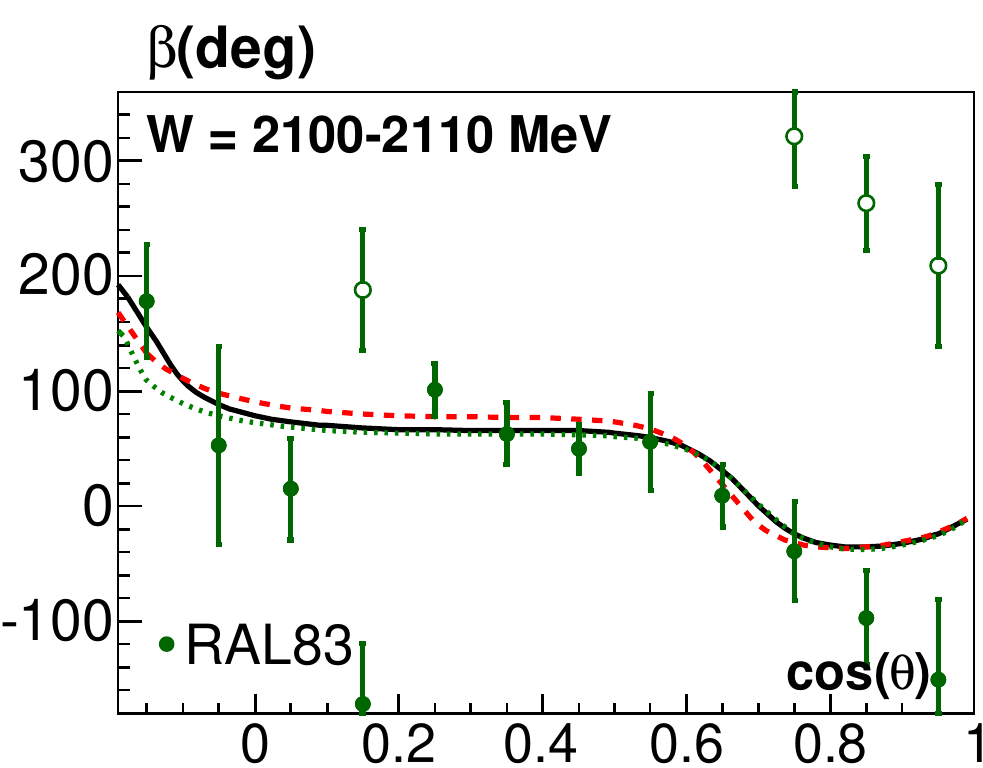}&
\hspace{-4mm}\includegraphics[width=0.195\textwidth,height=0.162\textwidth]{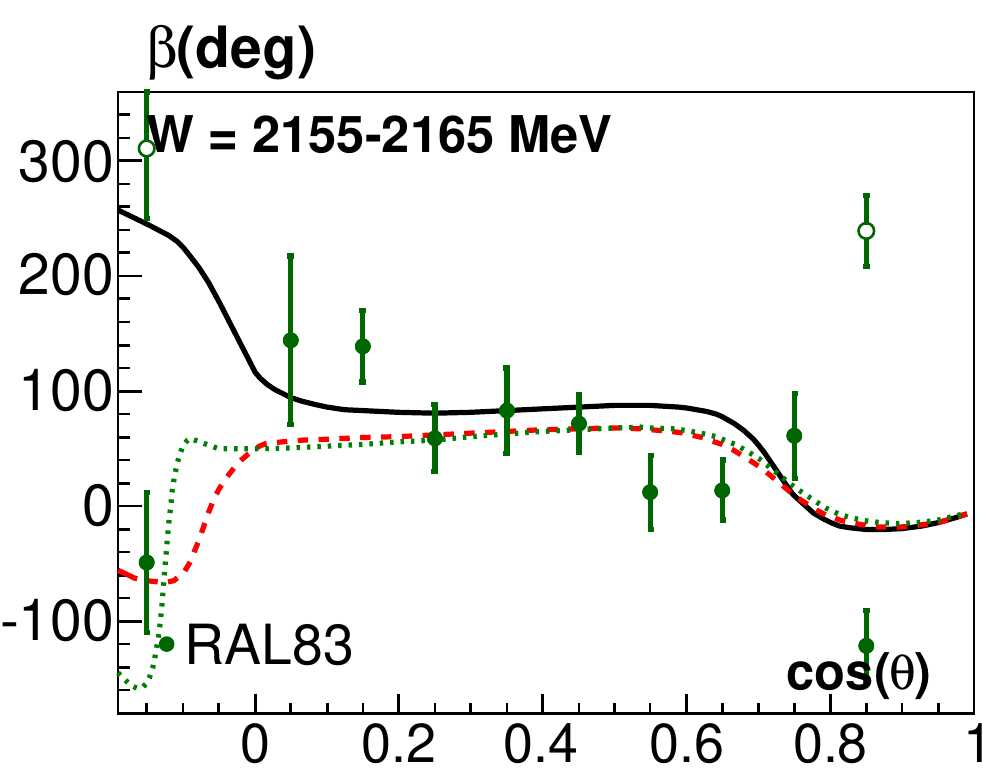}&
\end{tabular}
\end{center}
\caption{\label{pipKLambda_data}
Differential cross sections $d\sigma/d\Omega$, $\Lambda$ recoil polarization $P$, and
spin rotation angle $\beta$ for the reaction $\pi^- p \rightarrow K^0 \Lambda$ from
ANL75 (blue)~\cite{Knasel:1975rr} and RAL (black) \cite{Baker:1978qm,Saxon:1979xu,Bell:1983dm}.
Note that a few differential cross sections from \cite{Knasel:1975rr} fall into a single
energy window. The $\beta$ is 360-degree cyclic which leads to additional data points
shown by empty circles. The solid (black) line corresponds the $L+P$ fit, the dashed (red)
line corresponds the  fit from which the amplitudes of Fig.~\ref{pipKLambda_fit} are
deduced, the dotted (green) line corresponds to BnGa 2011-02 fit.
}
\end{figure*}

A detailed study showed that the data require angular momenta up to $l=3$ or even $l=4$ but
do not have the precision to determine all partial wave amplitudes~\cite{Anisovich:2014yza}.
Therefore we try to determine at least
the low-$l$ amplitudes, in particular $A_1^- (= S_{11})$, $A_0^+(= P_{11})$,
$A_1^+(= P_{13})$,  leading to
$J^P=1/2^-$, $1/2^+$, and $3/2^+$. The higher partial waves,
those above $A{_1}^-$, $A{_0}^+$, $A{_1}^+$, are taken from our
current BnGa fit.

Figure~\ref{pipKLambda_data} shows the data. The solid curves
represent the final BnGa fit. It reproduces the data with a
$\chi^2/N_{\rm data}= 570/916$. The number of free parameters is 75.

The fit returns the real and imaginary parts of amplitudes for the  $S_{11}$, $P_{11}$, and
$P_{13}$ partial waves. The $S_{11}$ and $P_{11}$ amplitudes are shown in Fig.~\ref{pipKLambda_fit},
the $P_{13}$ amplitude in Ref.~\cite{Anisovich:2014yza} only (since it could not be fit with the L+P method).
The solid line represents the L+P fit described below, and the energy-dependent
solution BnGa2011-02 is shown as error band.  Note that the higher
partial waves are constrained fixed to the BnGa solution, while the other lower amplitudes are free to adopt any values.

\subsection{\boldmath The multipoles for $\gamma p\to K^+\Lambda$}
\subsubsection{Formalism}

The amplitude for the reaction $\gamma p\to K^+\Lambda$ can be written in the form
\be
A&=&\omega^*J_\mu\varepsilon_\mu \omega' \;,
\ee
\begin{figure}[pt]
\begin{center}
\begin{tabular}{cc}
\hspace{-0.3cm}\includegraphics[width=0.21\textwidth]{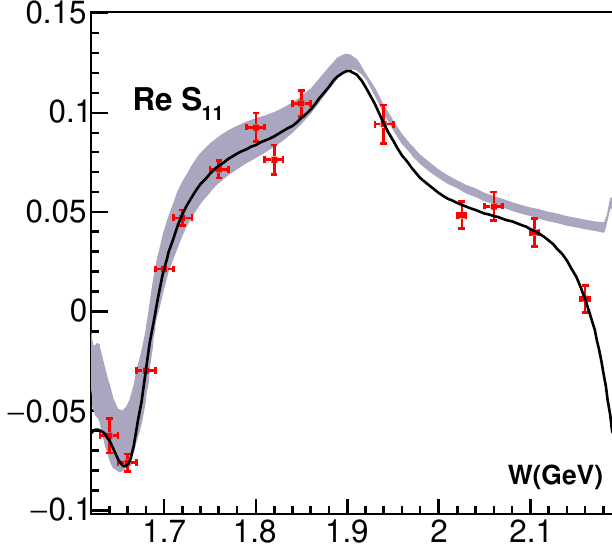}&
\hspace{-0.33cm}\includegraphics[width=0.21\textwidth]{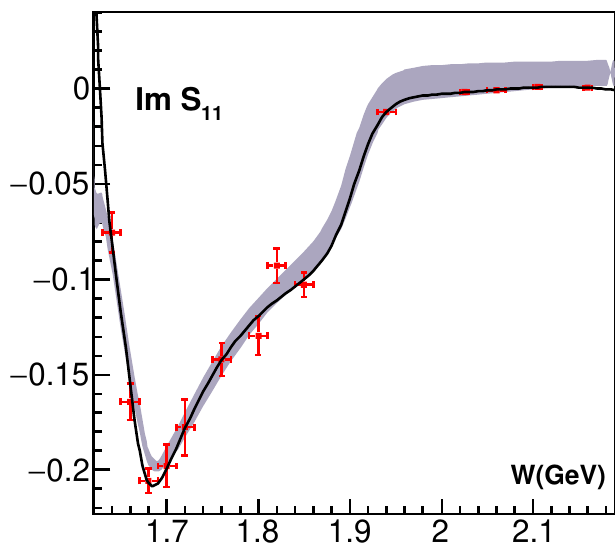}\\
\hspace{-0.3cm}\includegraphics[width=0.21\textwidth]{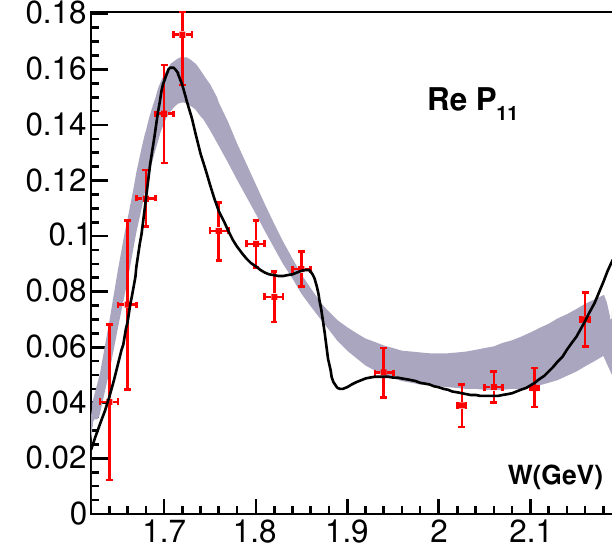}&
\hspace{-0.33cm}\includegraphics[width=0.21\textwidth]{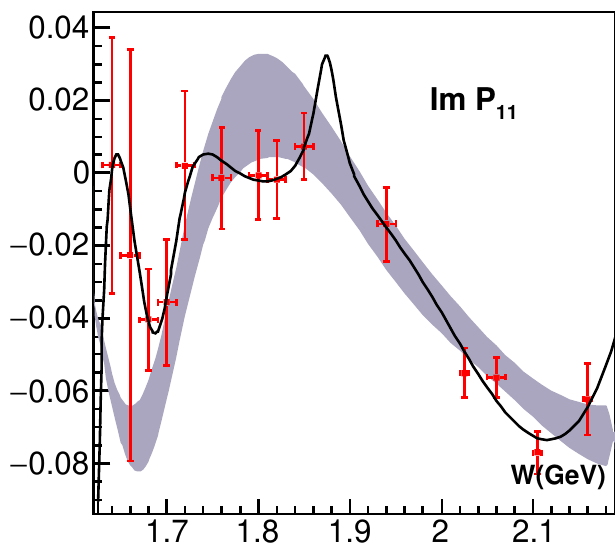}
\end{tabular}
\end{center}
\caption{\label{pipKLambda_fit}
Real and imaginary part of the (dimensionless) $S_{11}$ and $P_{11}$
waves~\cite{Anisovich:2014yza}.
The energy-dependent solution BnGa2011-02 is shown as error band. The band covers a variety of
solutions when the model space was altered. The solid curve represents a L+P fit. The narrow
structure in the $P_{11}$ wave is enforced by the photoproduction data.
 }
\end{figure}
where $\omega'$ and $\omega$ are spinors representing the baryon in the
initial and final state, $J_\mu$ is the electromagnetic current of the nucleon, and $\varepsilon_\mu$ characterizes the polarization of the photon. The amplitude can be expanded into four
invariant (CGLN) amplitudes $\mathcal F_i$~~\cite{Chew:1957tf}
\be
\label{mult_1}
&& J_\mu = \\&& i {\mathcal F_1}
 \sigma_\mu +{\mathcal F_2}
\frac{(\vec \sigma \vec q)} {|\vec k| |\vec q|} \varepsilon_{\mu i j}{\sigma_i k_j}
+i {\mathcal F_3} \frac{(\vec \sigma
\vec k)}{|\vec k| |\vec q|} q_\mu +i {\mathcal F_4} \frac{(\vec \sigma \vec q)}{\vec q^2} q_\mu
\;.\nonumber
\ee
where $\vec q$ is the momentum of the $\Lambda$ hyperon in the final state,
$\vec k$ is the momentum of the nucleon in the initial state,
calculated in  the center-of-mass system of the reaction,
and $\sigma_i$ are the Pauli matrices. These four functions $\mathcal F_i$ are functions of
the invariant mass and of $z$ with $z = (\vec k\vec q)/(|\vec  k||\vec q|) =\cos\theta$
and $\theta$ as the scattering angle.
 A determination of these four amplitudes requires
the measurement with sufficient accuracy of at least eight well chosen observables
\cite{Barker:1975bp,Fasano:1992es,Keaton:1996pe,Chiang:1996em,Sandorfi:2010uv}. For each slice in
energy and angle one phase remains undetermined. It needs to be fixed from other sources.
In $\pi^\pm p$ elastic scattering, the phase can be determined from the (calculable) Coulomb
interference. In hyperon production, one could try to fix the phase to the phase of
$t$-channel Kaon exchange. Once the $\mathcal F_i$ functions are known for each energy
and angle, the results of all experiments can be predicted.

The relations between the $\mathcal F_i$ functions and the observables can be found, e.g.,
in \cite{Sandorfi:2010uv}. For convenience, we give the expressions for the observables used
in the fits. The differential cross section $d\sigma/d\Omega$ and the
single polarization observables, the beam asymmetry $\Sigma$, the recoil asymmetry $P$, and the target asymmetry $T$, are given by \\
\begin{subequations}
\begin{align} \label{singlepol}
 \frac{d\sigma}{d\Omega}&=&\frac{k}{q} I = \frac{k}{q} \Re e[{\mathcal F_1} {\mathcal F_1^*} + {\mathcal F_2} {\mathcal F_2^*} -2 z {\mathcal F_2} {\mathcal F_1^*} + \\&&\hspace{-4mm}
    \hspace{-4mm}\frac{\sin^2(\theta)}{2}
             ({\mathcal F_3} {\mathcal F_3^*} + {\mathcal F_4} {\mathcal F_4^*} + 2 {\mathcal F_4} {\mathcal F_1^*} +
              2 {\mathcal F_3} {\mathcal F_2^*} + 2 z {\mathcal F_4} {\mathcal F_3^*})] \nonumber .\\
  \Sigma\; I &=&
     -\frac{\sin^2(\theta)}{2}\times\\&&\hspace{-4mm} \Re e[{\mathcal F_3} {\mathcal F_3^*} + {\mathcal F_4} {\mathcal F_4^*}+
             2{\mathcal F_4} {\mathcal F_1^*} + 2{\mathcal F_3} {\mathcal F_2^*} +2 z {\mathcal F_4} {\mathcal
             F_3^*}]\,,
             \nonumber \\
  P\;I &=&
         \sin(\theta) \Im m[(2 {\mathcal F_2^*} + {\mathcal F_3^*}+ z {\mathcal F_4^*}) {\mathcal F_1} + \\&&
          {\mathcal F_2^*}(z {\mathcal F_3} + {\mathcal F_4}) +
                          \sin^2(\theta) {\mathcal F_3^*} {\mathcal F_4}]\,, \nonumber\\
  T\;I &=&
           \sin(\theta)\Im m[{\mathcal F_1^*} {\mathcal F_3} - {\mathcal F_2^*} {\mathcal F_4} + \\&&
            z({\mathcal F_1^*} {\mathcal F_4} - {\mathcal F_2^*} {\mathcal F_3}) -
                          \sin^2(\theta) {\mathcal F_3^*} {\mathcal F_4}]\,, \nonumber
\end{align}
\begin{subequations}

The double polarization observables $O_{x'}$, $O_{z'}$ ($C_{x'}$, $C_{z'}$) define the spin transfer
from linearly (circularly) polarized photons to the $\Lambda$ hyperon
where the $z'$ axis is given by the meson direction. This is referred to as the {\it primed} frame. Experimentally, the data on the spin transfer
from polarized photons to the $\Lambda$ hyperon are sometimes presented in an {\it unprimed} frame, in which the photon momentum
is chosen as reference axis. Observables in the two frames are related by a simple rotation:
\be
 C_{x} &=& \sin(\theta) C_{z'} + \cos(\theta) C_{x'}\,,\nonumber\\
 C_{z} &=& \cos(\theta) C_{z'} - \sin(\theta) C_{x'}\,,\nonumber,
\ee
with similar relations holding for the quantities $O_{x'}$ and $O_{z'}$.

The double polarization observables $O_{x'}$, $O_{z'}$ ($C_{x'}$, $C_{z'}$)
can be written as
\end{subequations}
\begin{align} \label{twopol}
  O_{x'}\;I &=&\\
           &&\hspace{-4mm}\nonumber\sin(\theta) \Im m[{\mathcal F_2} {\mathcal F_3^*} - {\mathcal F_1} {\mathcal F_4^*}
                                         + z ({\mathcal F_2} {\mathcal F_4^*} - {\mathcal F_1} {\mathcal
                                         F_3^*})]\,,   \\
  O_{z'}\;I &=&
           -\sin^2(\theta) \Im m[{\mathcal F_1} {\mathcal F_3^*} + {\mathcal F_2} {\mathcal F_4^*}]\,,\\
  C_{x'}\;I &=& \sin(\theta) \Re e[{\mathcal F_2} {\mathcal F_2^*} - {\mathcal F_1} {\mathcal F_1^*} +
   {\mathcal F_2} {\mathcal F_3^*} - {\mathcal F_1} {\mathcal F_4^*}
                             + \nonumber \\&&
                             z ({\mathcal F_2} {\mathcal F_4^*} - {\mathcal F_1} {\mathcal F_3^*})]\,, \\
  C_{z'}\;I & =&  \Re e[-2 {\mathcal F_1} {\mathcal F_2^*} + z ({\mathcal F_1} {\mathcal F_1^*} +
   {\mathcal F_2} {\mathcal F_2^*})- \nonumber \\&&
                              \sin^2(\theta) ({\mathcal F_1} {\mathcal F_3^*} + {\mathcal F_2} {\mathcal
                              F_4^*})]\,.
\end{align}
\end{subequations}

When the ${\mathcal F_i}$ are known with sufficient statistical accuracy they can be
expanded -- for each slice in energy -- into power series using Legendre polynomials
$P_{L}(z)$ and their derivatives:
\setcounter{equation}{9} 
\begin{subequations}\label{Legendreexpansion}
\begin{align}
&{\mathcal F_1}(W,z) &= \nonumber
 \sum^{\infty}_{L=0} [LM_{L+}+E_{L+}]
P^{\prime}_{L+1}(z) + \\&& [(L+1)M_{L-} + E_{L-}] P^{\prime}_{L-1}(z) \;,  \\& {\mathcal
F_2}(W,z)& = \sum^{\infty}_{L=1} [(L+1)M_{L+}+LM_{L-}] P^{\prime}_{L}(z)   \;, \\&
{\mathcal F_3}(W,z)& =  \sum^{\infty}_{L=1} [E_{L+}-M_{L+}] P^{\prime\prime}_{L+1}(z) +
\nonumber \\&&[E_{L-}
+ M_{L-}] P^{\prime\prime}_{L-1}(z)\;, \\& {\mathcal
 F_4} (W,z) &= \sum^{\infty}_{L=2} [M_{L+}
- E_{L+} - M_{L-} -E_{L-}] P^{\prime\prime}_{L}(z).\nonumber \\[-3ex]
\label{mult_2}
\end{align}
\end{subequations}
Here, $L$ corresponds to the orbital angular momentum in the
$ K^+\Lambda$ system,  $W$ is the total energy, $P_L(z)$ are
Legendre polynomials with ,
and $E_{L\pm}$ and $M_{L\pm}$ are electric and magnetic multipoles
describing transitions to states with $J=L\pm 1/2$.
$M_{0+}$ or $E_{1-}$ multipoles do not exist. Processes due to meson exchanges in the $t$ channel
may provide significant contributions to the reaction. They may demand high-order multipoles.
The minmal $L$ required to describe the data can be determined by polynomial expansions
of the data~\cite{Wunderlich:2016imj}.
A more direct approach is to insert the ${\mathcal F_i}$ functions (eqns. \ref{Legendreexpansion})
into the expressions for the observables (eqns. \ref{singlepol} and \ref{twopol}) and
to truncate the expansion at an appropriate value of $L$ \cite{Wunderlich:2014xya}. The
observables are now functions of the invariant mass and the scattering angle, and the fit parameters
are the electric and magnetic multipoles. In this method, the number of observables required to
get the full information might be reduced if the number of contributing higher partial waves is not too big. But still, high precision is mandatory for the expansion.
\begin{figure}[pt]
\begin{center}
\begin{tabular}{cc}
\hspace{-2mm}\includegraphics[width=0.25\textwidth]{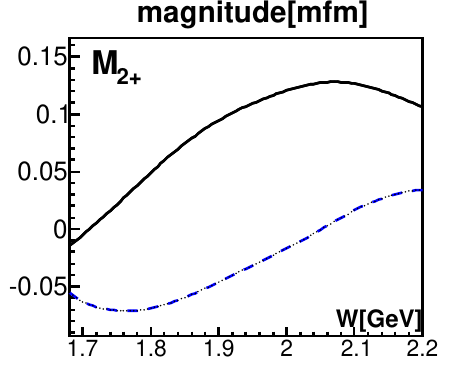}&
\hspace{-6mm}\includegraphics[width=0.25\textwidth]{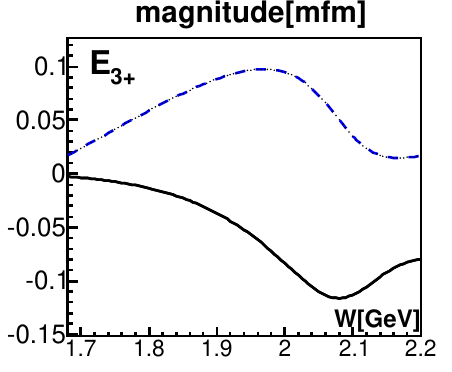}\\
\hspace{-2mm}\includegraphics[width=0.25\textwidth]{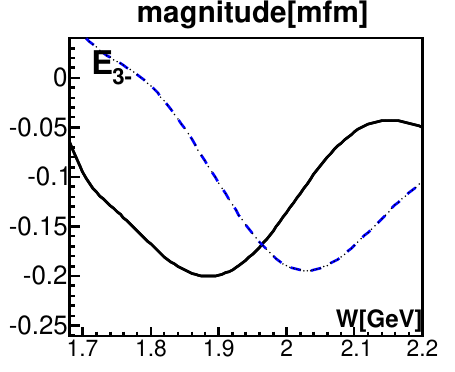}&
\hspace{-6mm}\includegraphics[width=0.25\textwidth]{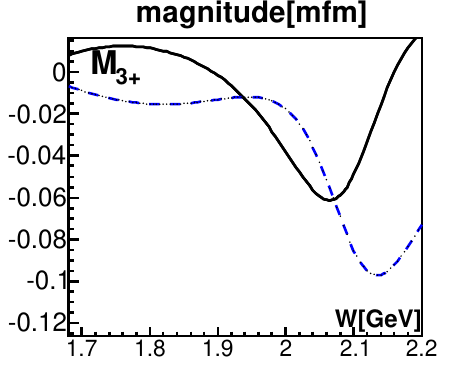}\\
\hspace{-2mm}\includegraphics[width=0.25\textwidth]{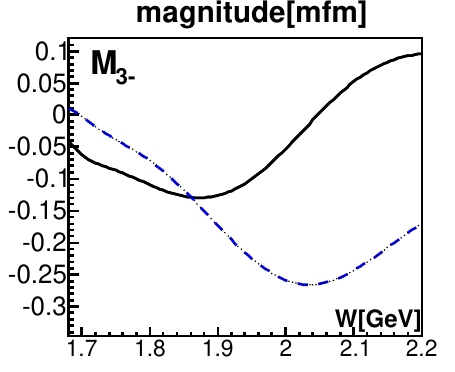}&
\hspace{-6mm}\includegraphics[width=0.25\textwidth]{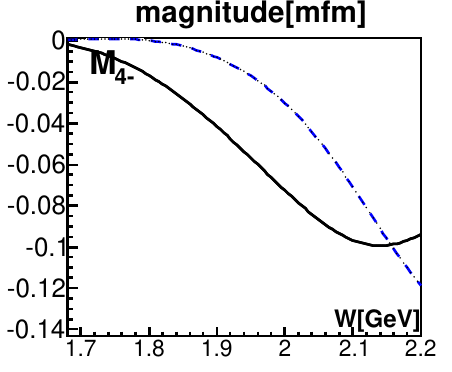}\\
\hspace{-2mm}\includegraphics[width=0.25\textwidth]{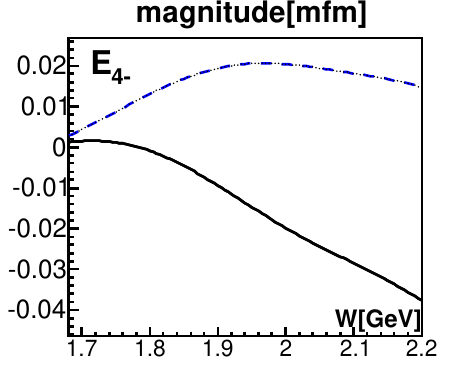}&
\hspace{-6mm}\includegraphics[width=0.25\textwidth]{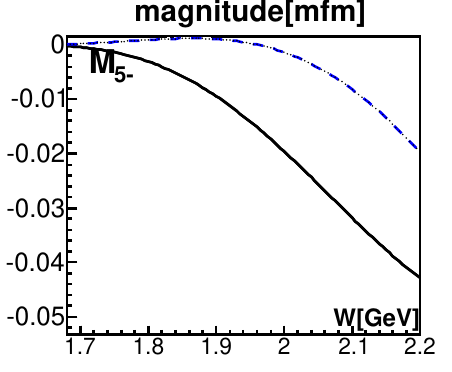}
\end{tabular}
\end{center}
\caption{\label{fig:high-l}(Color online) High-$L$ multipoles from BnGa
fits to a large body of pion and photo-induced reactions. These multipoles
are imposed in the fits to the data Figs.~\ref{fig:data-g1}-\ref{fig:data-g4}.
Show real and imaginary part of
$M_{2+}$, $M_{3-}$, $E_{3-}$, $M_{3+}$, $E_{3+}$,  $M_{4-}$, $E_{4-}$, $M_{5-}$.
The solid (black) line is the real part of the multipole,
dash (blue) line is the imaginary part of the multipole. The multipoles are given in mfm
(milli-fermi=attometer).
}
\end{figure}

\begin{figure*}[pt]
\begin{tabular}{cccccc}
\hspace{-3mm}\includegraphics[width=0.165\textwidth,height=0.17\textwidth]{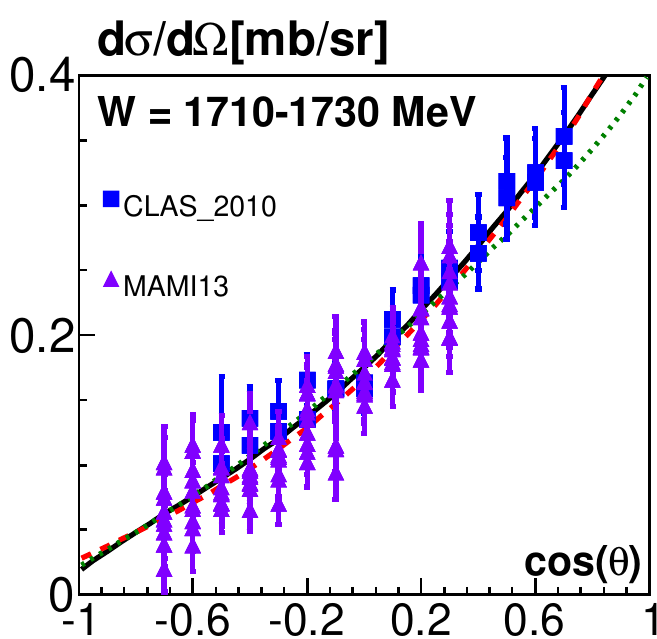}&
\hspace{-4mm}\includegraphics[width=0.165\textwidth,height=0.17\textwidth]{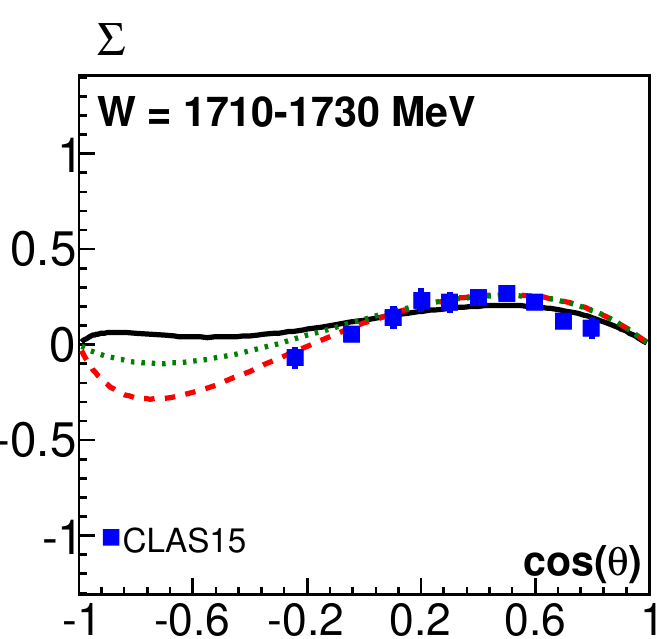}&
\hspace{-4mm}\includegraphics[width=0.165\textwidth,height=0.17\textwidth]{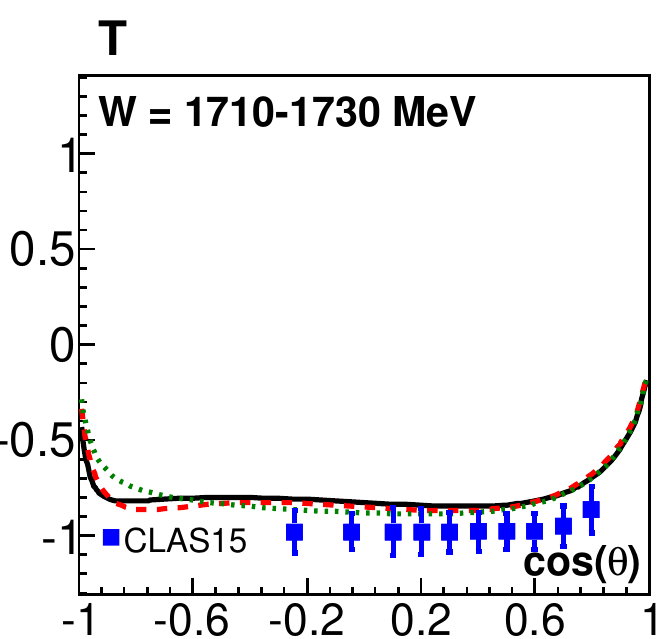}&
\hspace{-4mm}\includegraphics[width=0.165\textwidth,height=0.17\textwidth]{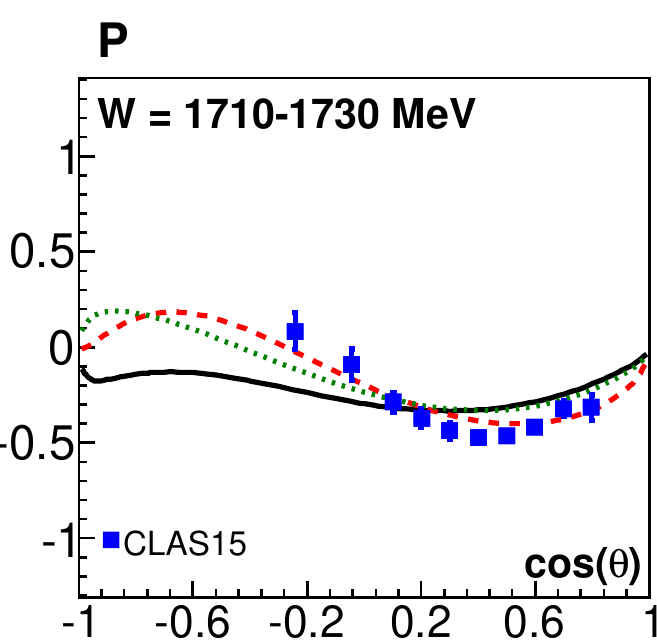}&
\hspace{-4mm}\includegraphics[width=0.165\textwidth,height=0.17\textwidth]{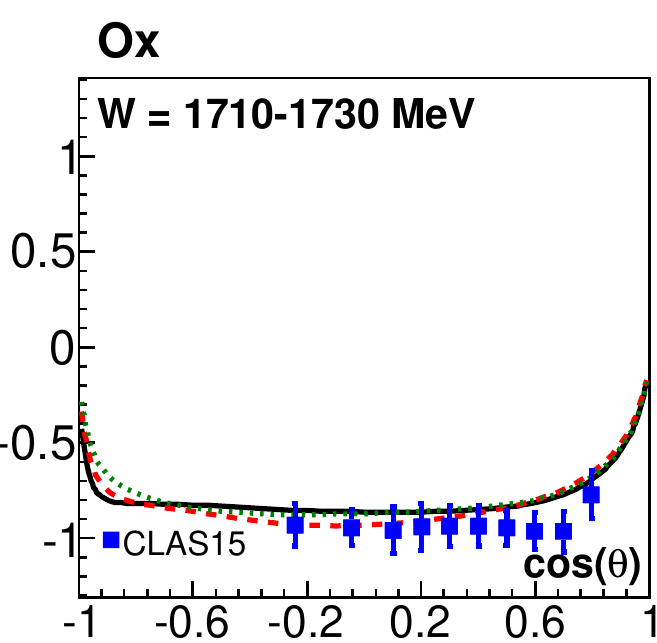}&
\hspace{-4mm}\includegraphics[width=0.165\textwidth,height=0.17\textwidth]{Ox_no_prime_observables1720.pdf}\\
\hspace{-3mm}\includegraphics[width=0.165\textwidth,height=0.17\textwidth]{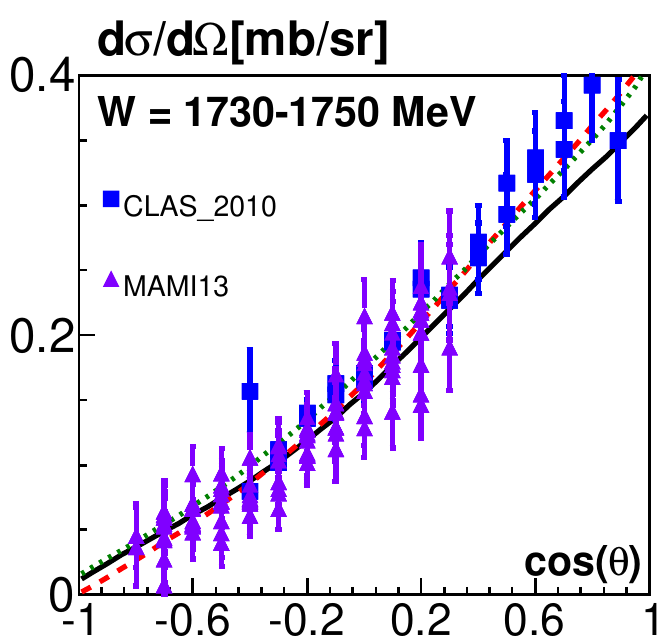}&
\hspace{-4mm}\includegraphics[width=0.165\textwidth,height=0.17\textwidth]{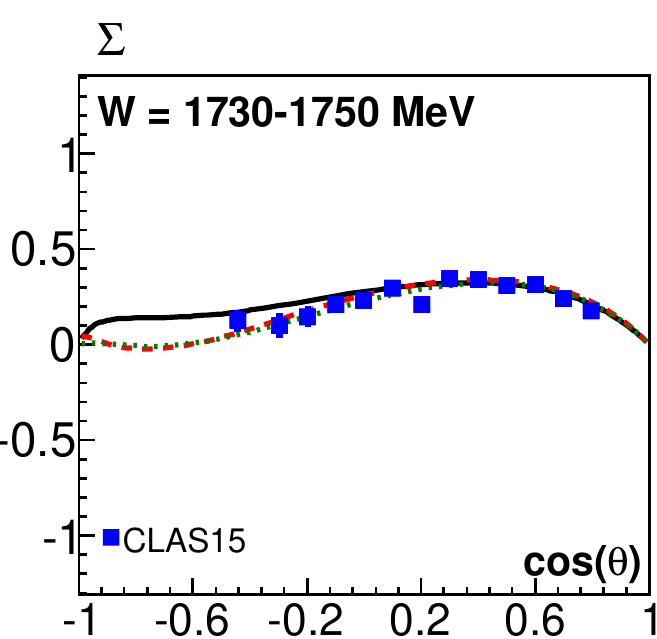}&
\hspace{-4mm}\includegraphics[width=0.165\textwidth,height=0.17\textwidth]{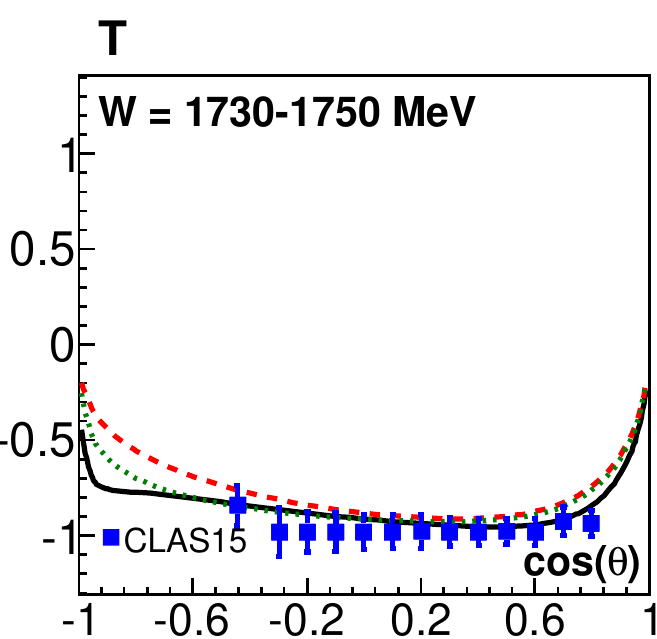}&
\hspace{-4mm}\includegraphics[width=0.165\textwidth,height=0.17\textwidth]{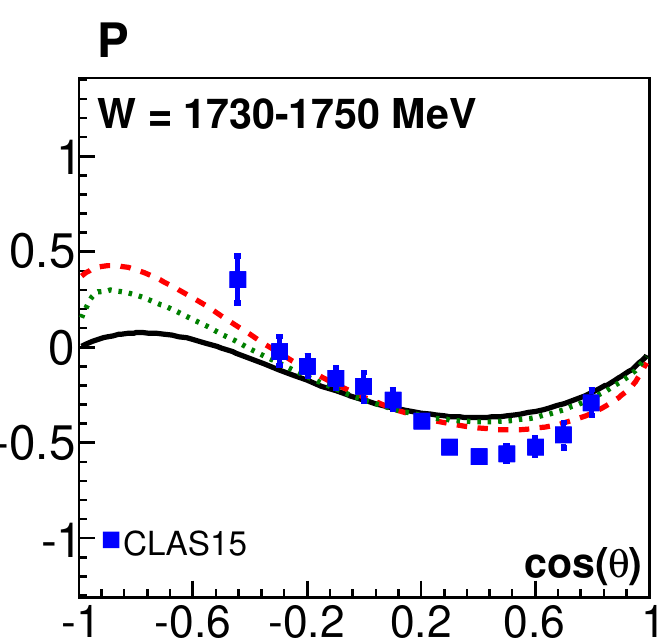}&
\hspace{-4mm}\includegraphics[width=0.165\textwidth,height=0.17\textwidth]{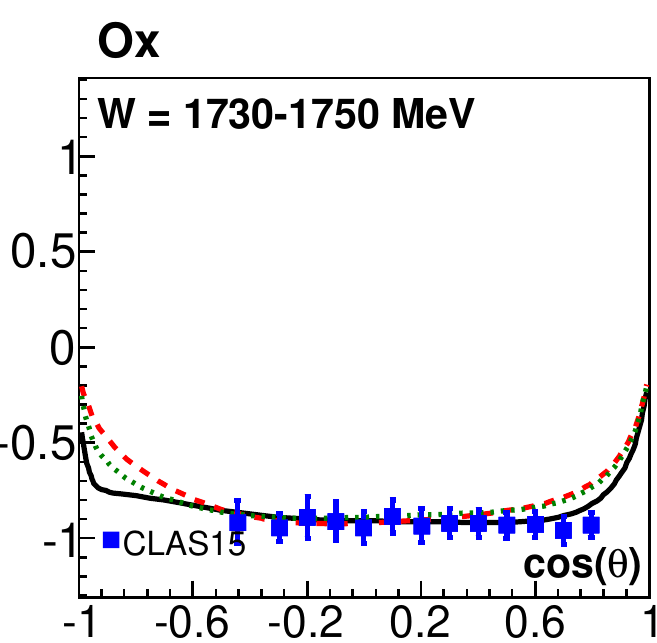}&
\hspace{-4mm}\includegraphics[width=0.165\textwidth,height=0.17\textwidth]{Ox_no_prime_observables1740.pdf}\\
\hspace{-3mm}\includegraphics[width=0.165\textwidth,height=0.17\textwidth]{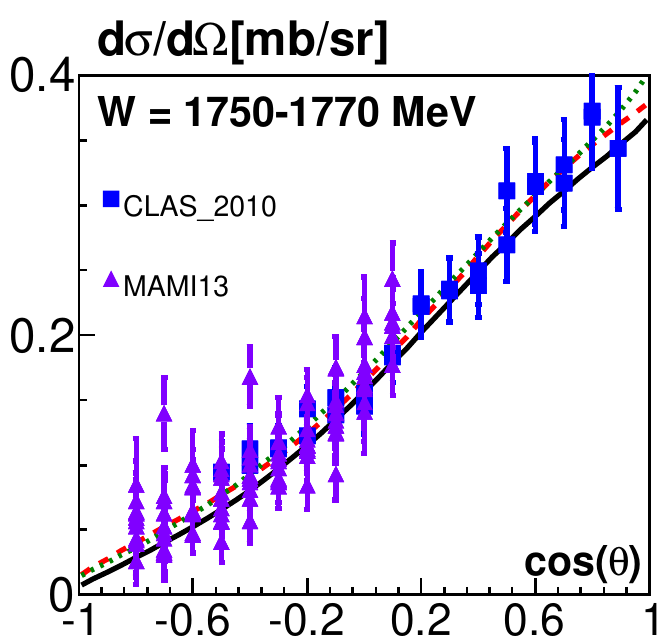}&
\hspace{-4mm}\includegraphics[width=0.165\textwidth,height=0.17\textwidth]{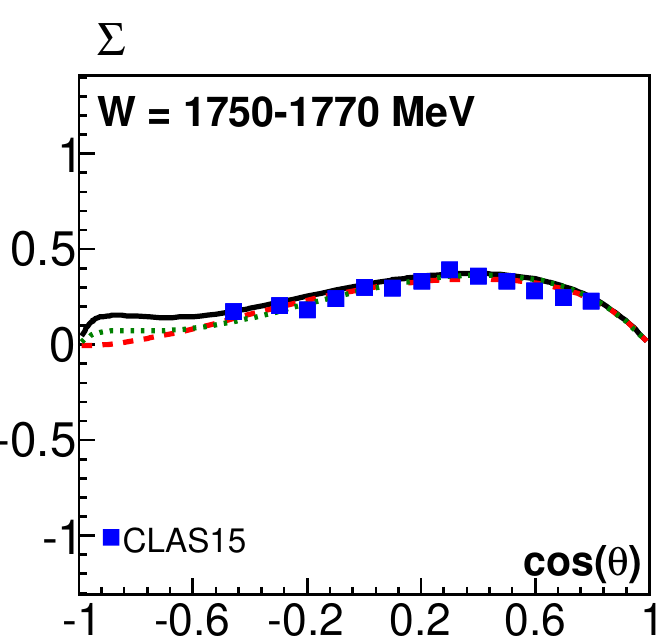}&
\hspace{-4mm}\includegraphics[width=0.165\textwidth,height=0.17\textwidth]{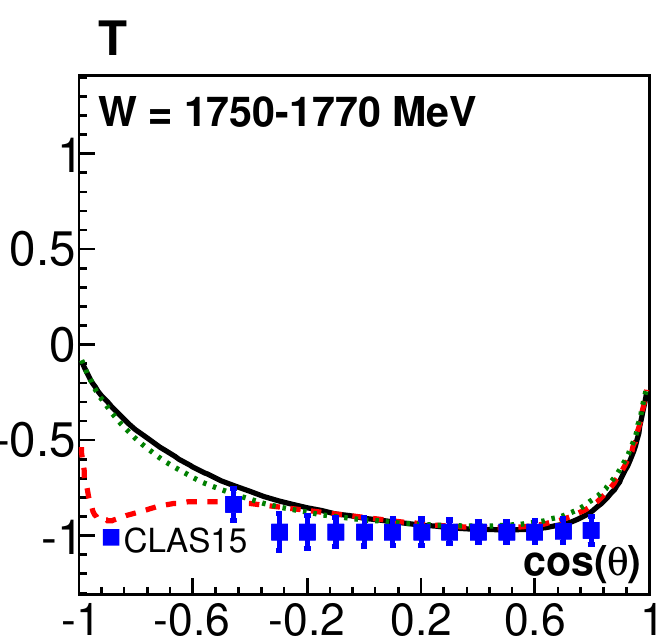}&
\hspace{-4mm}\includegraphics[width=0.165\textwidth,height=0.17\textwidth]{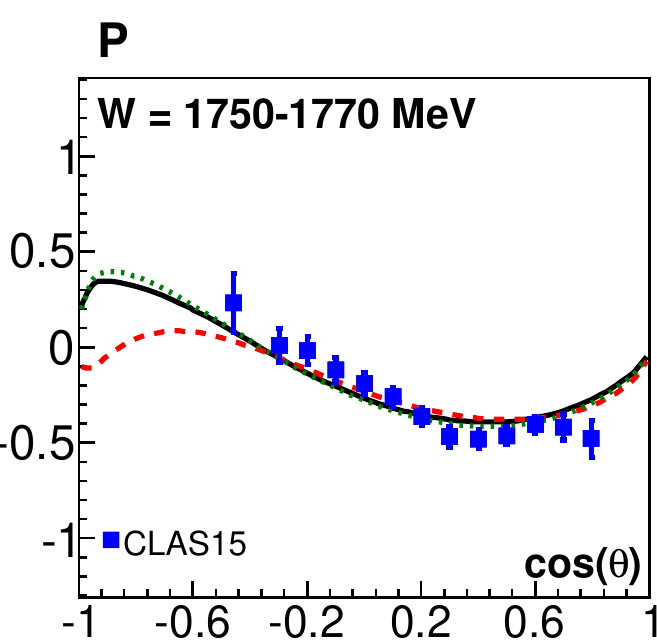}&
\hspace{-4mm}\includegraphics[width=0.165\textwidth,height=0.17\textwidth]{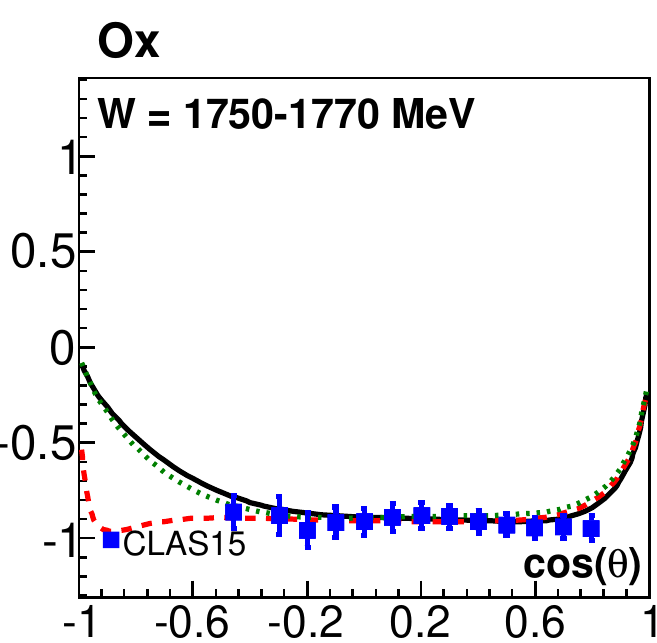}&
\hspace{-4mm}\includegraphics[width=0.165\textwidth,height=0.17\textwidth]{Ox_no_prime_observables1760.pdf}\\
\hspace{-3mm}\includegraphics[width=0.165\textwidth,height=0.17\textwidth]{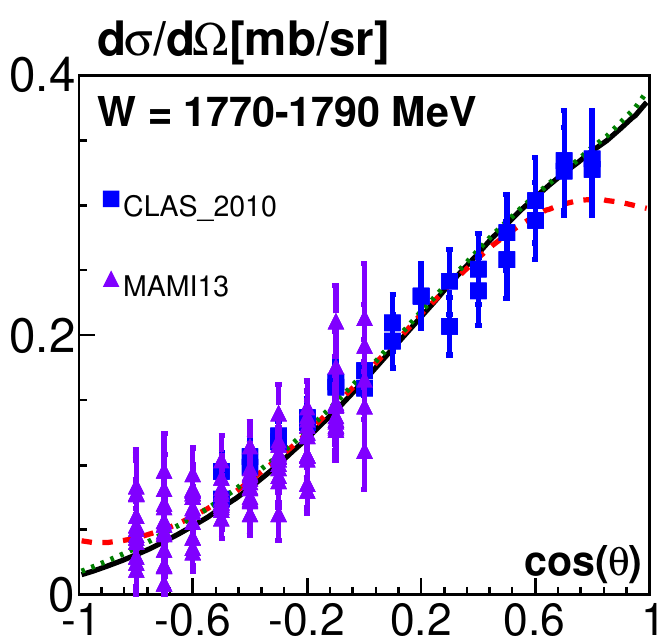}&
\hspace{-4mm}\includegraphics[width=0.165\textwidth,height=0.17\textwidth]{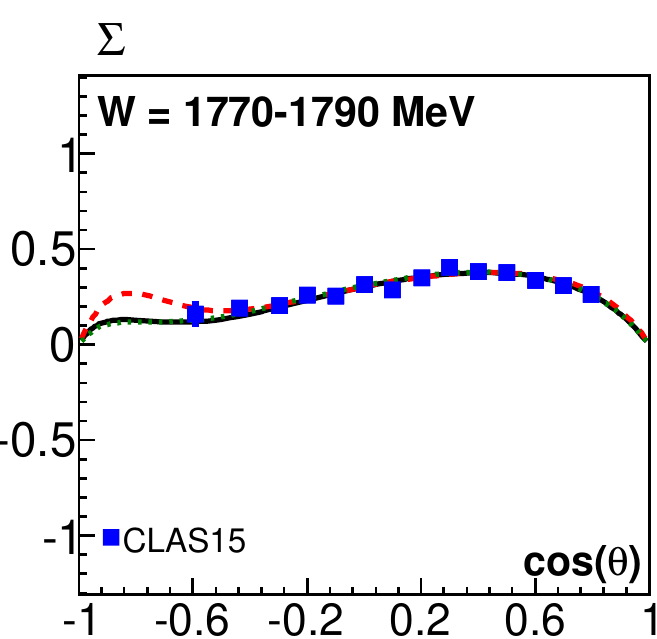}&
\hspace{-4mm}\includegraphics[width=0.165\textwidth,height=0.17\textwidth]{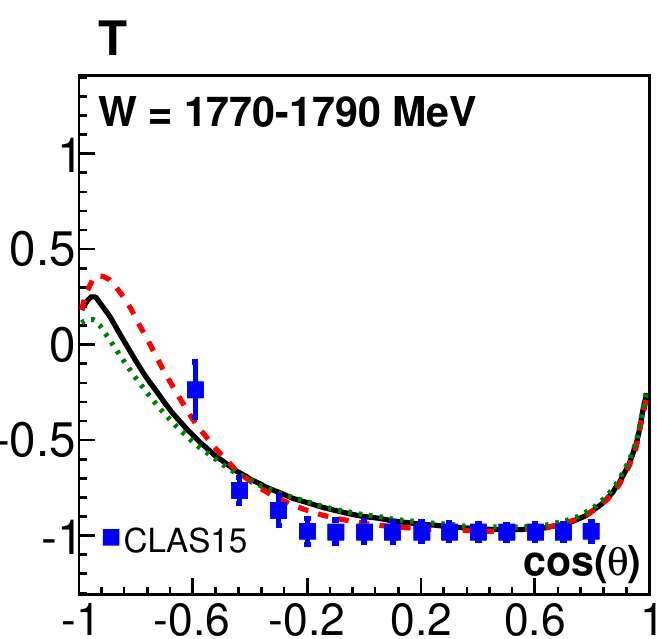}&
\hspace{-4mm}\includegraphics[width=0.165\textwidth,height=0.17\textwidth]{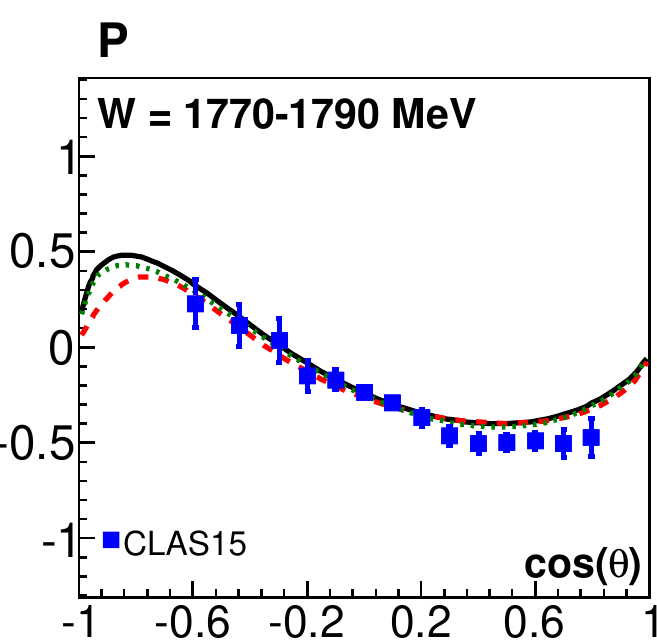}&
\hspace{-4mm}\includegraphics[width=0.165\textwidth,height=0.17\textwidth]{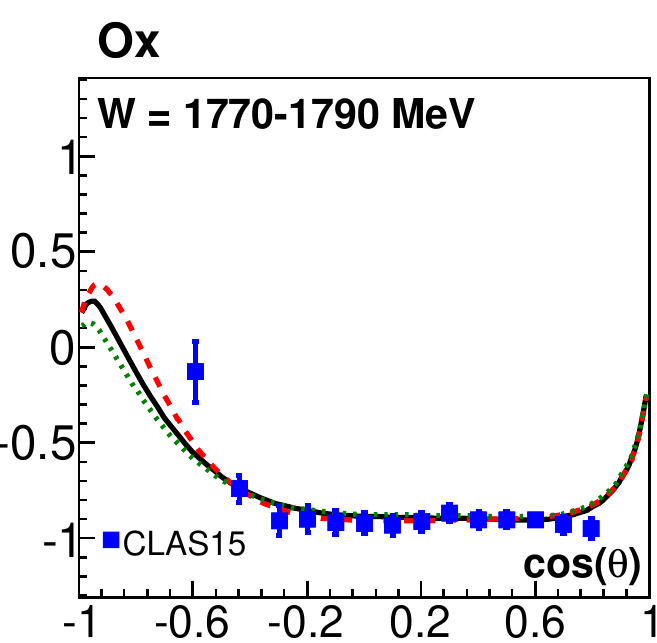}&
\hspace{-4mm}\includegraphics[width=0.165\textwidth,height=0.17\textwidth]{Ox_no_prime_observables1780.pdf}\\
\hspace{-3mm}\includegraphics[width=0.165\textwidth,height=0.17\textwidth]{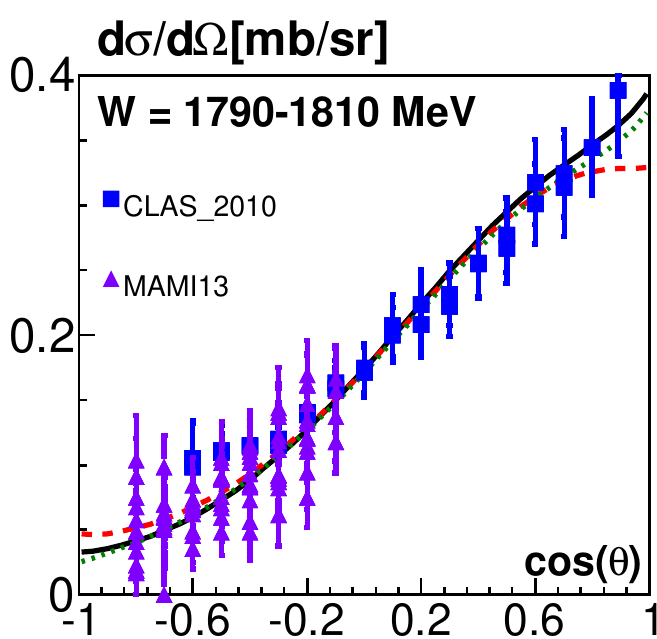}&
\hspace{-4mm}\includegraphics[width=0.165\textwidth,height=0.17\textwidth]{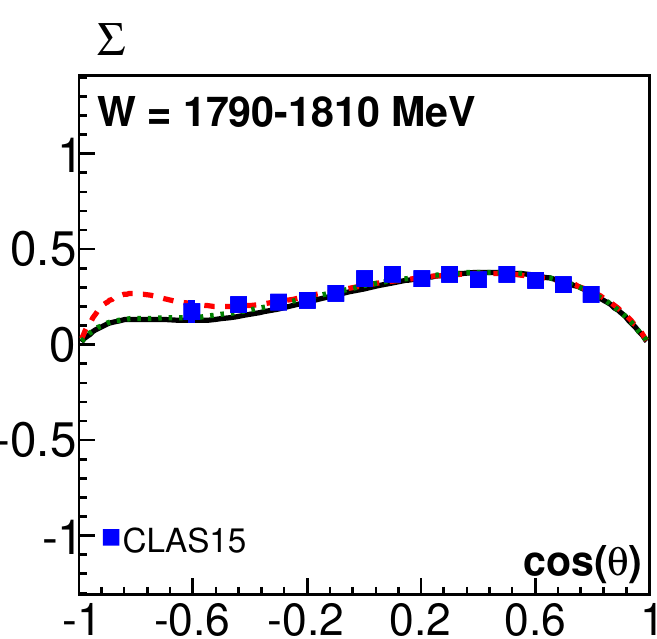}&
\hspace{-4mm}\includegraphics[width=0.165\textwidth,height=0.17\textwidth]{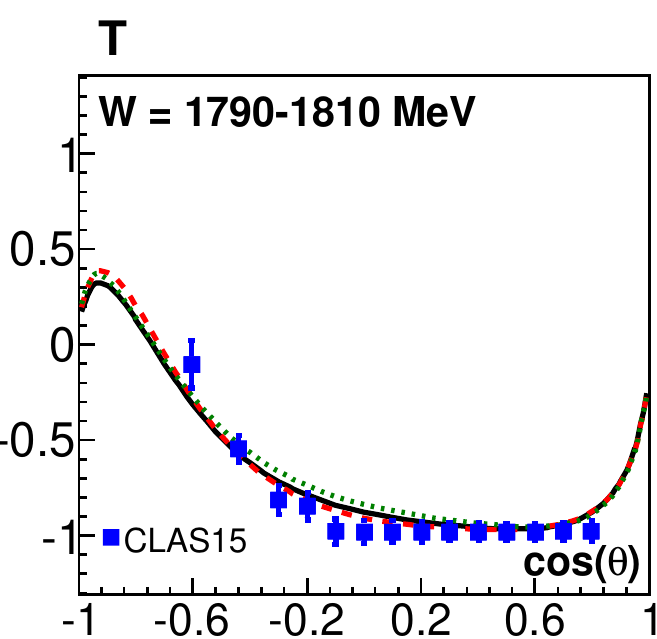}&
\hspace{-4mm}\includegraphics[width=0.165\textwidth,height=0.17\textwidth]{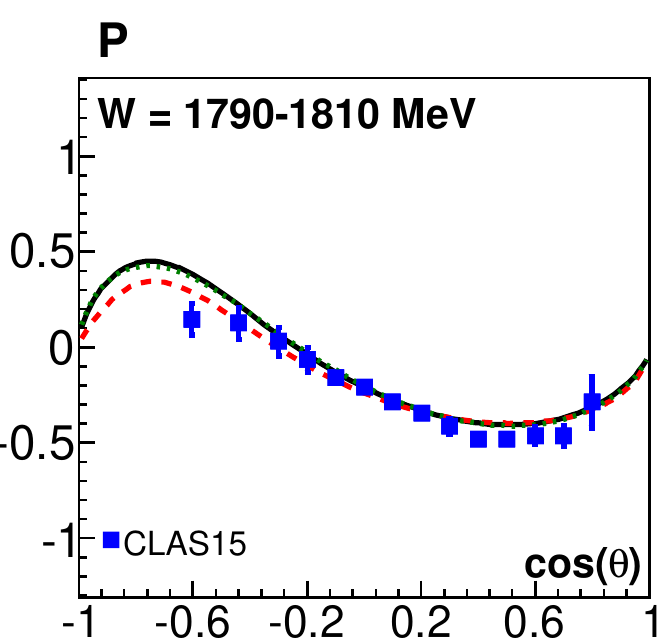}&
\hspace{-4mm}\includegraphics[width=0.165\textwidth,height=0.17\textwidth]{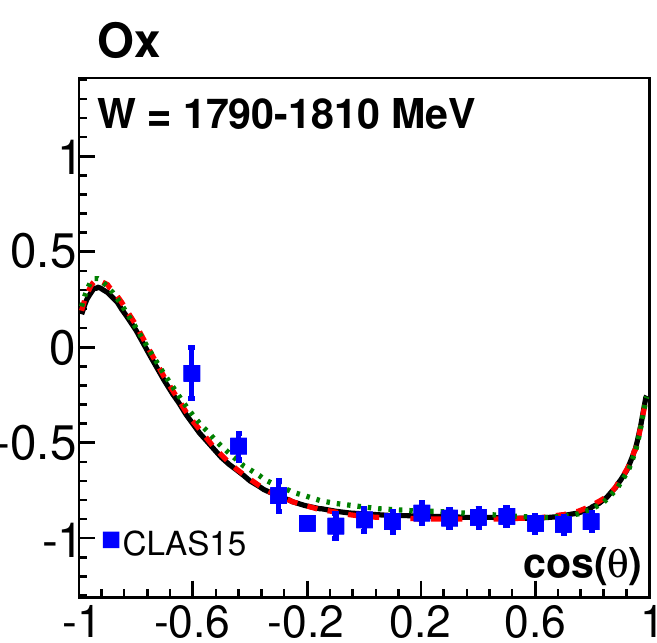}&
\hspace{-4mm}\includegraphics[width=0.165\textwidth,height=0.17\textwidth]{Ox_no_prime_observables1800.pdf}\\
\hspace{-3mm}\includegraphics[width=0.165\textwidth,height=0.17\textwidth]{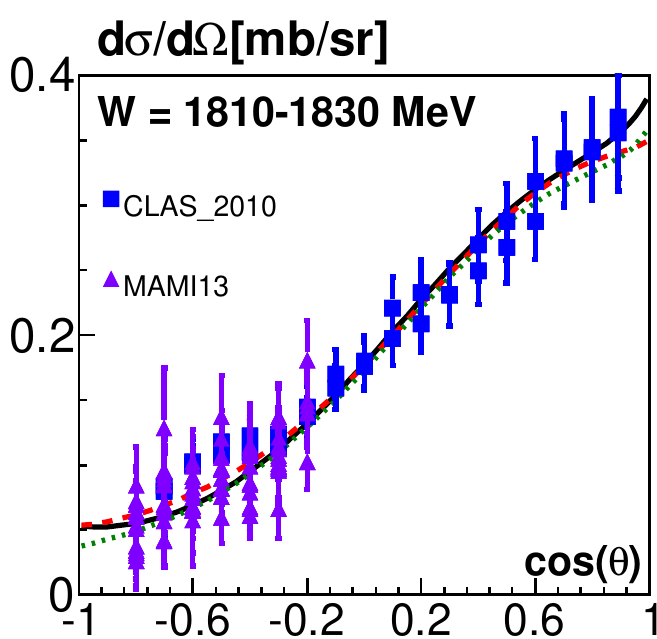}&
\hspace{-4mm}\includegraphics[width=0.165\textwidth,height=0.17\textwidth]{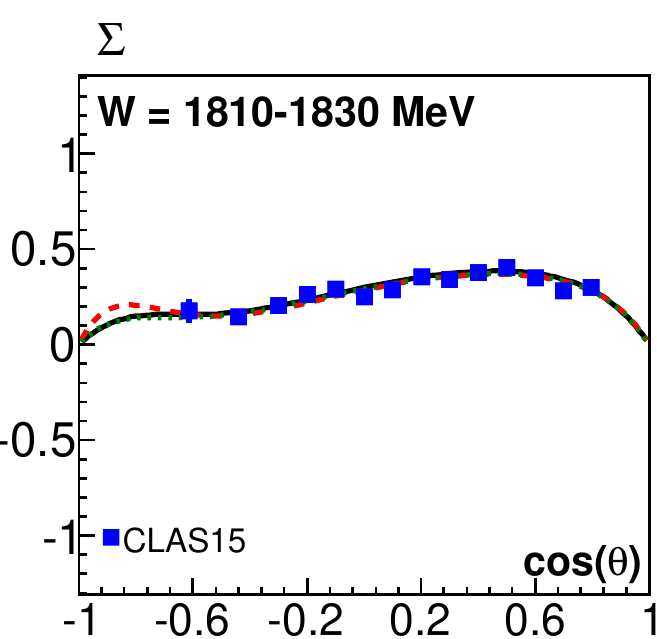}&
\hspace{-4mm}\includegraphics[width=0.165\textwidth,height=0.17\textwidth]{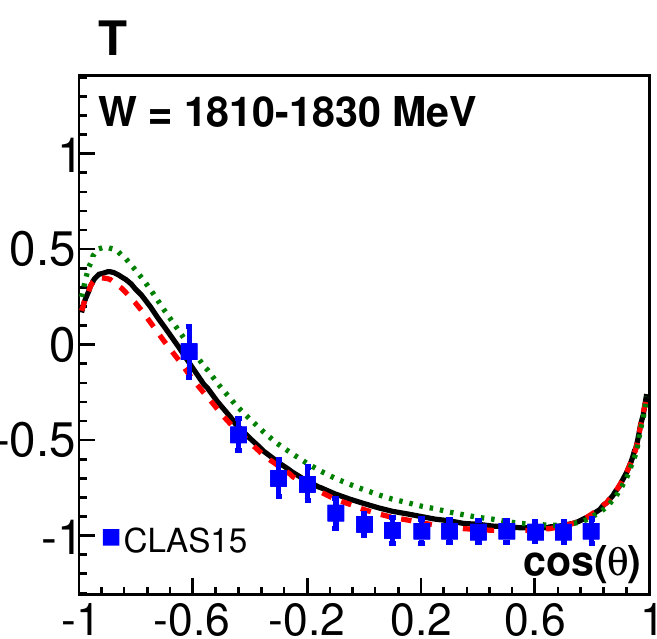}&
\hspace{-4mm}\includegraphics[width=0.165\textwidth,height=0.17\textwidth]{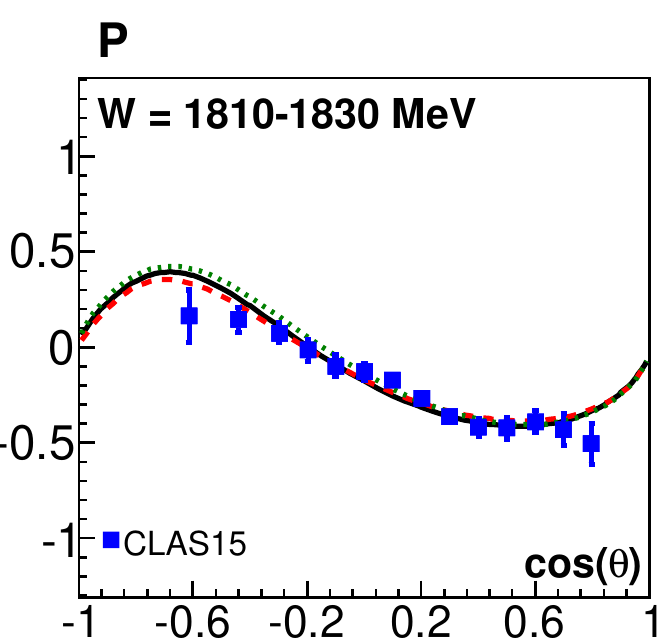}&
\hspace{-4mm}\includegraphics[width=0.165\textwidth,height=0.17\textwidth]{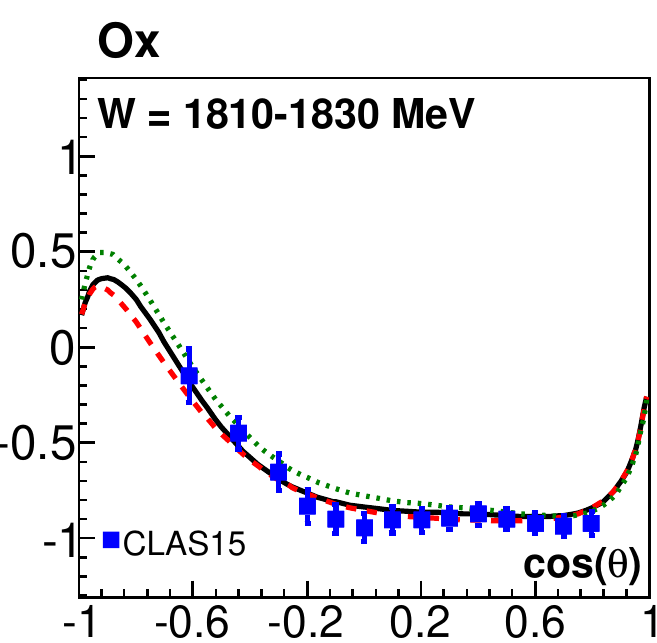}&
\hspace{-4mm}\includegraphics[width=0.165\textwidth,height=0.17\textwidth]{Ox_no_prime_observables1820.pdf}\\
\hspace{-3mm}\includegraphics[width=0.165\textwidth,height=0.17\textwidth]{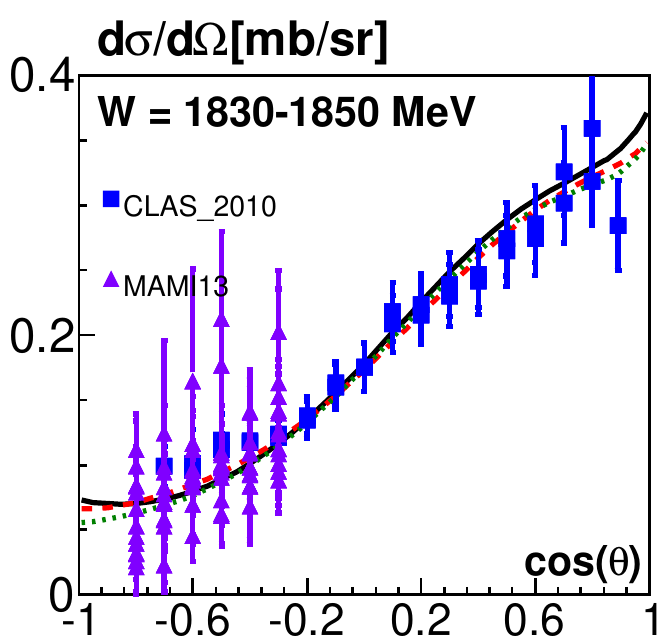}&
\hspace{-4mm}\includegraphics[width=0.165\textwidth,height=0.17\textwidth]{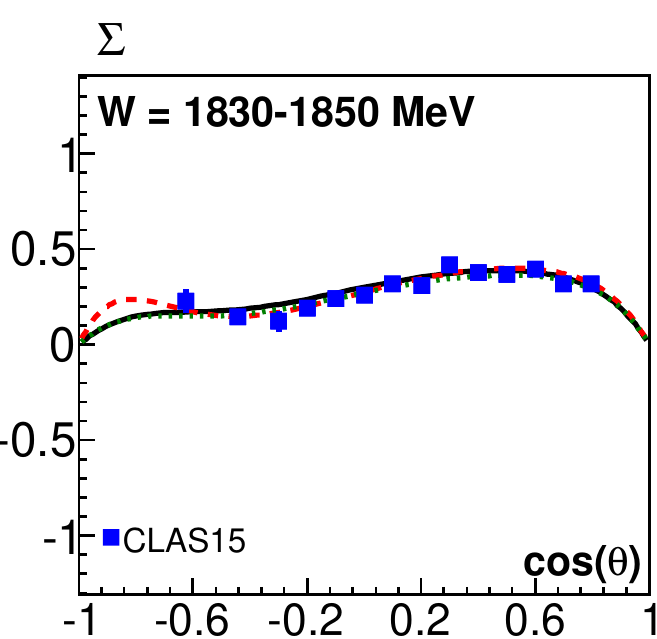}&
\hspace{-4mm}\includegraphics[width=0.165\textwidth,height=0.17\textwidth]{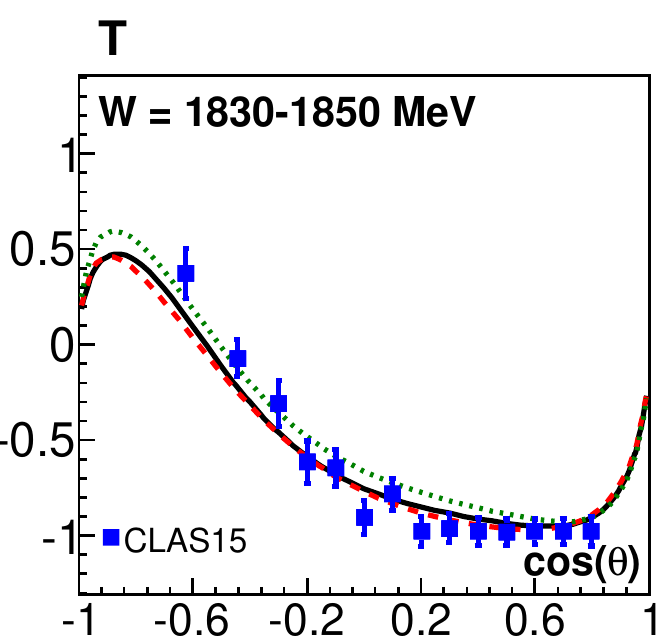}&
\hspace{-4mm}\includegraphics[width=0.165\textwidth,height=0.17\textwidth]{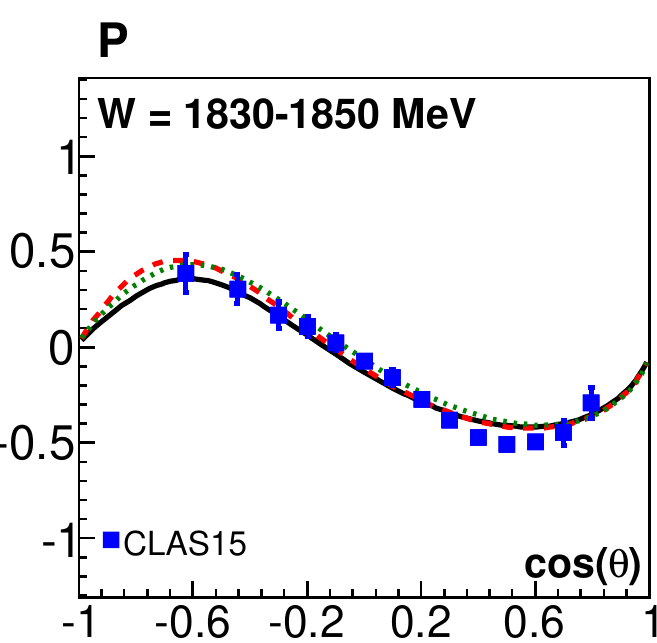}&
\hspace{-4mm}\includegraphics[width=0.165\textwidth,height=0.17\textwidth]{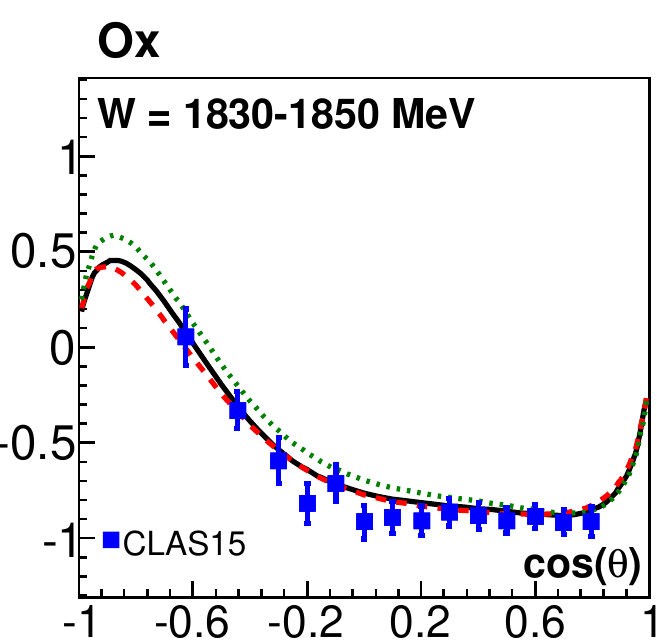}&
\hspace{-4mm}\includegraphics[width=0.165\textwidth,height=0.17\textwidth]{Ox_no_prime_observables1840.pdf}\\
\end{tabular}
\caption{\label{fig:data-g1}(Color online) Fit to the data on $d\sigma/d\Omega$:~\cite{McCracken:2009ra},
$P$~\cite{McCracken:2009ra}, and $\Sigma$, $T$, $O_x$, $O_z$~\cite{Paterson:2016vmc} for $\gamma p\to K^+\Lambda$  reaction
for the mass range from 1710 to 1850\,MeV.
 The solid (black) line corresponds the $L+P$ fit, the dashed (red)
line corresponds to fit used to
determine the multipoles of Fig.~\ref{fig:mult}., the dotted (green) line corresponds to BnGa fit.
}
\end{figure*}
\begin{figure*}[pt]
\begin{tabular}{cccccc}
\hspace{-3mm}\includegraphics[width=0.165\textwidth,height=0.17\textwidth]{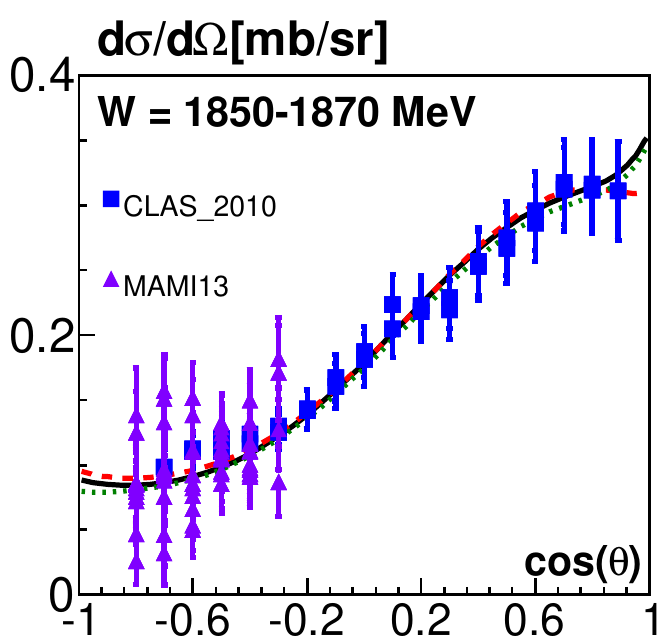}&
\hspace{-4mm}\includegraphics[width=0.165\textwidth,height=0.17\textwidth]{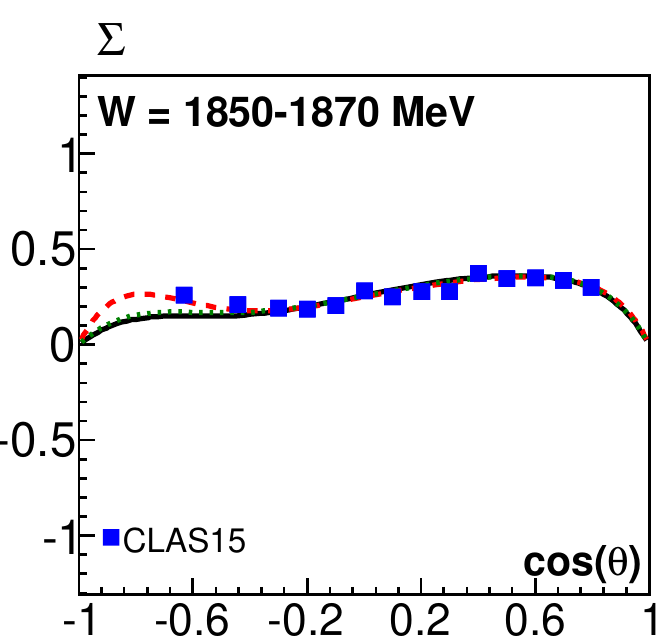}&
\hspace{-4mm}\includegraphics[width=0.165\textwidth,height=0.17\textwidth]{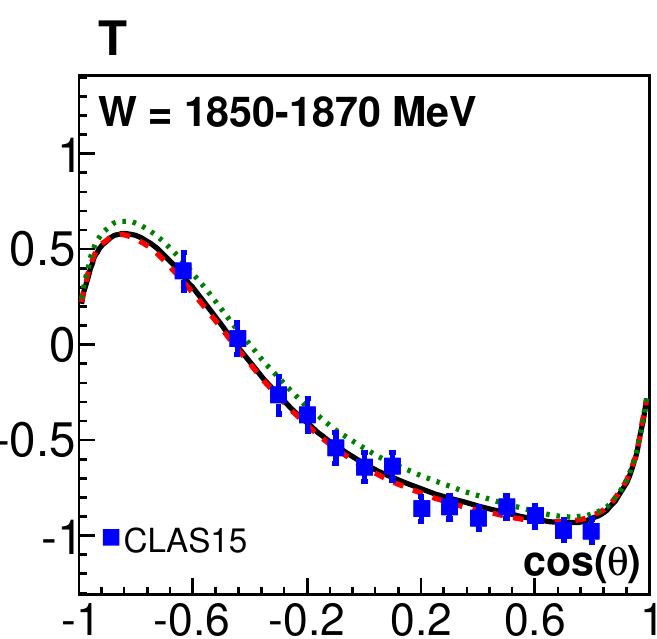}&
\hspace{-4mm}\includegraphics[width=0.165\textwidth,height=0.17\textwidth]{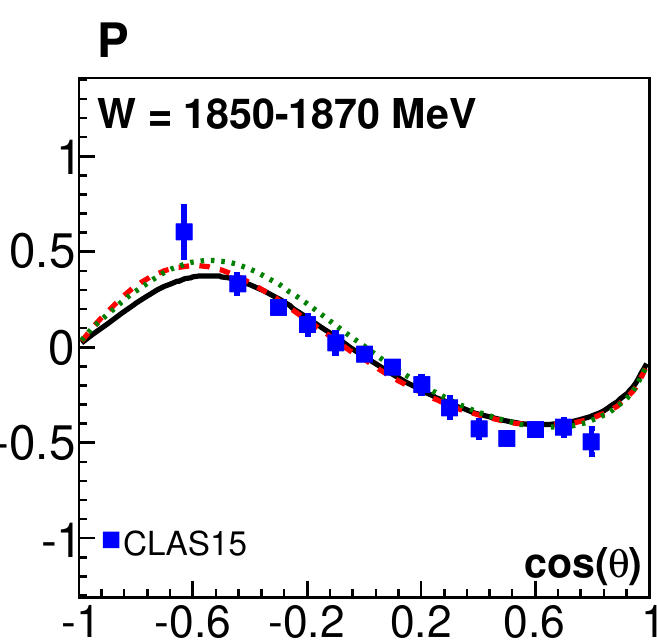}&
\hspace{-4mm}\includegraphics[width=0.165\textwidth,height=0.17\textwidth]{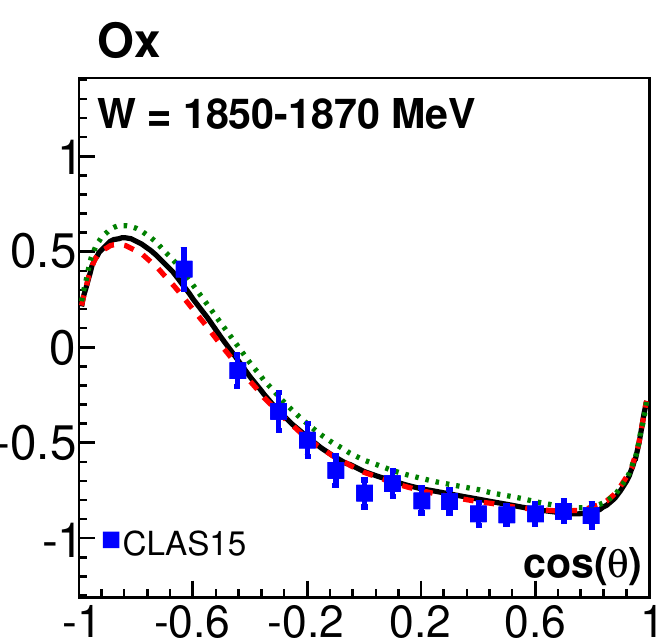}&
\hspace{-4mm}\includegraphics[width=0.165\textwidth,height=0.17\textwidth]{Ox_no_prime_observables1860.pdf}\\
\hspace{-3mm}\includegraphics[width=0.165\textwidth,height=0.17\textwidth]{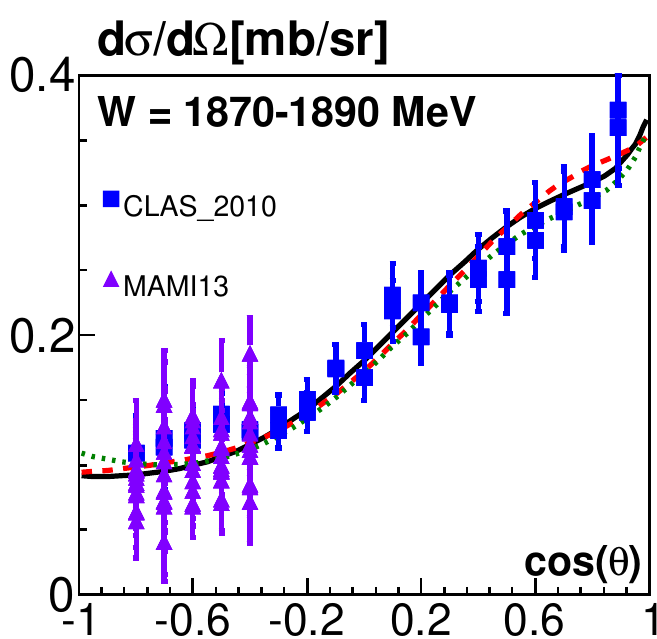}&
\hspace{-4mm}\includegraphics[width=0.165\textwidth,height=0.17\textwidth]{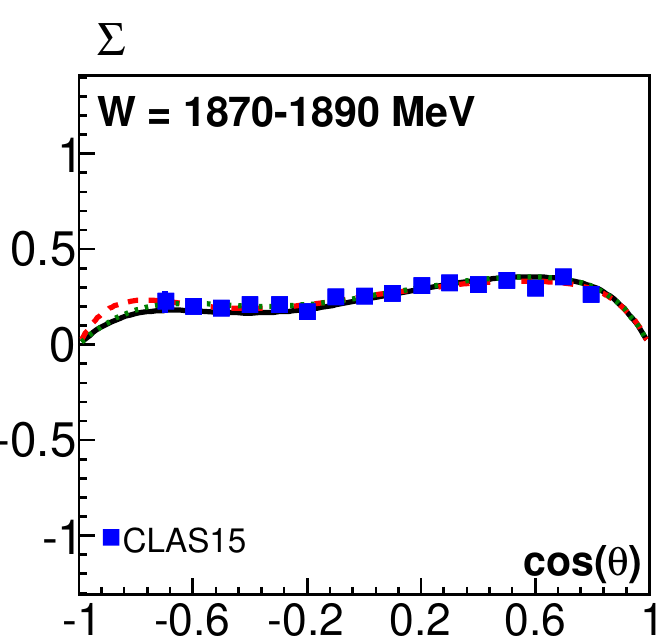}&
\hspace{-4mm}\includegraphics[width=0.165\textwidth,height=0.17\textwidth]{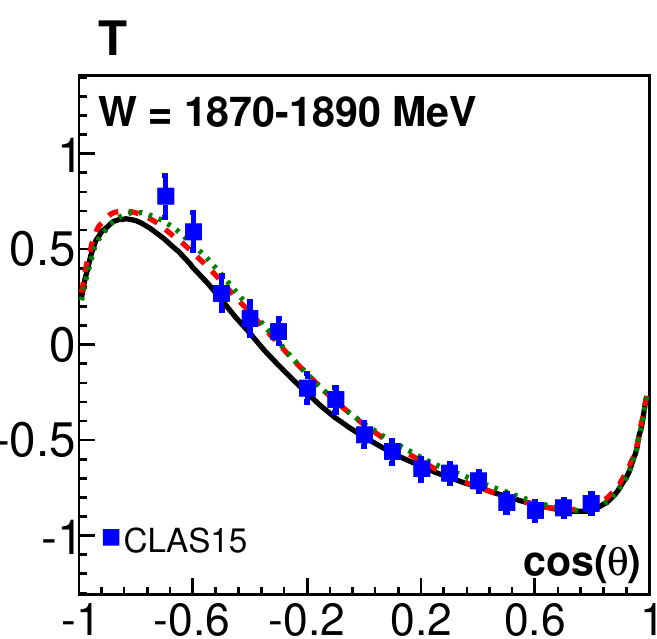}&
\hspace{-4mm}\includegraphics[width=0.165\textwidth,height=0.17\textwidth]{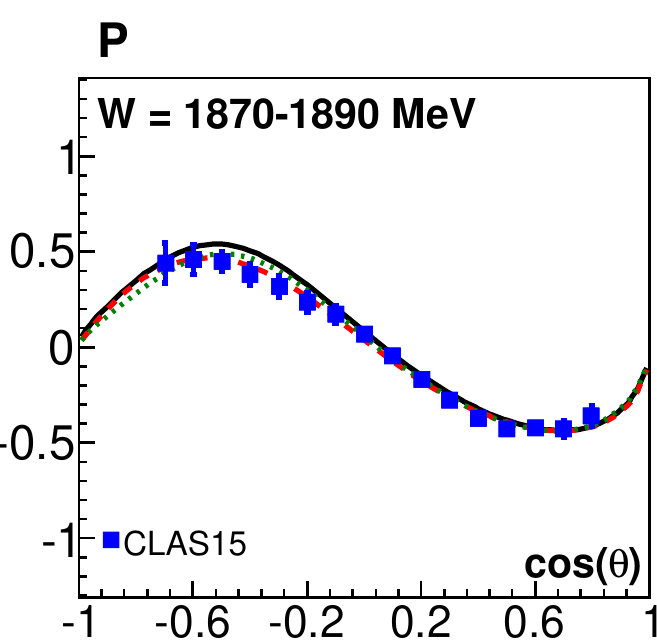}&
\hspace{-4mm}\includegraphics[width=0.165\textwidth,height=0.17\textwidth]{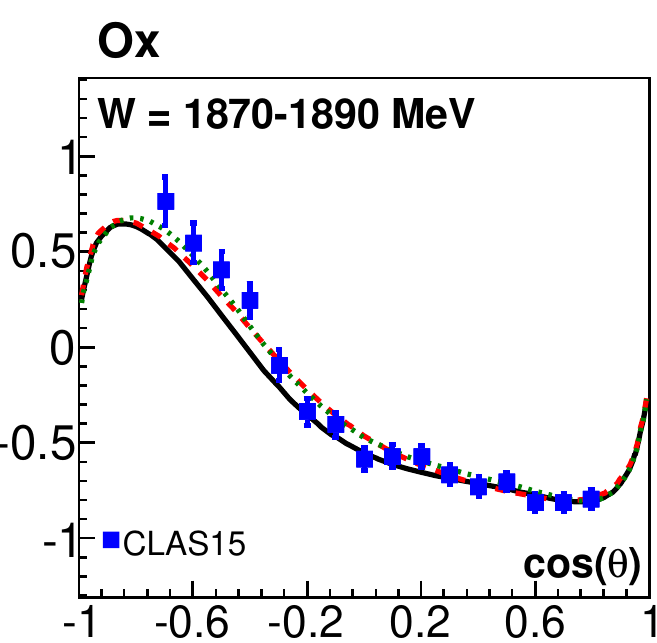}&
\hspace{-4mm}\includegraphics[width=0.165\textwidth,height=0.17\textwidth]{Ox_no_prime_observables1880.pdf}\\
\hspace{-3mm}\includegraphics[width=0.165\textwidth,height=0.17\textwidth]{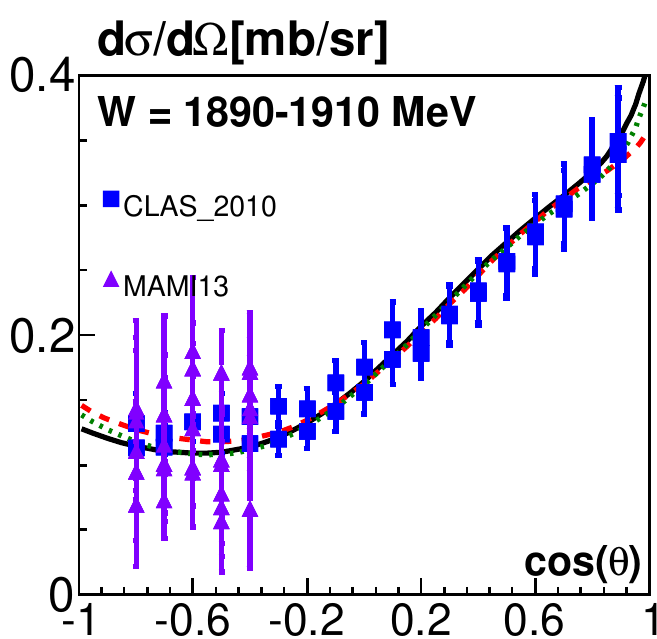}&
\hspace{-4mm}\includegraphics[width=0.165\textwidth,height=0.17\textwidth]{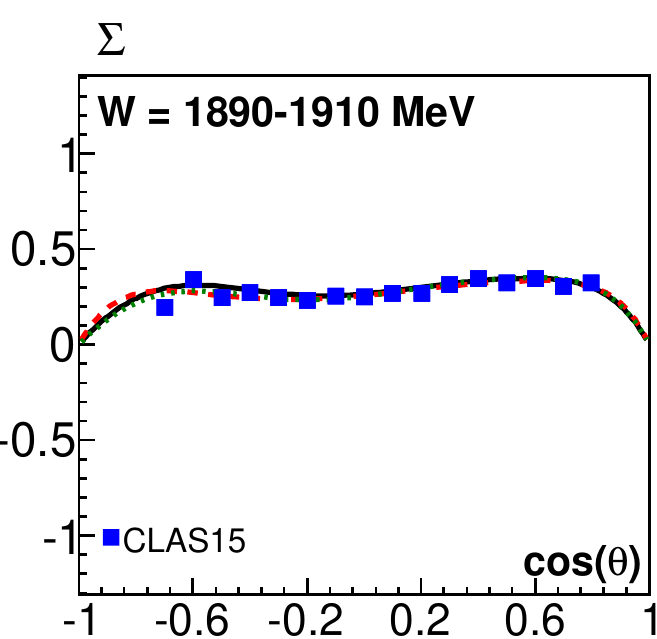}&
\hspace{-4mm}\includegraphics[width=0.165\textwidth,height=0.17\textwidth]{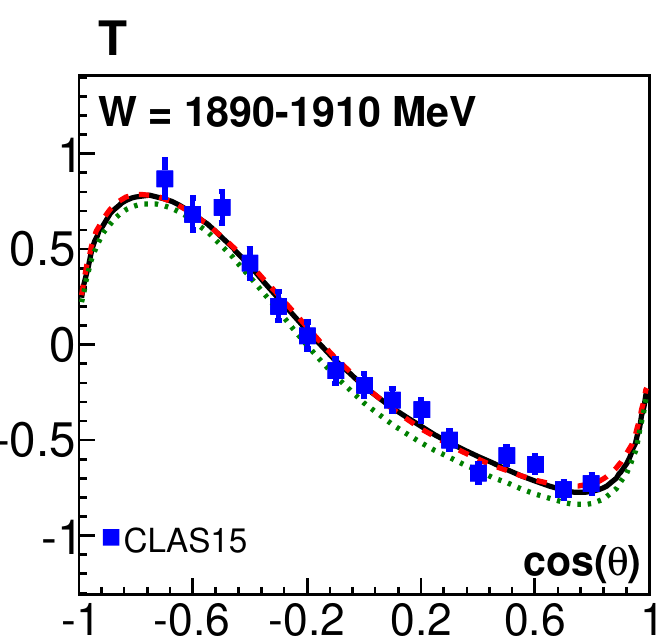}&
\hspace{-4mm}\includegraphics[width=0.165\textwidth,height=0.17\textwidth]{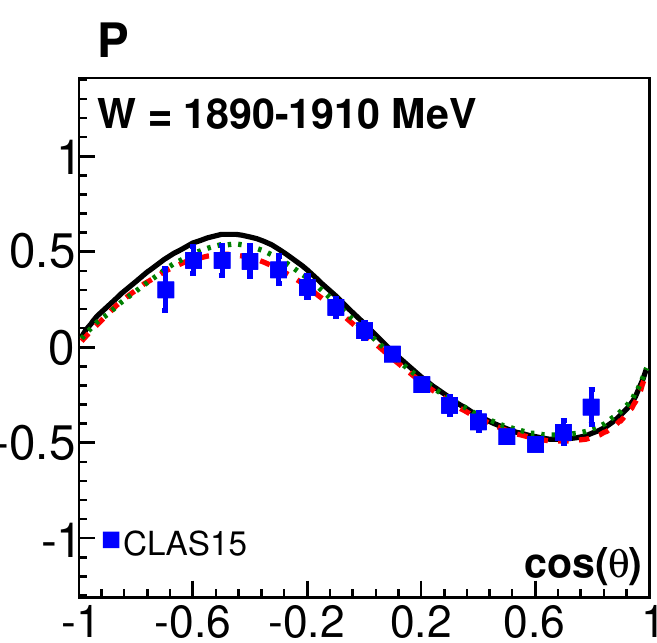}&
\hspace{-4mm}\includegraphics[width=0.165\textwidth,height=0.17\textwidth]{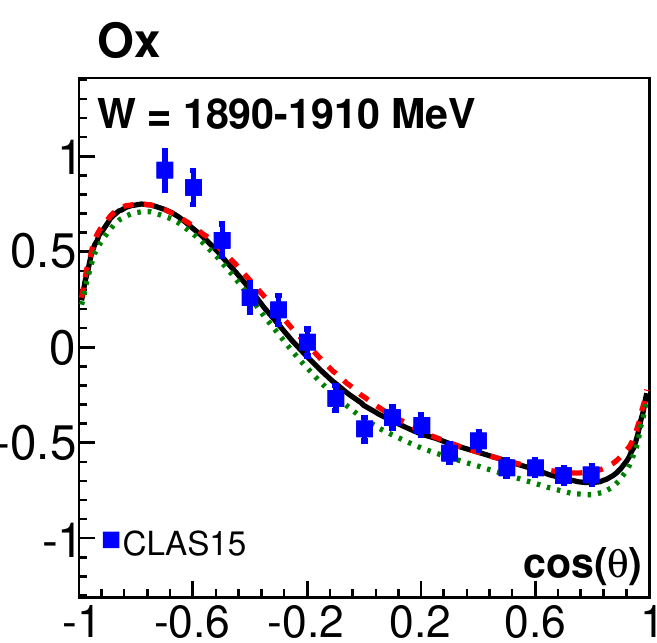}&
\hspace{-4mm}\includegraphics[width=0.165\textwidth,height=0.17\textwidth]{Ox_no_prime_observables1900.pdf}\\
\hspace{-3mm}\includegraphics[width=0.165\textwidth,height=0.17\textwidth]{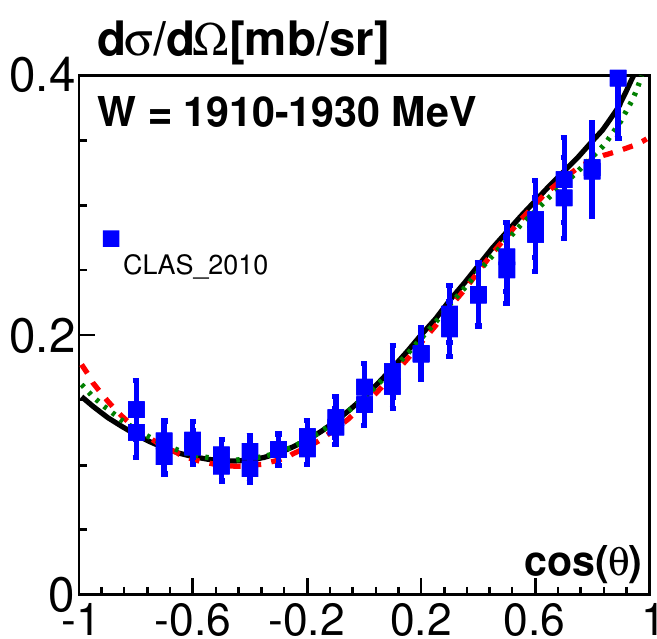}&
\hspace{-4mm}\includegraphics[width=0.165\textwidth,height=0.17\textwidth]{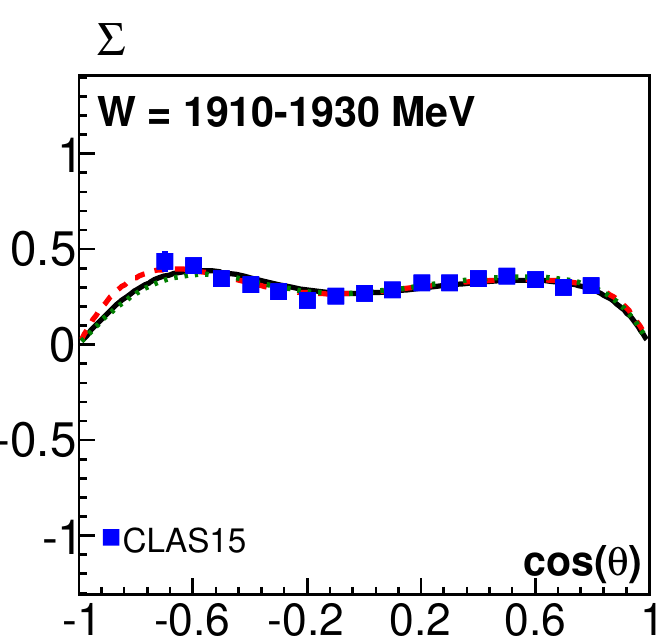}&
\hspace{-4mm}\includegraphics[width=0.165\textwidth,height=0.17\textwidth]{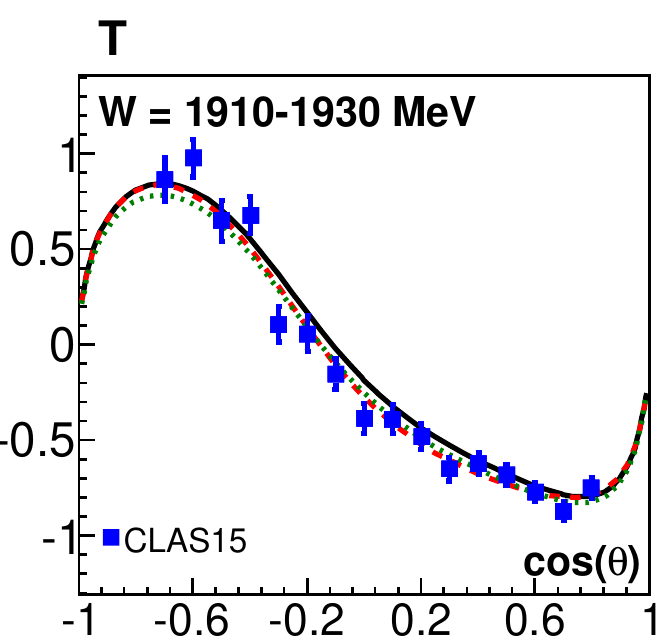}&
\hspace{-4mm}\includegraphics[width=0.165\textwidth,height=0.17\textwidth]{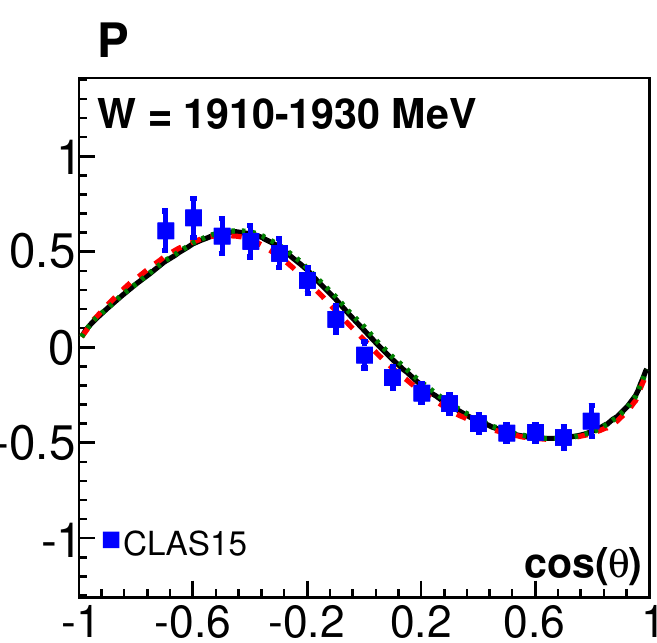}&
\hspace{-4mm}\includegraphics[width=0.165\textwidth,height=0.17\textwidth]{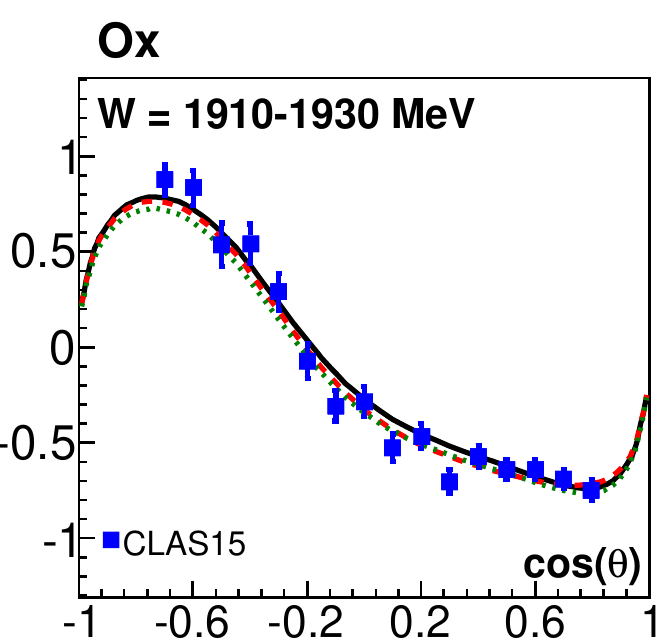}&
\hspace{-4mm}\includegraphics[width=0.165\textwidth,height=0.17\textwidth]{Ox_no_prime_observables1920.pdf}\\
\hspace{-3mm}\includegraphics[width=0.165\textwidth,height=0.17\textwidth]{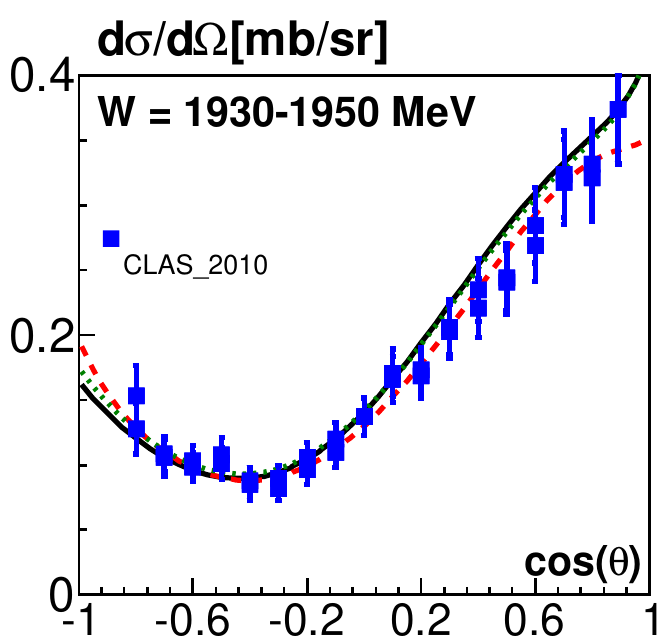}&
\hspace{-4mm}\includegraphics[width=0.165\textwidth,height=0.17\textwidth]{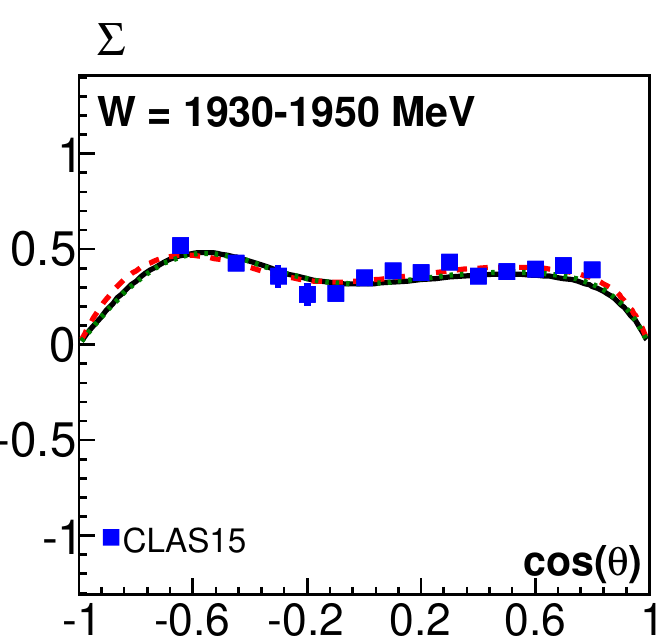}&
\hspace{-4mm}\includegraphics[width=0.165\textwidth,height=0.17\textwidth]{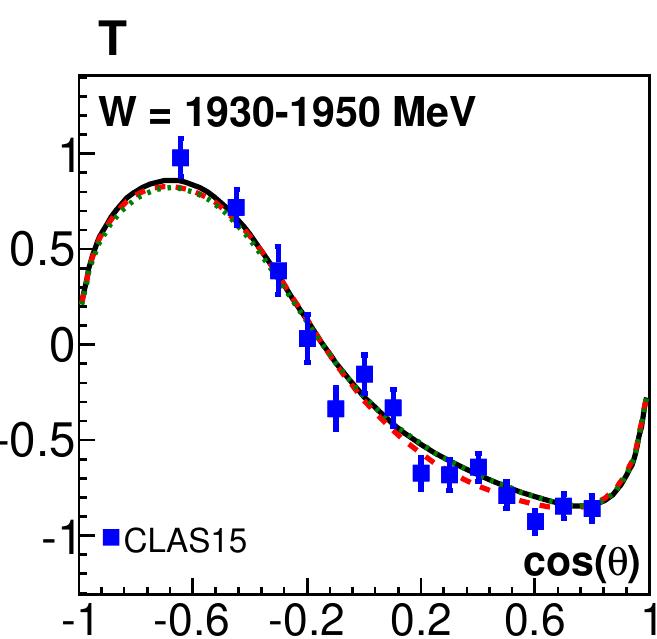}&
\hspace{-4mm}\includegraphics[width=0.165\textwidth,height=0.17\textwidth]{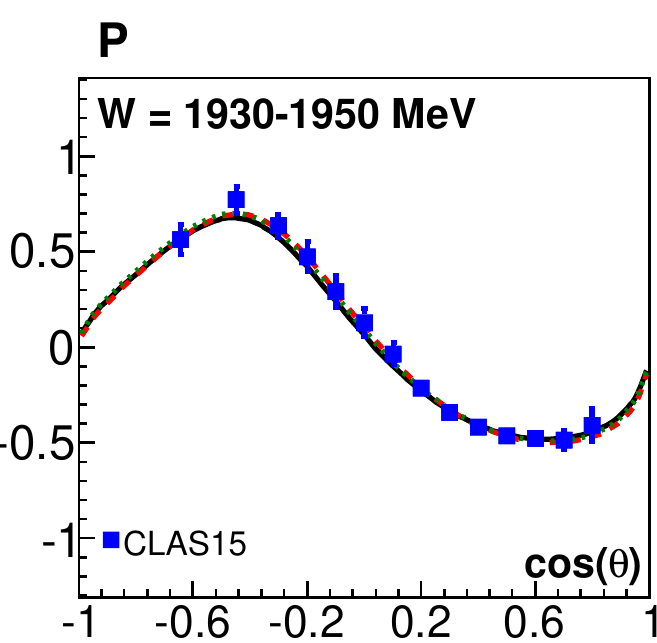}&
\hspace{-4mm}\includegraphics[width=0.165\textwidth,height=0.17\textwidth]{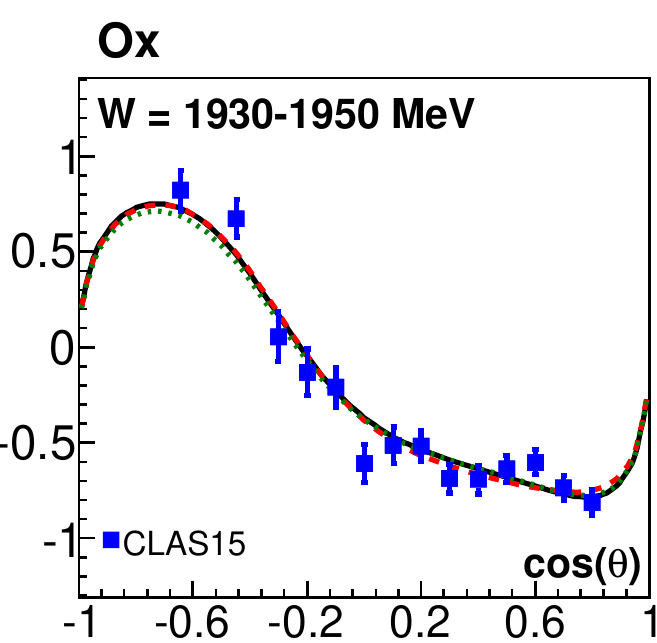}&
\hspace{-4mm}\includegraphics[width=0.165\textwidth,height=0.17\textwidth]{Ox_no_prime_observables1940.pdf}\\
\hspace{-3mm}\includegraphics[width=0.165\textwidth,height=0.17\textwidth]{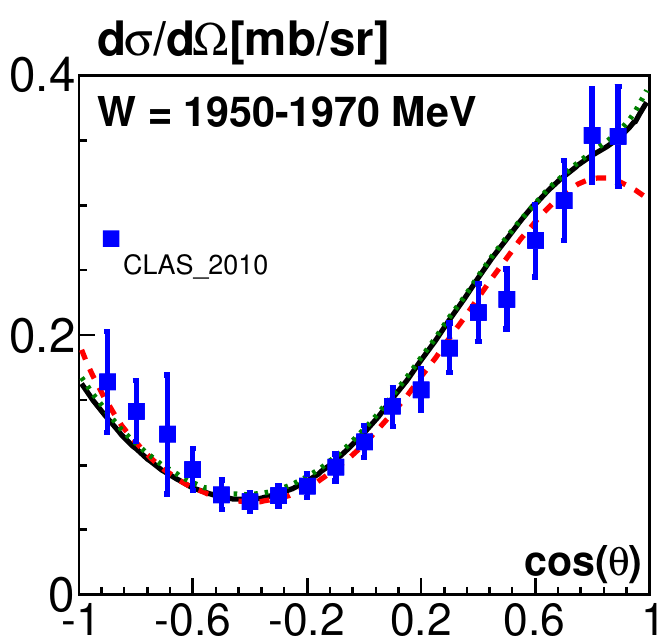}&
\hspace{-4mm}\includegraphics[width=0.165\textwidth,height=0.17\textwidth]{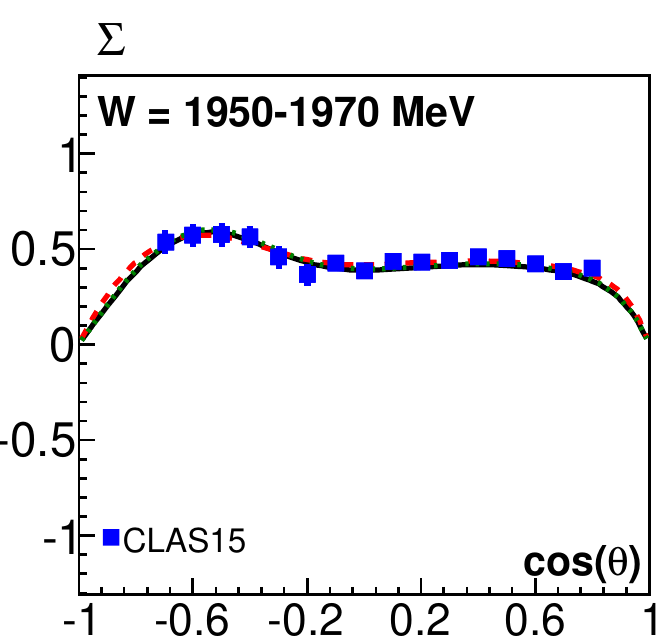}&
\hspace{-4mm}\includegraphics[width=0.165\textwidth,height=0.17\textwidth]{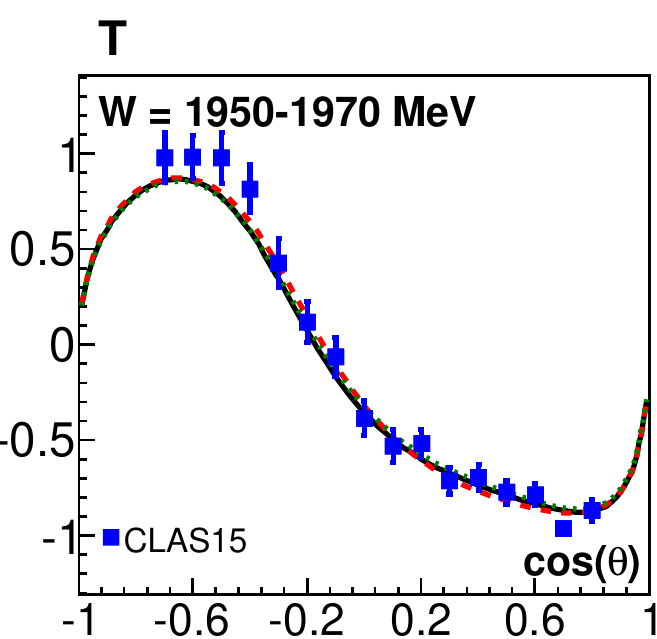}&
\hspace{-4mm}\includegraphics[width=0.165\textwidth,height=0.17\textwidth]{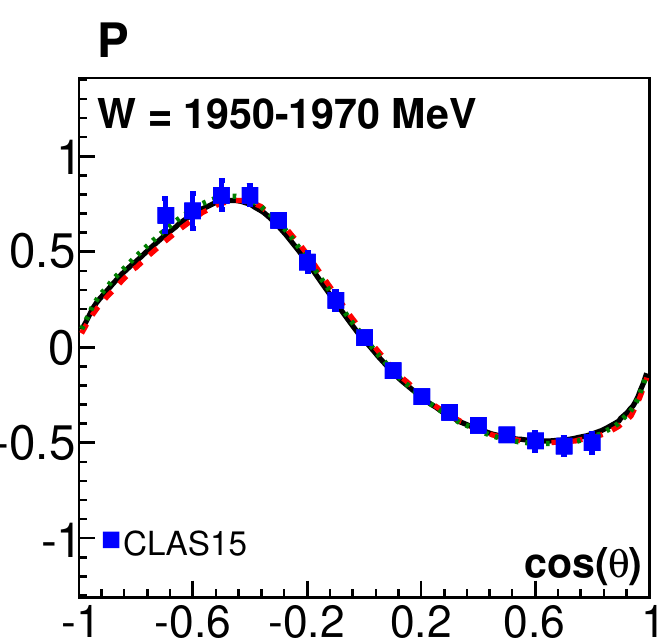}&
\hspace{-4mm}\includegraphics[width=0.165\textwidth,height=0.17\textwidth]{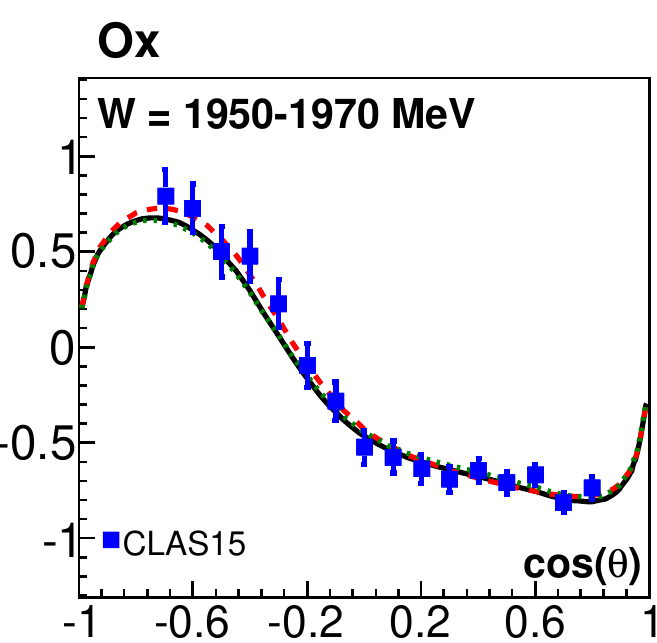}&
\hspace{-4mm}\includegraphics[width=0.165\textwidth,height=0.17\textwidth]{Ox_no_prime_observables1960.pdf}\\
\hspace{-3mm}\includegraphics[width=0.165\textwidth,height=0.17\textwidth]{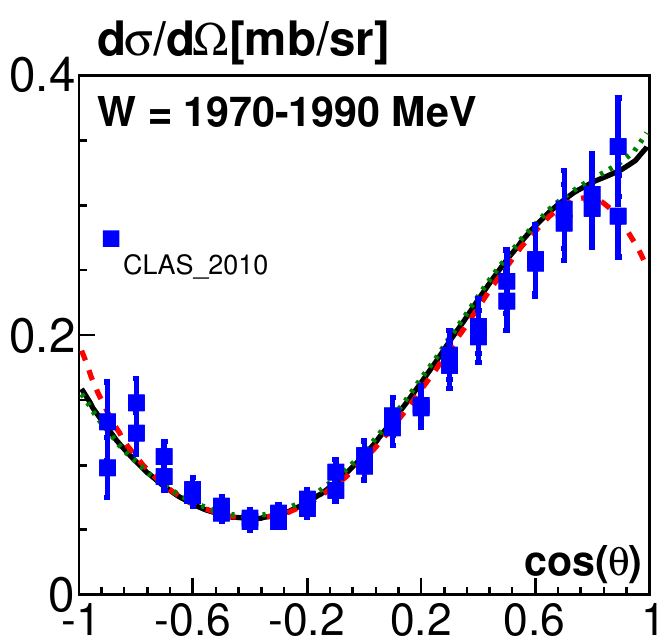}&
\hspace{-4mm}\includegraphics[width=0.165\textwidth,height=0.17\textwidth]{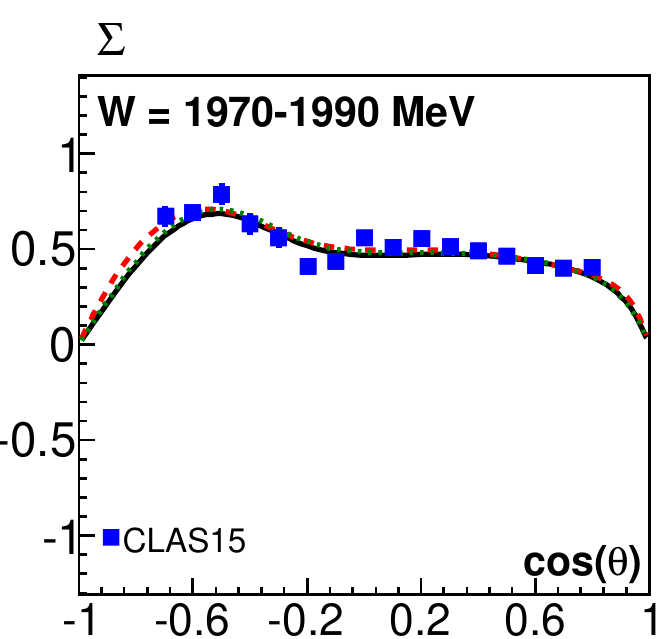}&
\hspace{-4mm}\includegraphics[width=0.165\textwidth,height=0.17\textwidth]{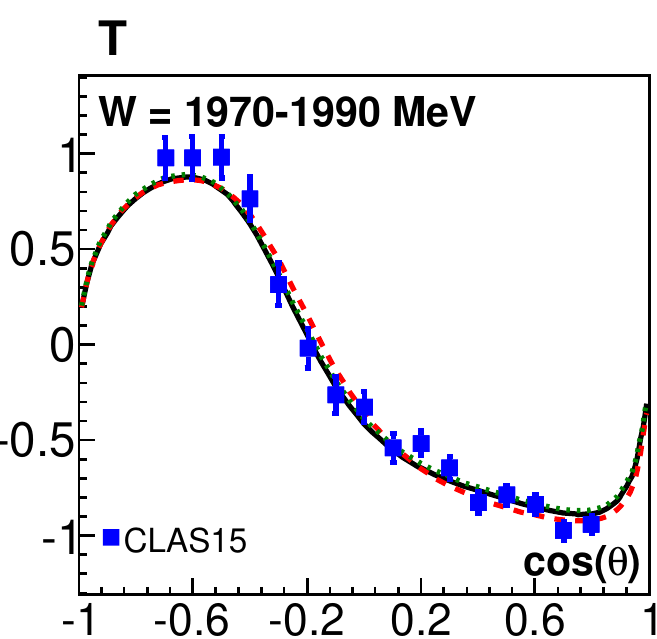}&
\hspace{-4mm}\includegraphics[width=0.165\textwidth,height=0.17\textwidth]{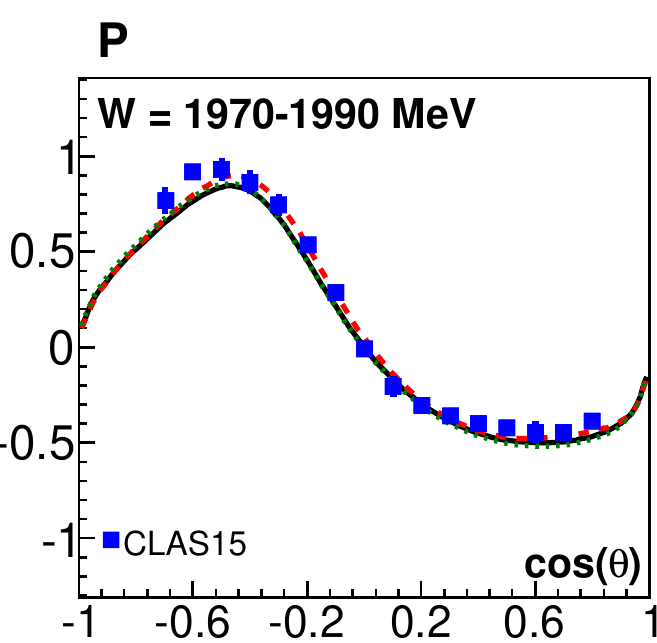}&
\hspace{-4mm}\includegraphics[width=0.165\textwidth,height=0.17\textwidth]{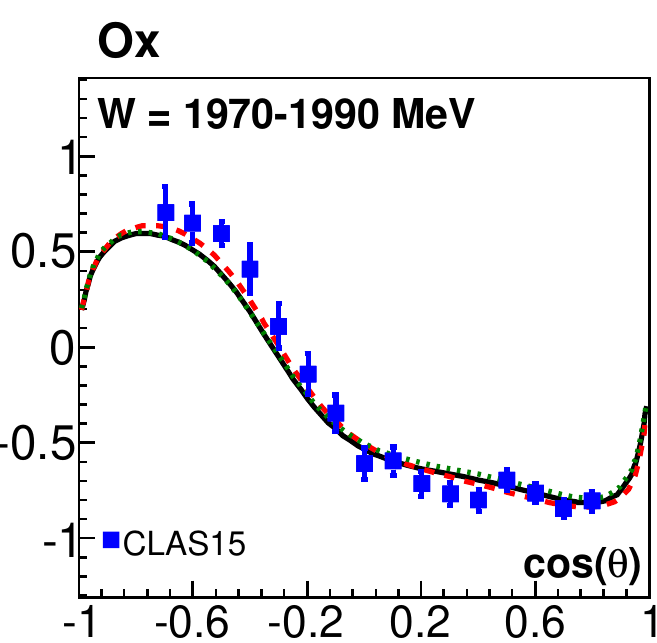}&
\hspace{-4mm}\includegraphics[width=0.165\textwidth,height=0.17\textwidth]{Ox_no_prime_observables1980.pdf}\\
\end{tabular}
\caption{\label{fig:data-g2}(Color online) Fit to the data on $d\sigma/d\Omega$:~\cite{McCracken:2009ra},
$P$~\cite{McCracken:2009ra}, and $\Sigma$, $T$, $O_x$, $O_z$~\cite{Paterson:2016vmc}  for $\gamma p\to K^+\Lambda$  reaction
for the mass range from 1850 to 1990\,MeV.
The solid (black) line corresponds the $L+P$ fit, the dashed (red)
line corresponds to fit used to
determine the multipoles of Fig.~\ref{fig:mult}., the dotted (green) line corresponds to BnGa fit.
}
\end{figure*}

\begin{figure*}[pt]
\begin{tabular}{cccccc}
\hspace{-3mm}\includegraphics[width=0.165\textwidth,height=0.17\textwidth]{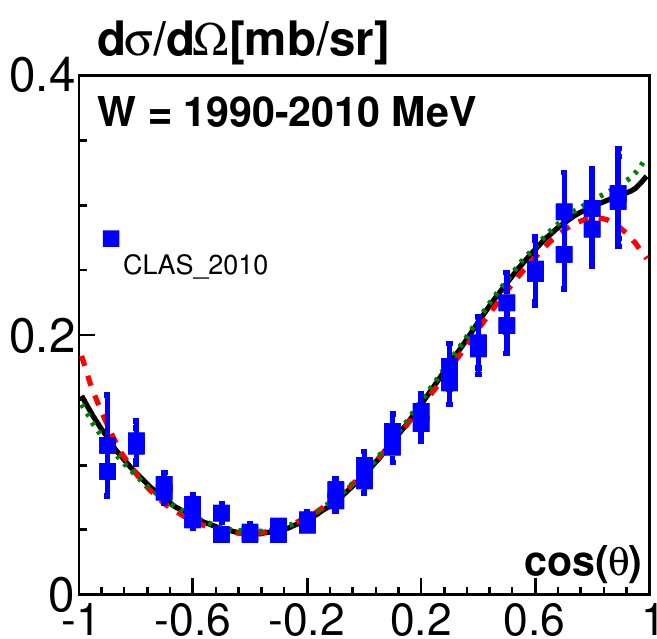}&
\hspace{-4mm}\includegraphics[width=0.165\textwidth,height=0.17\textwidth]{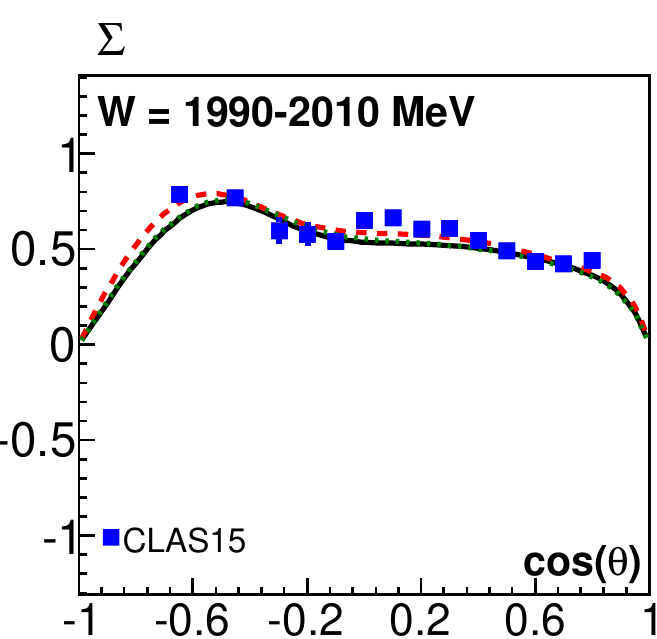}&
\hspace{-4mm}\includegraphics[width=0.165\textwidth,height=0.17\textwidth]{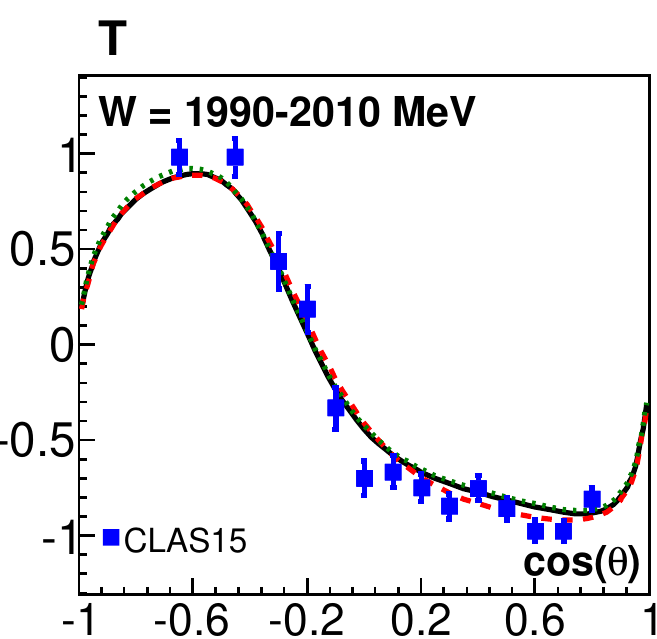}&
\hspace{-4mm}\includegraphics[width=0.165\textwidth,height=0.17\textwidth]{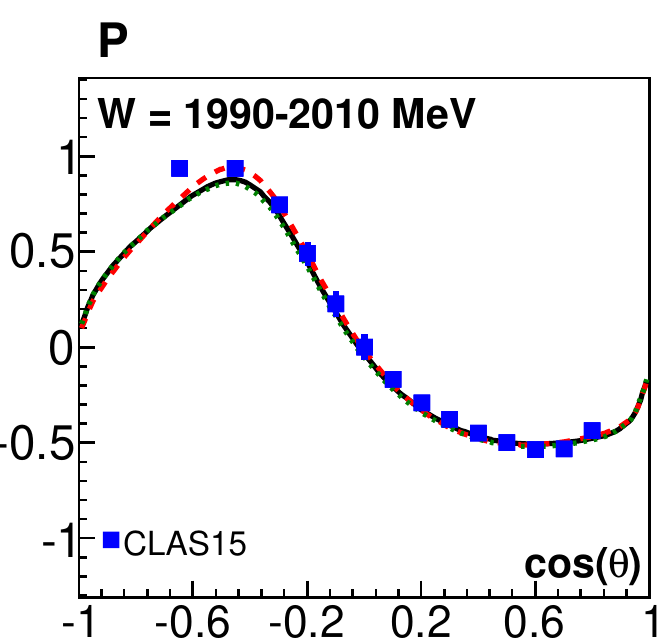}&
\hspace{-4mm}\includegraphics[width=0.165\textwidth,height=0.17\textwidth]{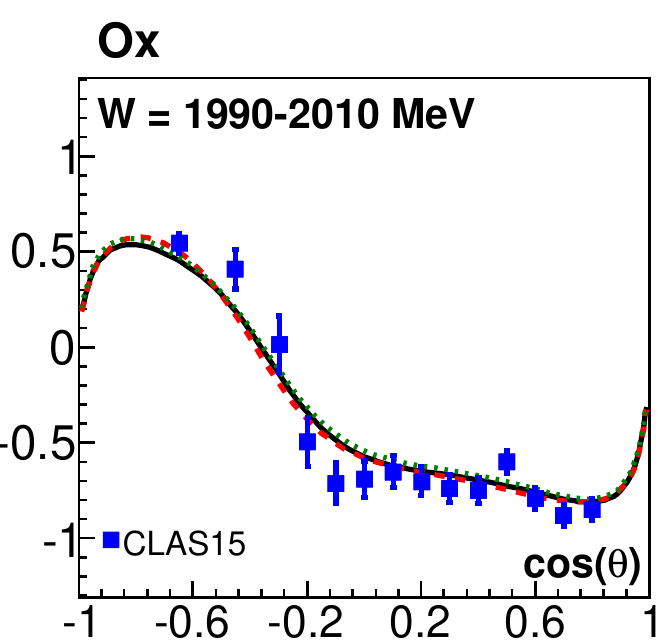}&
\hspace{-4mm}\includegraphics[width=0.165\textwidth,height=0.17\textwidth]{Ox_no_prime_observables2000.pdf}\\
\hspace{-3mm}\includegraphics[width=0.165\textwidth,height=0.17\textwidth]{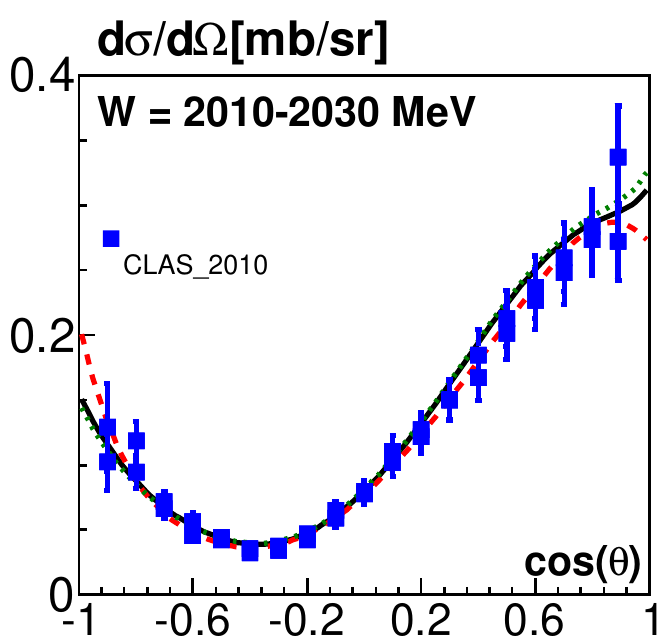}&
\hspace{-4mm}\includegraphics[width=0.165\textwidth,height=0.17\textwidth]{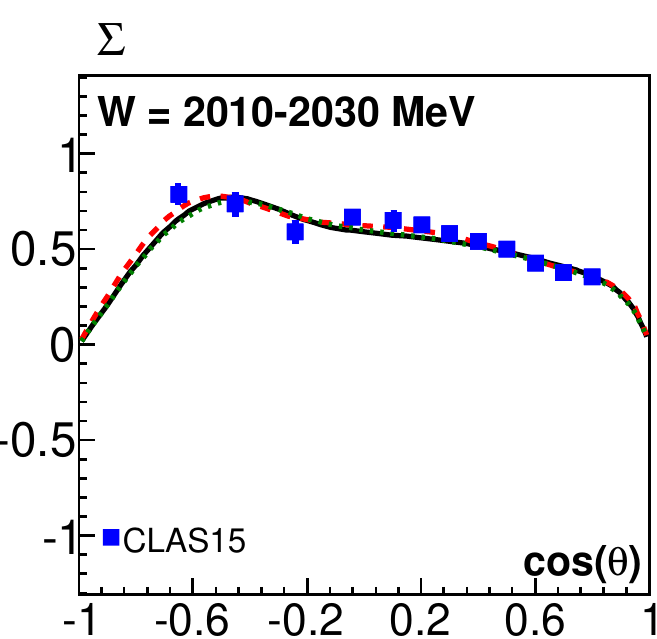}&
\hspace{-4mm}\includegraphics[width=0.165\textwidth,height=0.17\textwidth]{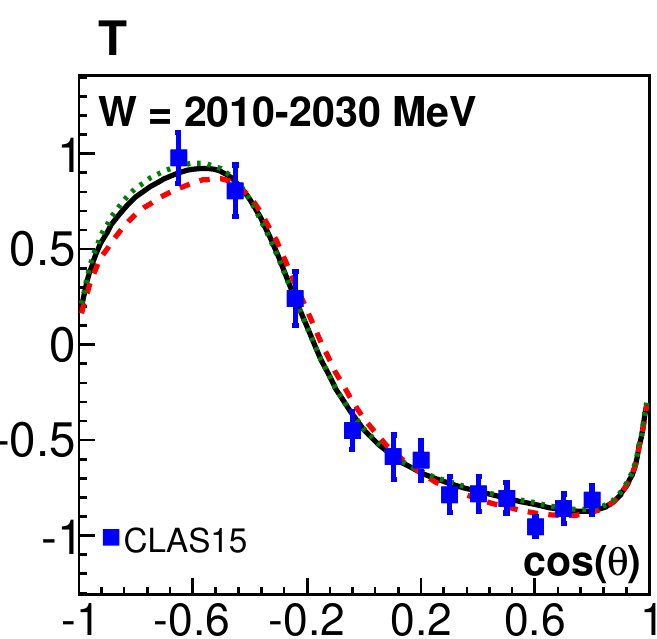}&
\hspace{-4mm}\includegraphics[width=0.165\textwidth,height=0.17\textwidth]{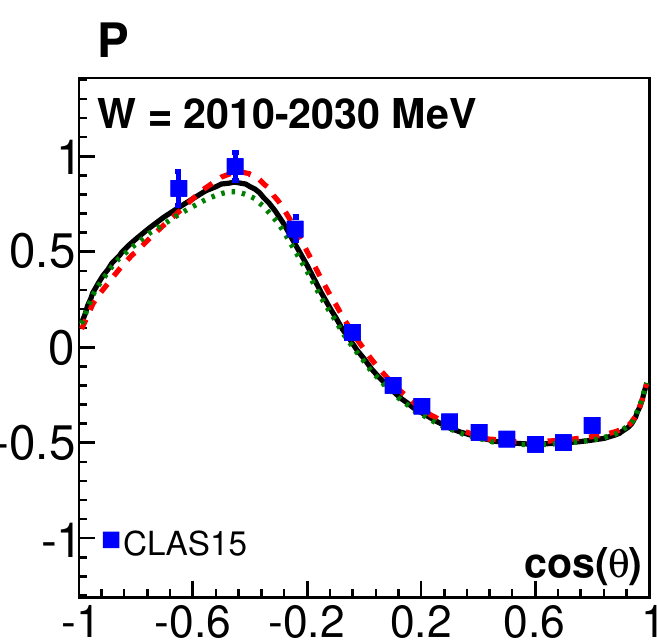}&
\hspace{-4mm}\includegraphics[width=0.165\textwidth,height=0.17\textwidth]{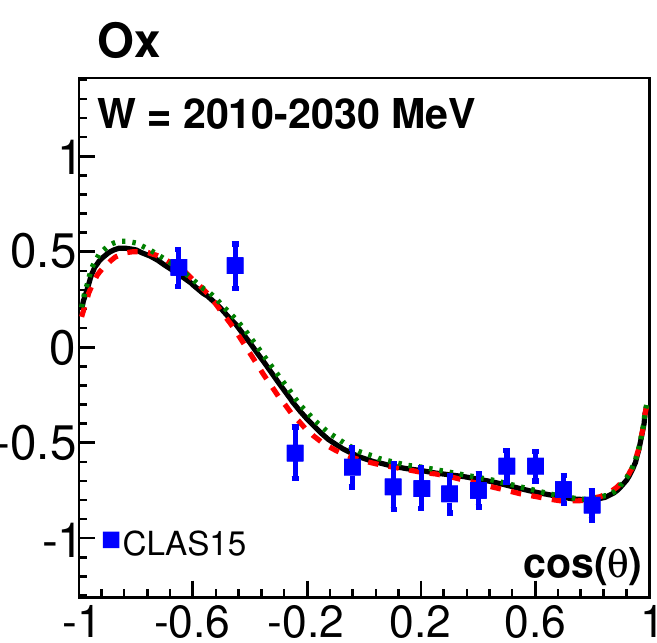}&
\hspace{-4mm}\includegraphics[width=0.165\textwidth,height=0.17\textwidth]{Ox_no_prime_observables2020.pdf}\\
\hspace{-3mm}\includegraphics[width=0.165\textwidth,height=0.17\textwidth]{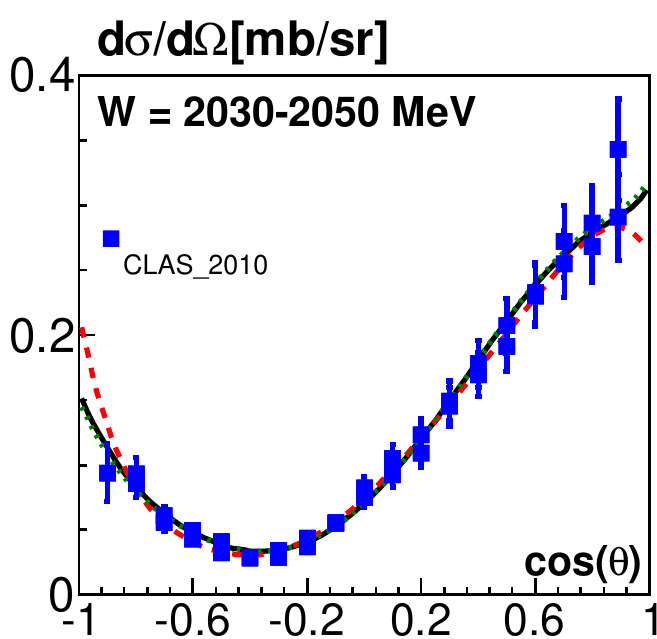}&
\hspace{-4mm}\includegraphics[width=0.165\textwidth,height=0.17\textwidth]{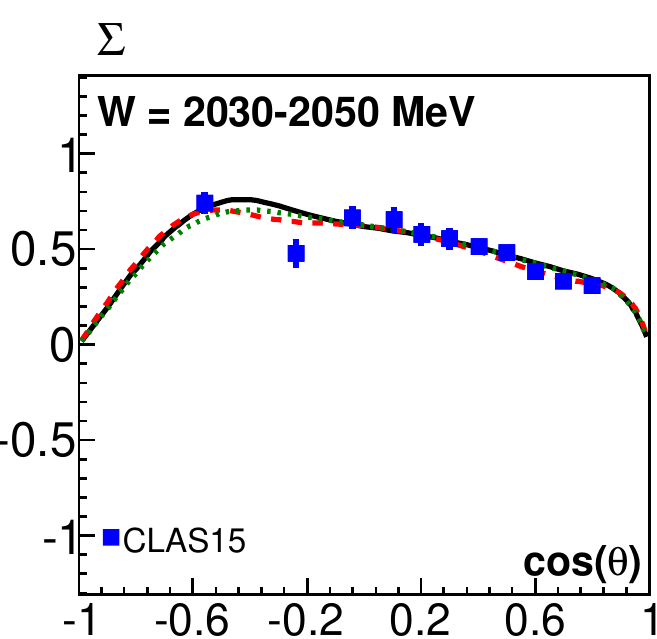}&
\hspace{-4mm}\includegraphics[width=0.165\textwidth,height=0.17\textwidth]{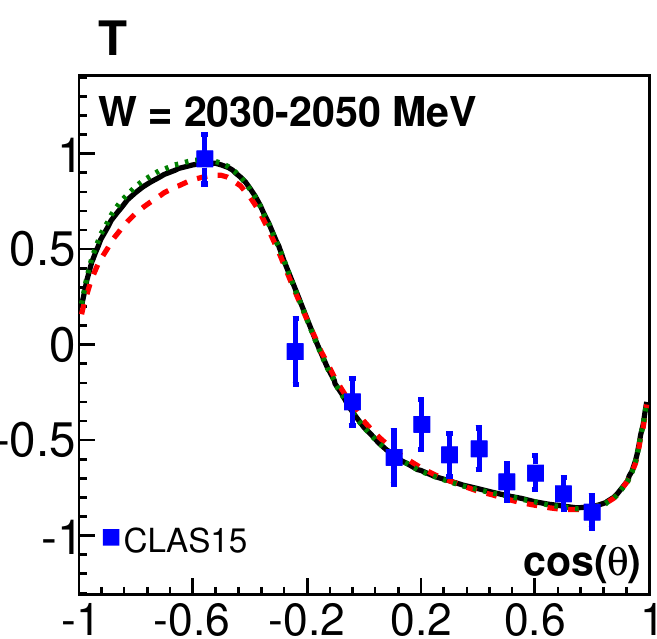}&
\hspace{-4mm}\includegraphics[width=0.165\textwidth,height=0.17\textwidth]{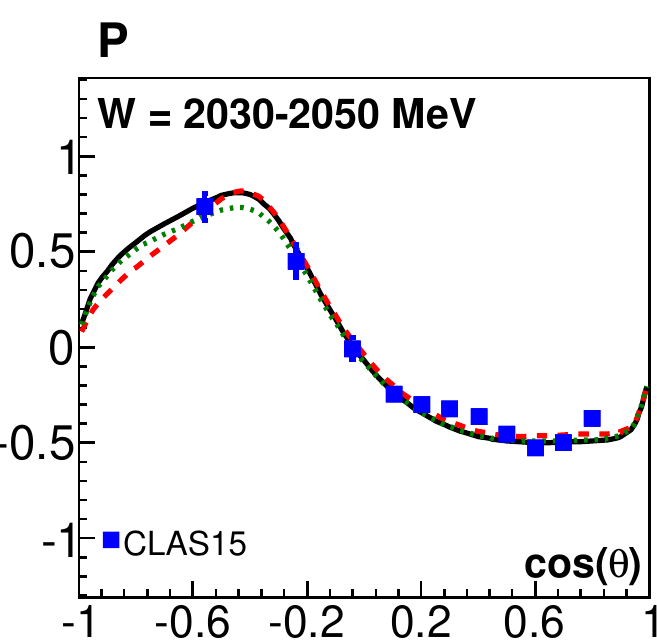}&
\hspace{-4mm}\includegraphics[width=0.165\textwidth,height=0.17\textwidth]{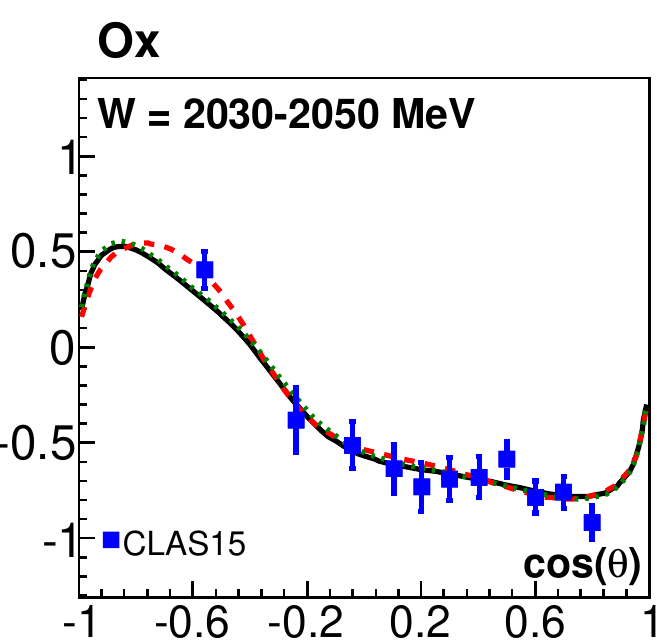}&
\hspace{-4mm}\includegraphics[width=0.165\textwidth,height=0.17\textwidth]{Ox_no_prime_observables2040.pdf}\\
\hspace{-3mm}\includegraphics[width=0.165\textwidth,height=0.17\textwidth]{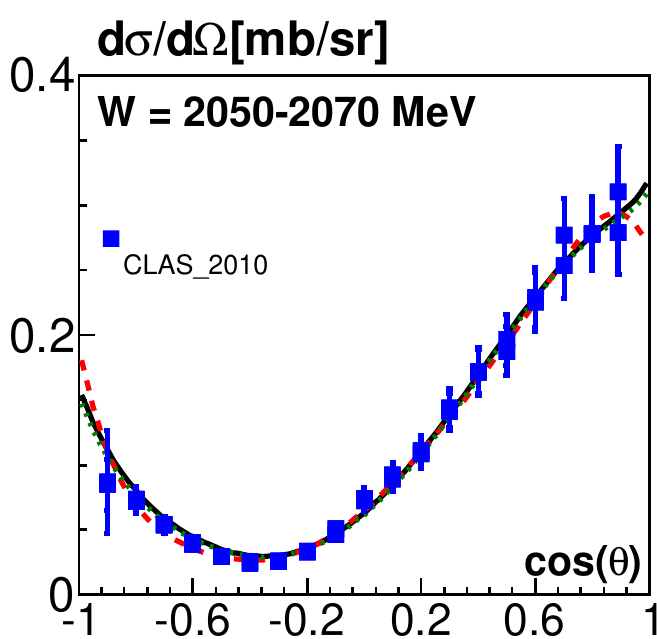}&
\hspace{-4mm}\includegraphics[width=0.165\textwidth,height=0.17\textwidth]{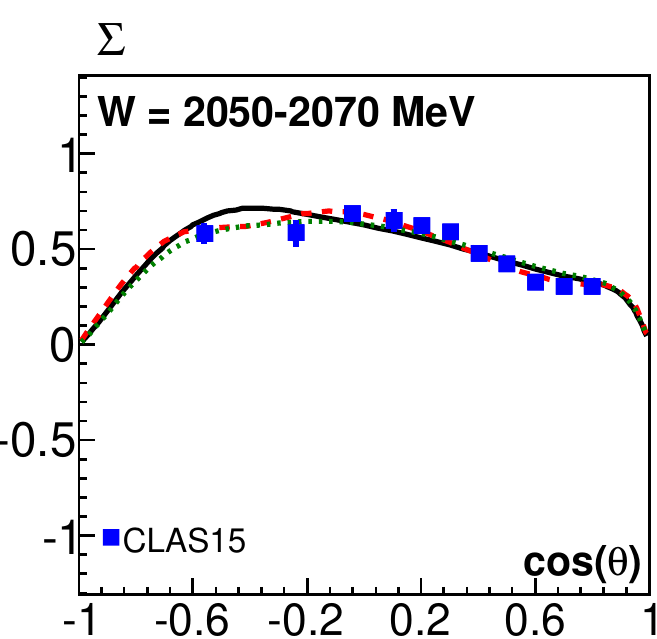}&
\hspace{-4mm}\includegraphics[width=0.165\textwidth,height=0.17\textwidth]{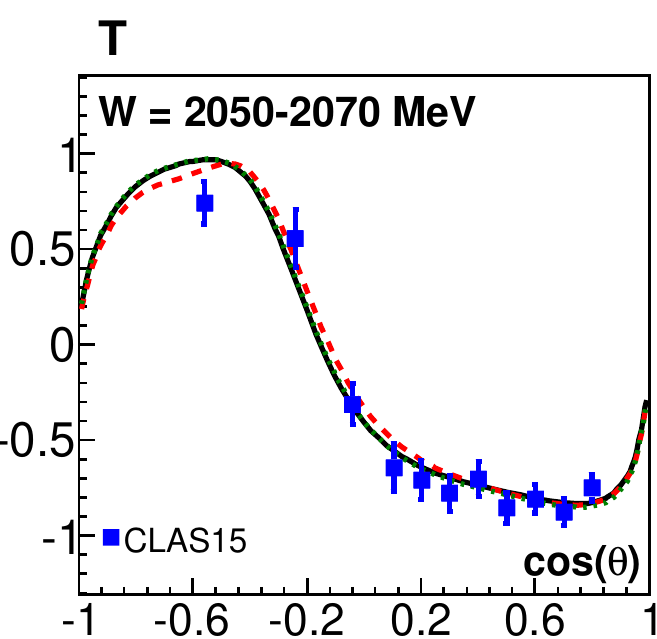}&
\hspace{-4mm}\includegraphics[width=0.165\textwidth,height=0.17\textwidth]{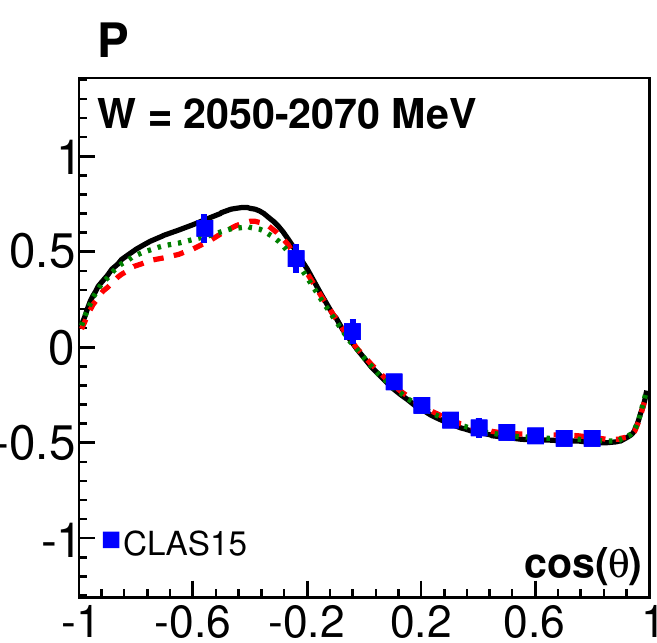}&
\hspace{-4mm}\includegraphics[width=0.165\textwidth,height=0.17\textwidth]{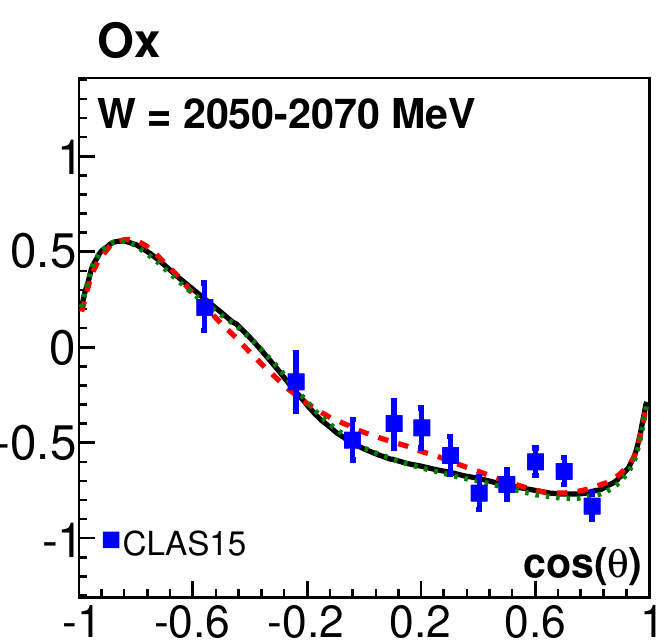}&
\hspace{-4mm}\includegraphics[width=0.165\textwidth,height=0.17\textwidth]{Ox_no_prime_observables2060.pdf}\\
\hspace{-3mm}\includegraphics[width=0.165\textwidth,height=0.17\textwidth]{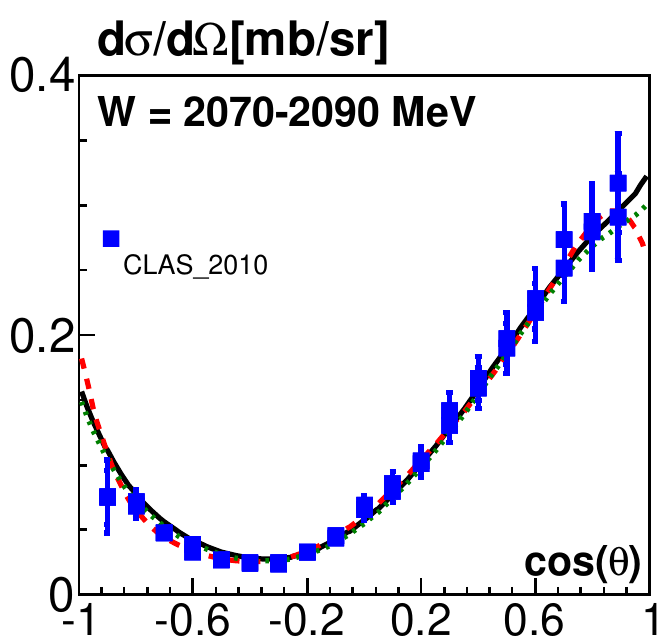}&
\hspace{-4mm}\includegraphics[width=0.165\textwidth,height=0.17\textwidth]{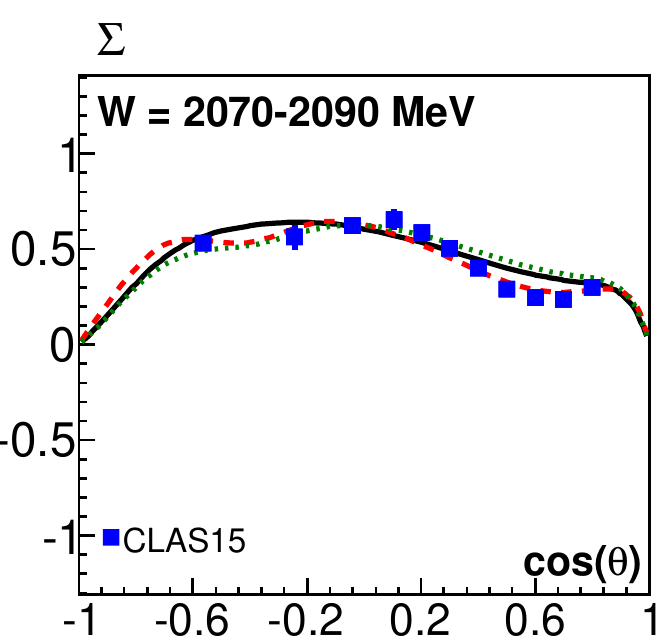}&
\hspace{-4mm}\includegraphics[width=0.165\textwidth,height=0.17\textwidth]{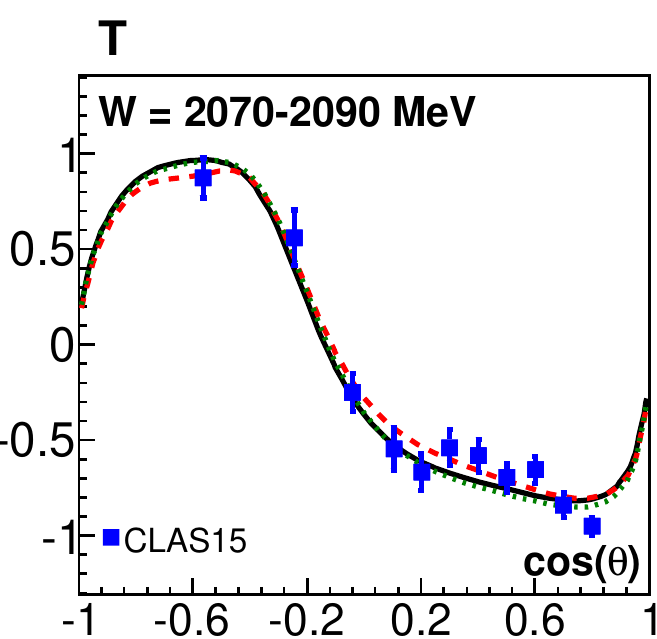}&
\hspace{-4mm}\includegraphics[width=0.165\textwidth,height=0.17\textwidth]{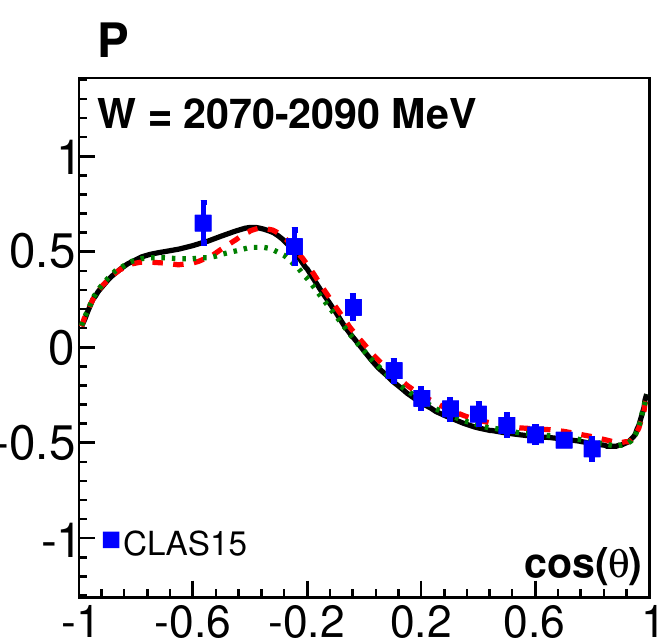}&
\hspace{-4mm}\includegraphics[width=0.165\textwidth,height=0.17\textwidth]{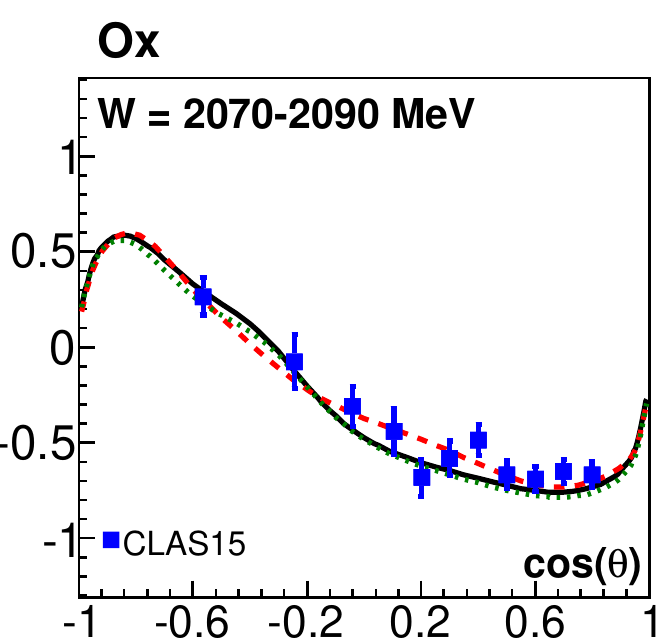}&
\hspace{-4mm}\includegraphics[width=0.165\textwidth,height=0.17\textwidth]{Ox_no_prime_observables2080.pdf}\\
\hspace{-3mm}\includegraphics[width=0.165\textwidth,height=0.17\textwidth]{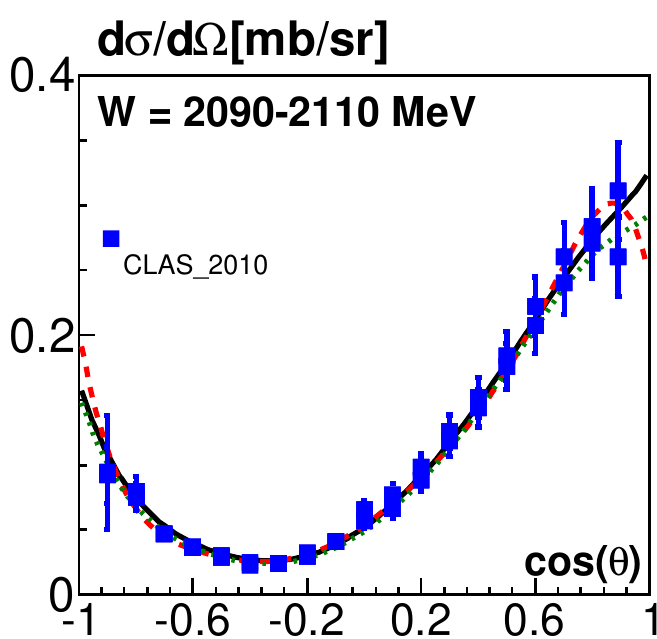}&
\hspace{-4mm}\includegraphics[width=0.165\textwidth,height=0.17\textwidth]{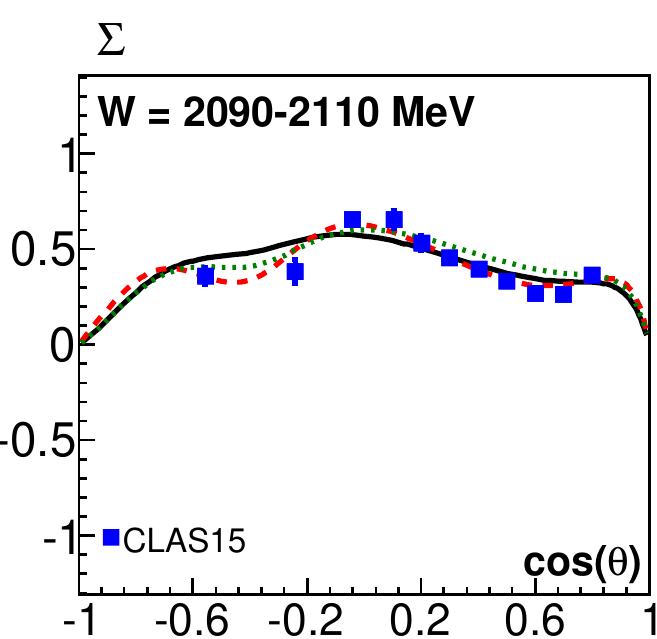}&
\hspace{-4mm}\includegraphics[width=0.165\textwidth,height=0.17\textwidth]{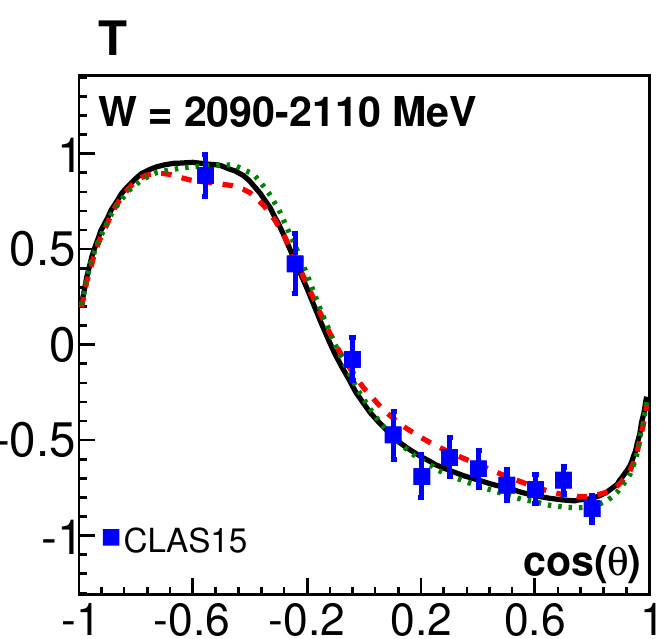}&
\hspace{-4mm}\includegraphics[width=0.165\textwidth,height=0.17\textwidth]{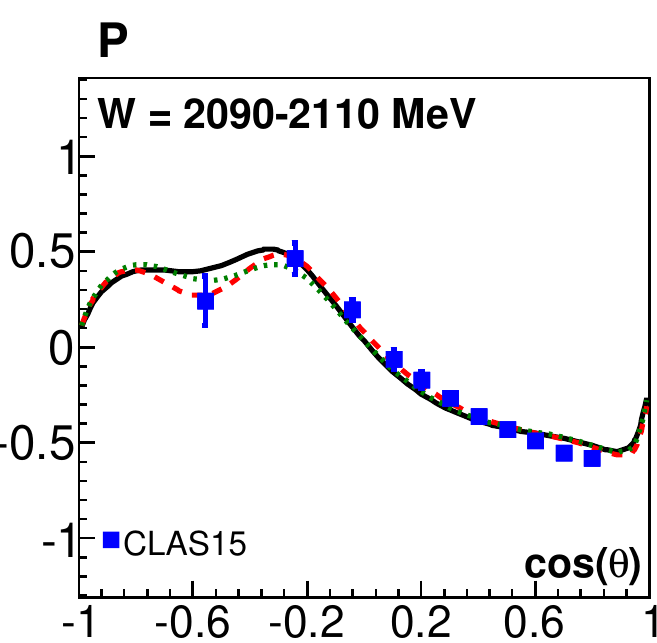}&
\hspace{-4mm}\includegraphics[width=0.165\textwidth,height=0.17\textwidth]{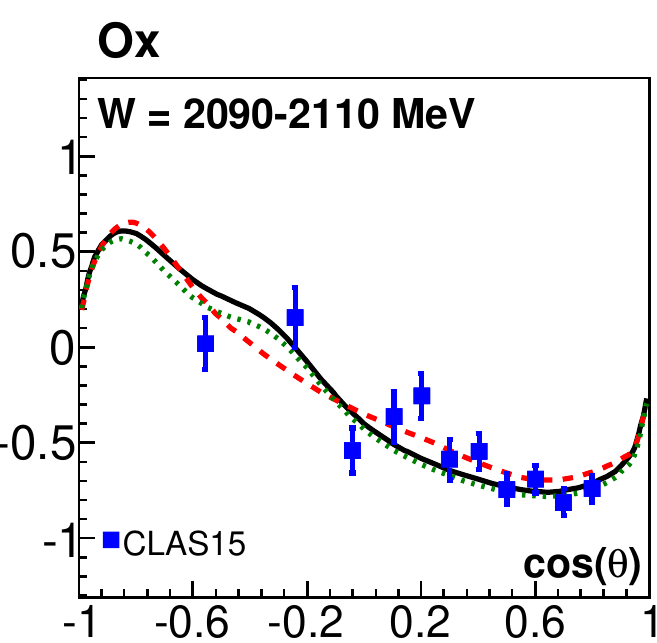}&
\hspace{-4mm}\includegraphics[width=0.165\textwidth,height=0.17\textwidth]{Ox_no_prime_observables2100.pdf}\\
\hspace{-3mm}\includegraphics[width=0.165\textwidth,height=0.17\textwidth]{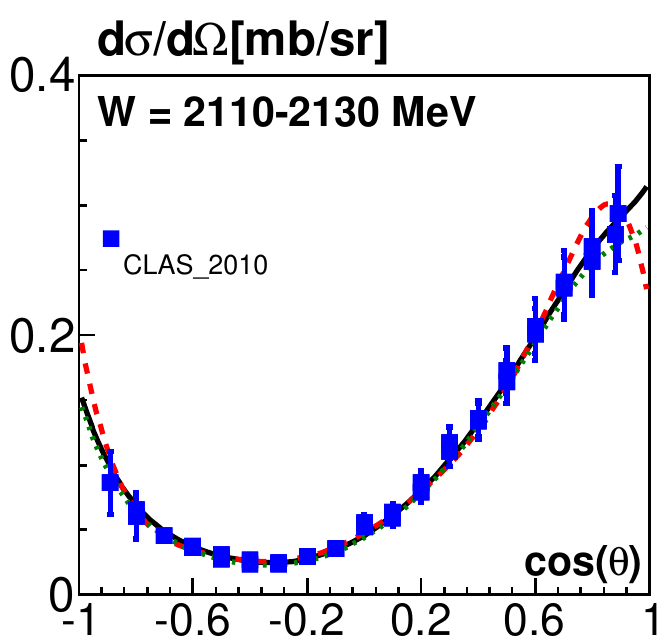}&
\hspace{-4mm}\includegraphics[width=0.165\textwidth,height=0.17\textwidth]{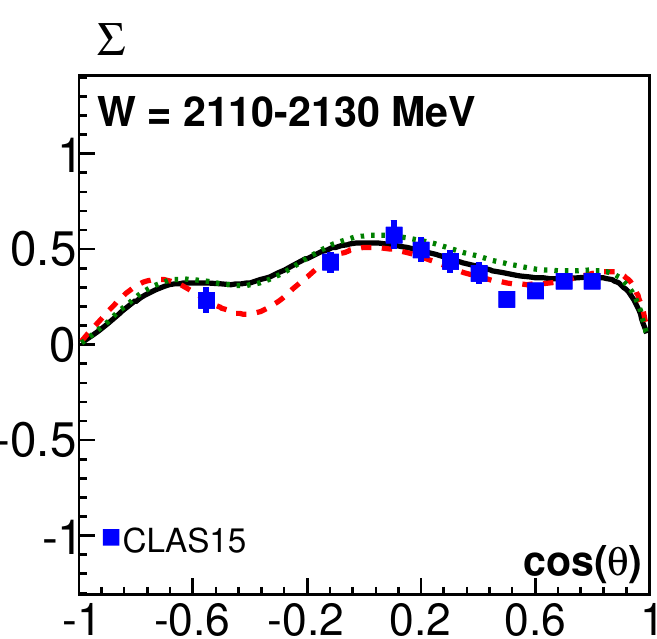}&
\hspace{-4mm}\includegraphics[width=0.165\textwidth,height=0.17\textwidth]{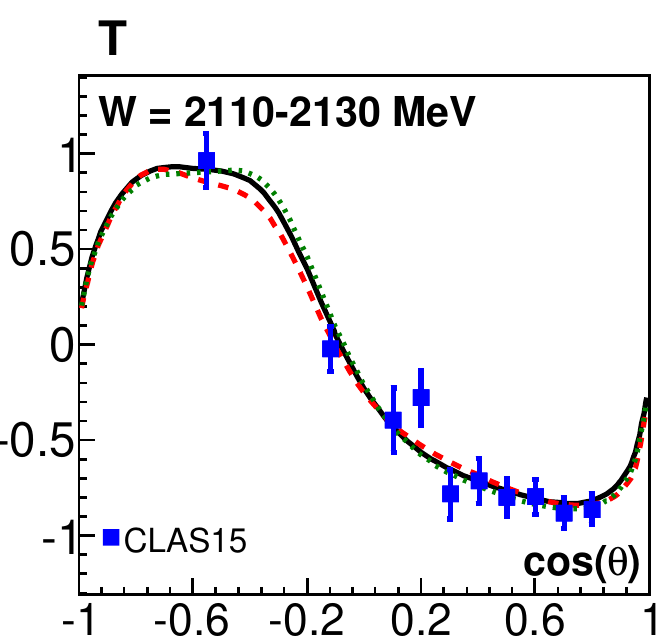}&
\hspace{-4mm}\includegraphics[width=0.165\textwidth,height=0.17\textwidth]{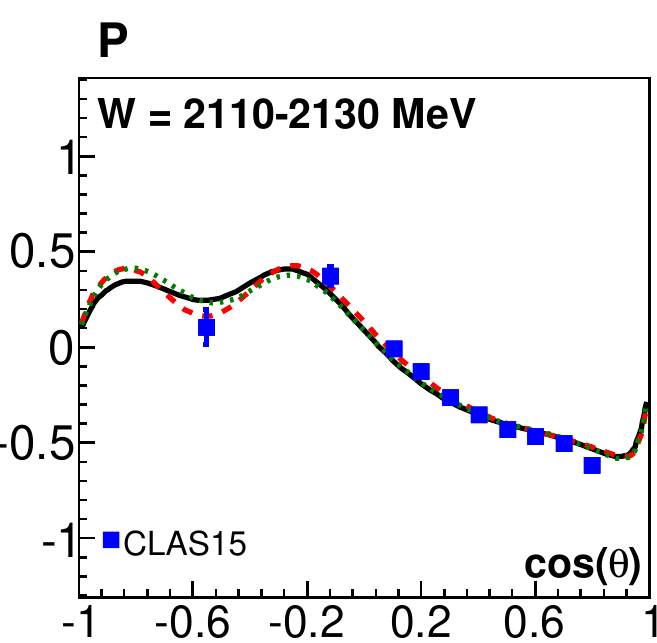}&
\hspace{-4mm}\includegraphics[width=0.165\textwidth,height=0.17\textwidth]{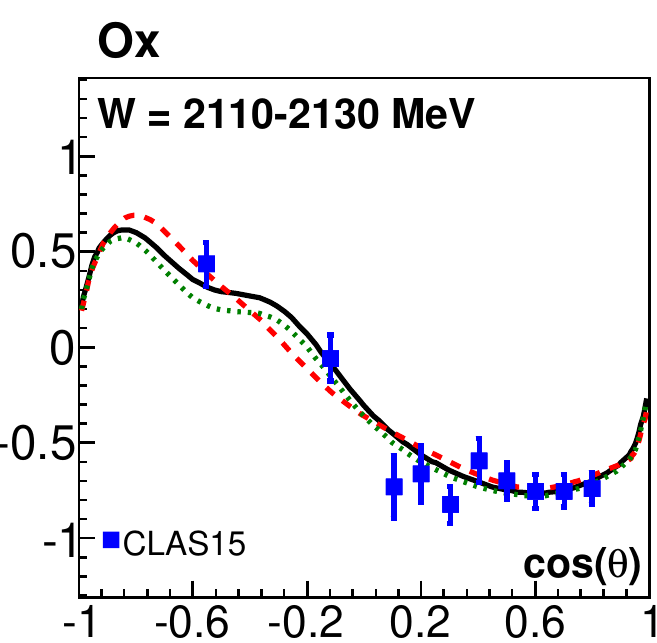}&
\hspace{-4mm}\includegraphics[width=0.165\textwidth,height=0.17\textwidth]{Ox_no_prime_observables2120.pdf}\\
\end{tabular}
\caption{\label{fig:data-g3}(Color online) Fit to the data on $d\sigma/d\Omega$:~\cite{McCracken:2009ra},
$P$~\cite{McCracken:2009ra}, and $\Sigma$, $T$, $O_x$, $O_z$~\cite{Paterson:2016vmc}  for $\gamma p\to K^+\Lambda$  reaction
for the mass range from 1990 to 2130\,MeV.
The solid (black) line corresponds the $L+P$ fit, the dashed (red)
line corresponds to fit used to
determine the multipoles of Fig.~\ref{fig:mult}., the dotted (green) line corresponds to BnGa fit.}
\end{figure*}

\begin{figure*}[pt]
\begin{tabular}{cccccc}
\hspace{-3mm}\includegraphics[width=0.165\textwidth,height=0.17\textwidth]{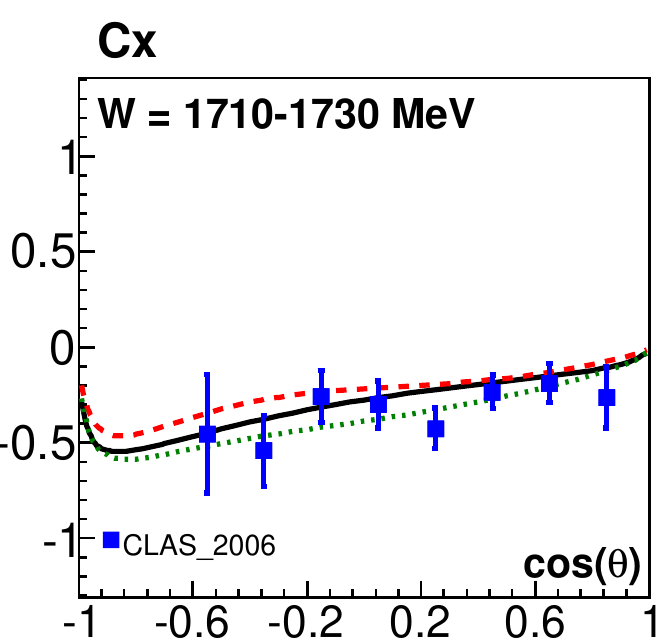}&
\hspace{-4mm}\includegraphics[width=0.165\textwidth,height=0.17\textwidth]{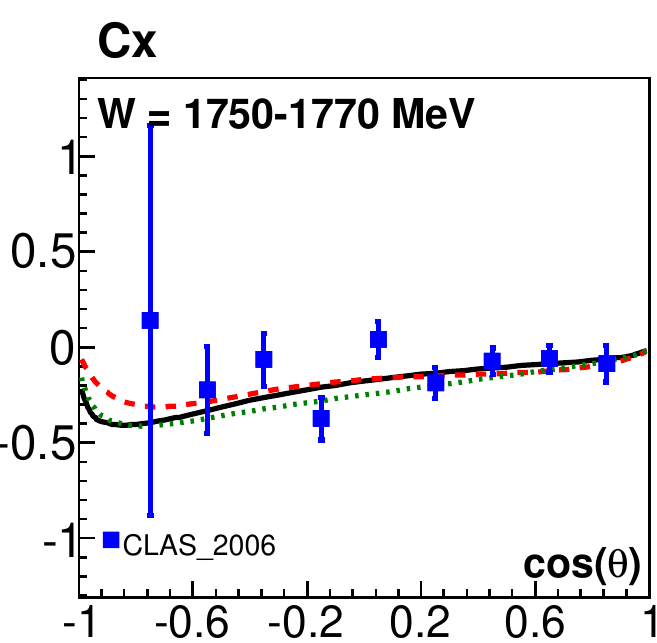}&
\hspace{-4mm}\includegraphics[width=0.165\textwidth,height=0.17\textwidth]{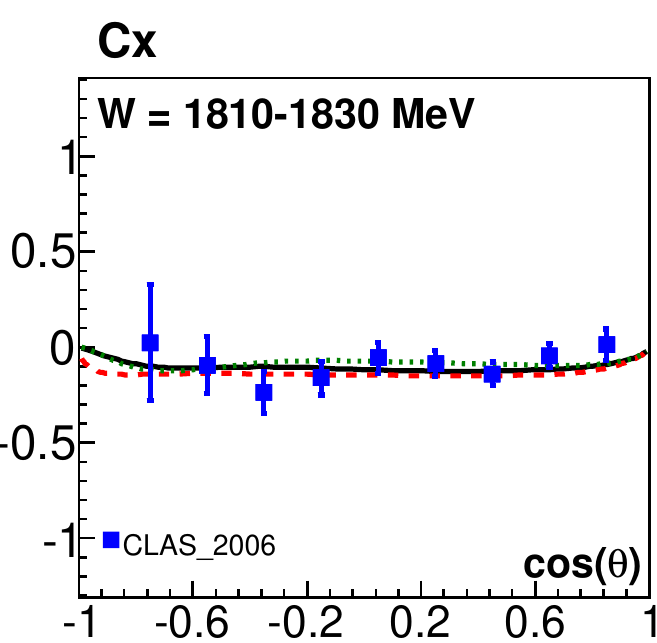}&
\hspace{-4mm}\includegraphics[width=0.165\textwidth,height=0.17\textwidth]{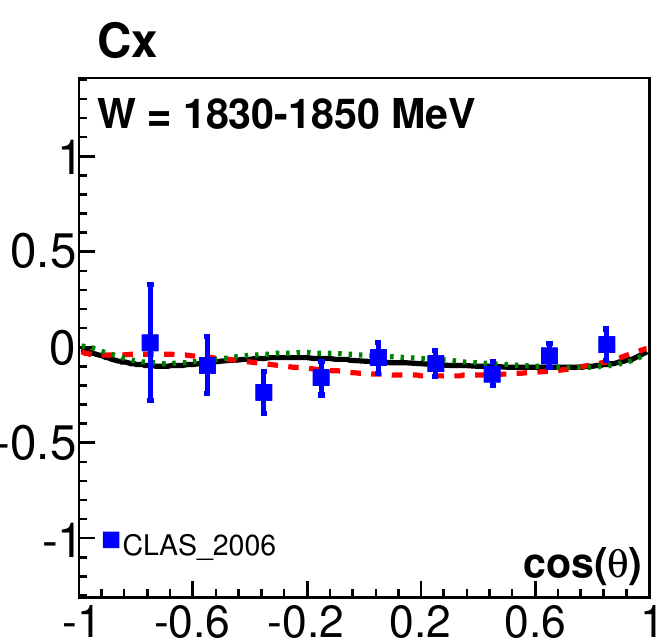}&
\hspace{-4mm}\includegraphics[width=0.165\textwidth,height=0.17\textwidth]{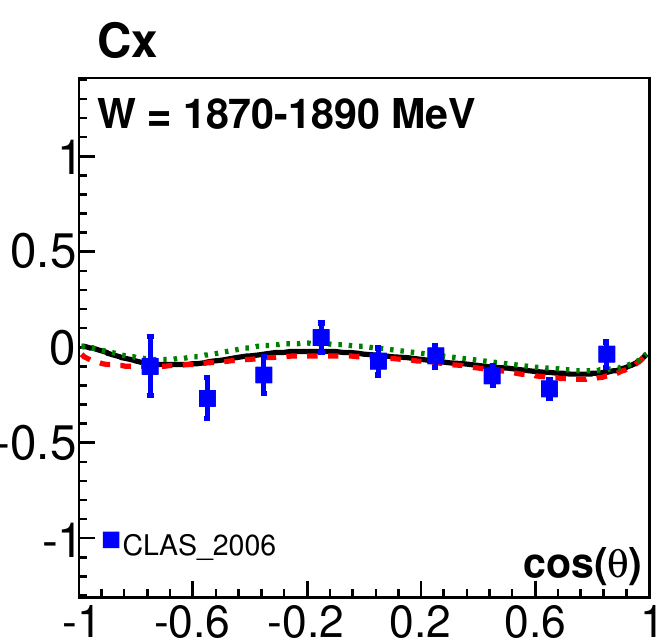}&
\hspace{-4mm}\includegraphics[width=0.165\textwidth,height=0.17\textwidth]{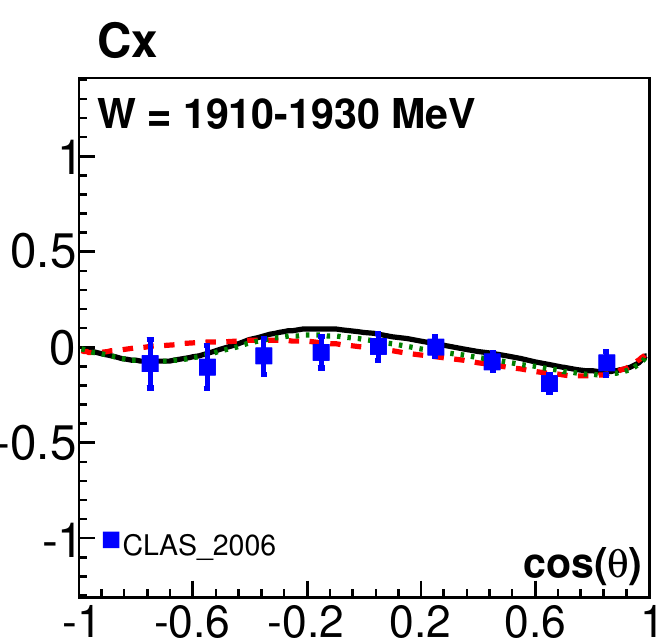}\\
\hspace{-3mm}\includegraphics[width=0.165\textwidth,height=0.17\textwidth]{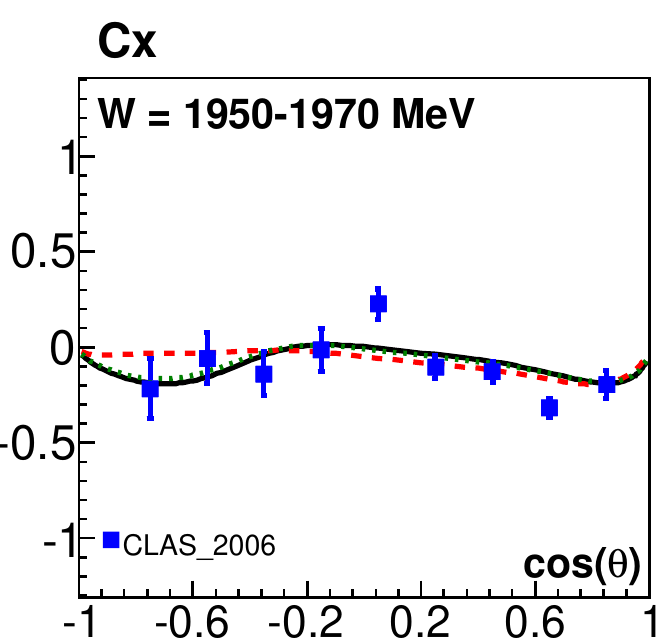}&
\hspace{-4mm}\includegraphics[width=0.165\textwidth,height=0.17\textwidth]{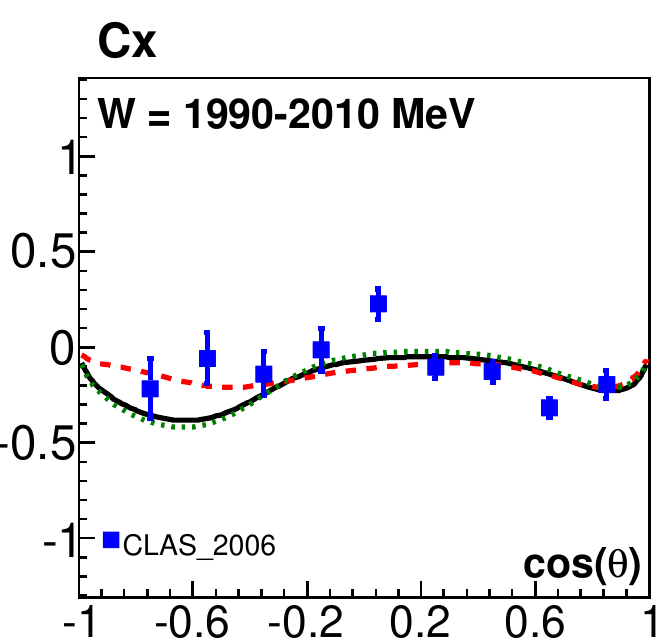}&
\hspace{-4mm}\includegraphics[width=0.165\textwidth,height=0.17\textwidth]{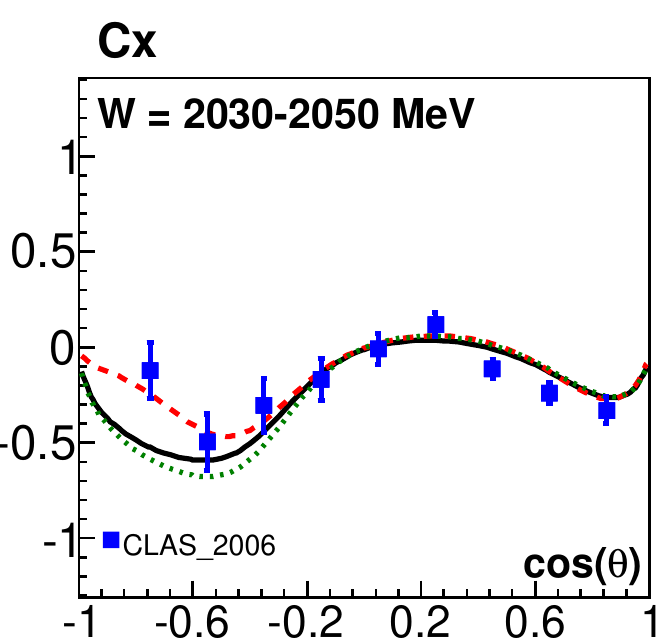}&
\hspace{-4mm}\includegraphics[width=0.165\textwidth,height=0.17\textwidth]{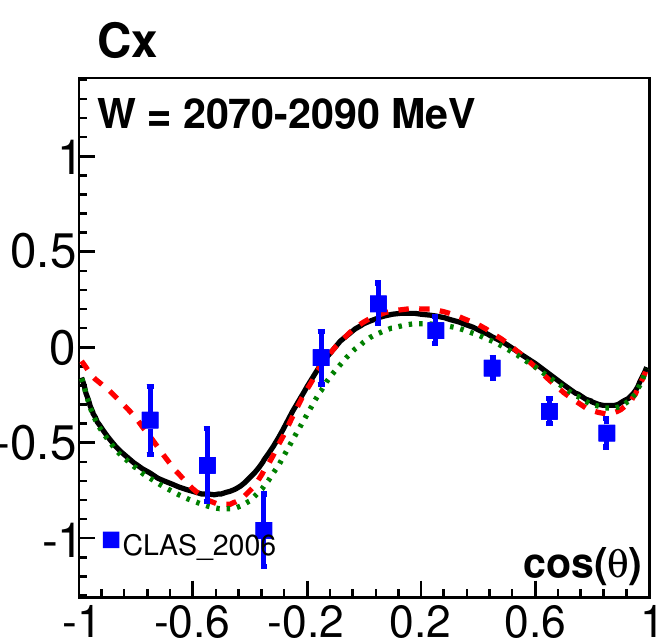}&
\hspace{-4mm}\includegraphics[width=0.165\textwidth,height=0.17\textwidth]{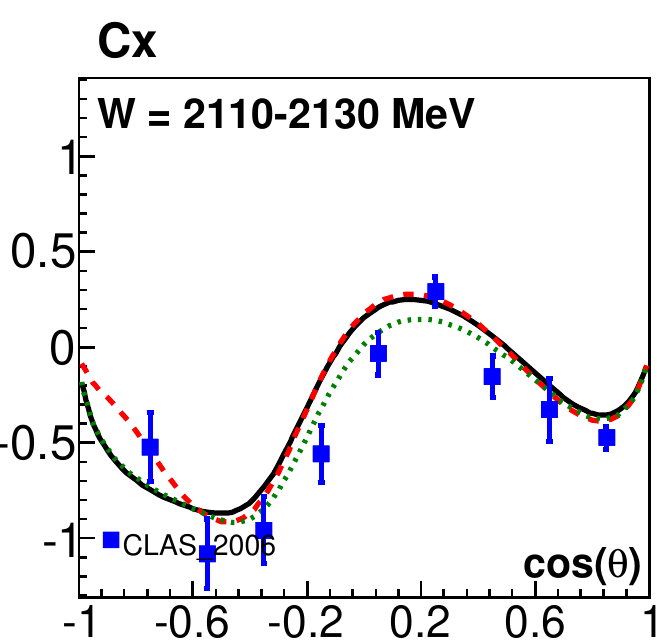}&
\hspace{-4mm}\includegraphics[width=0.165\textwidth,height=0.17\textwidth]{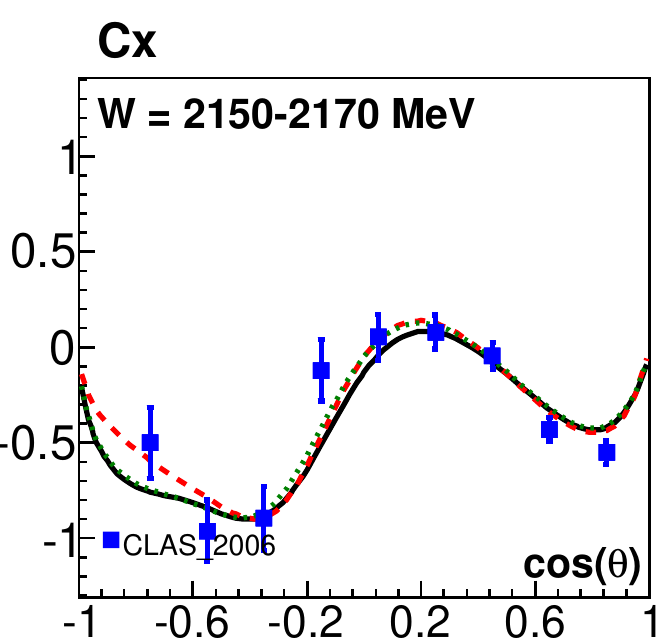}\\
\hspace{-3mm}\includegraphics[width=0.165\textwidth,height=0.17\textwidth]{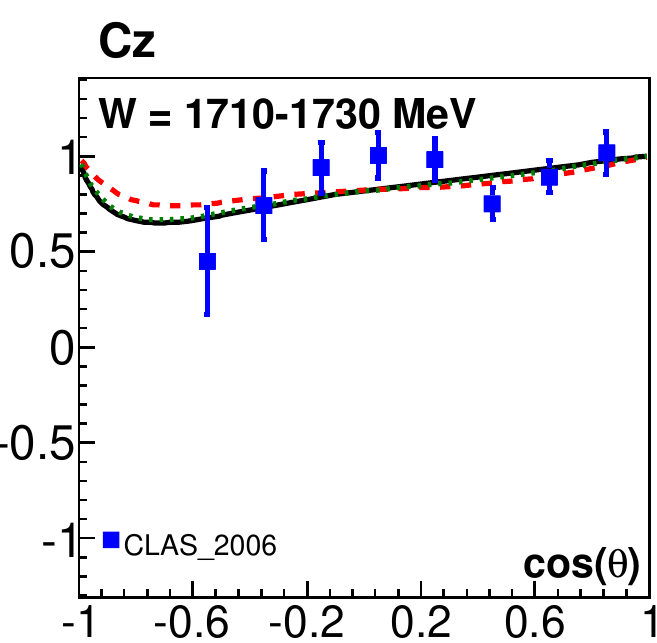}&
\hspace{-4mm}\includegraphics[width=0.165\textwidth,height=0.17\textwidth]{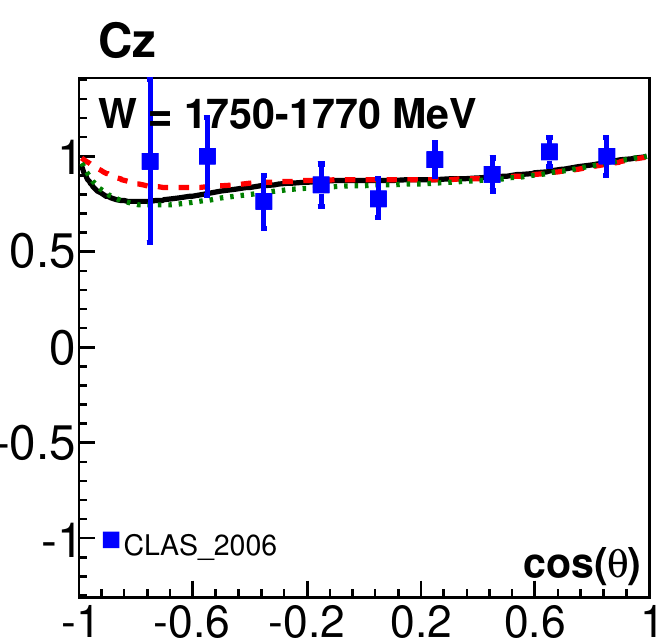}&
\hspace{-4mm}\includegraphics[width=0.165\textwidth,height=0.17\textwidth]{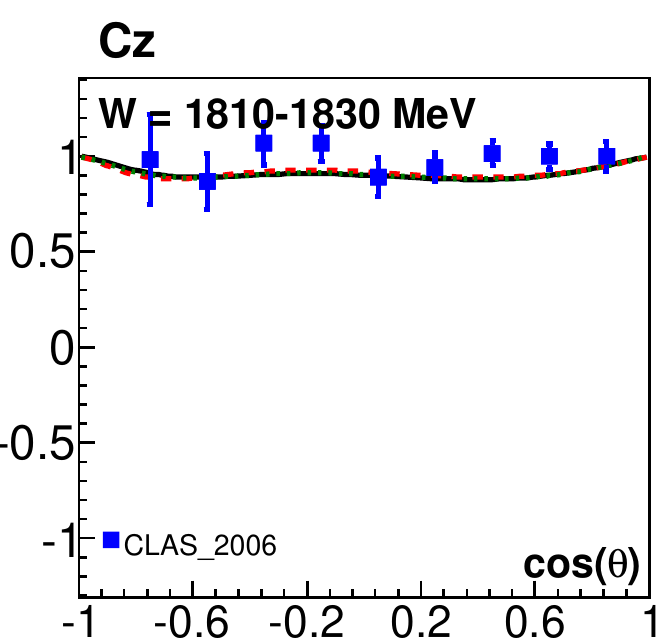}&
\hspace{-4mm}\includegraphics[width=0.165\textwidth,height=0.17\textwidth]{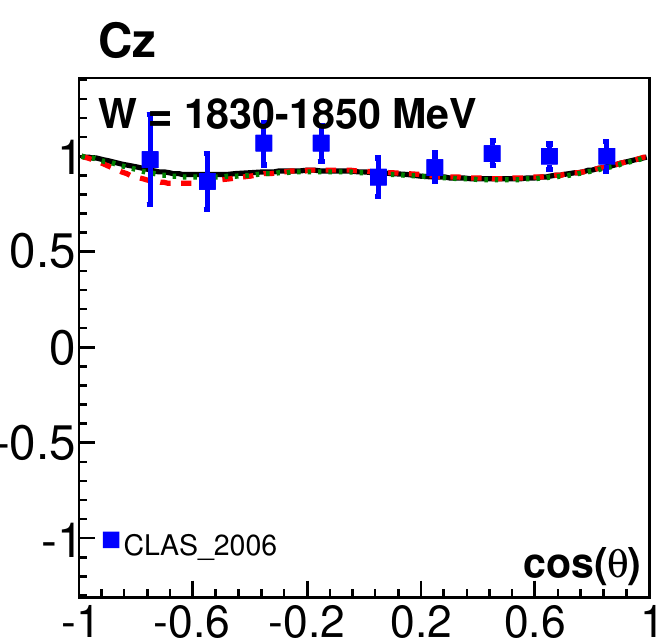}&
\hspace{-4mm}\includegraphics[width=0.165\textwidth,height=0.17\textwidth]{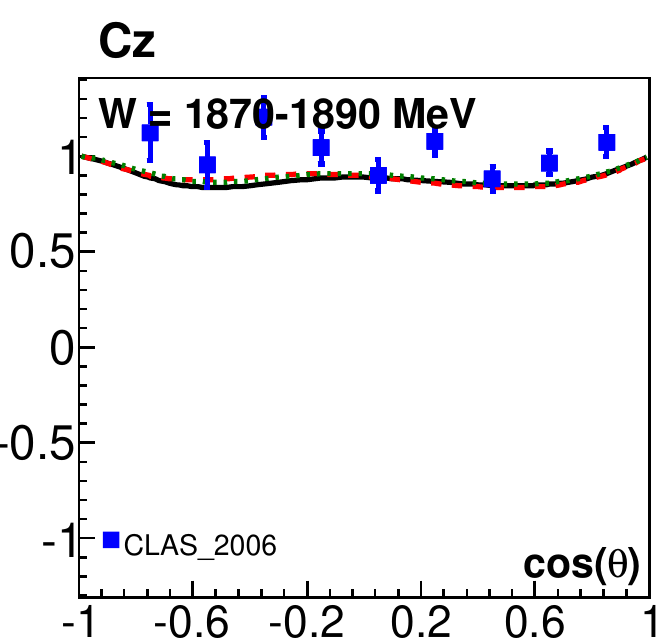}&
\hspace{-4mm}\includegraphics[width=0.165\textwidth,height=0.17\textwidth]{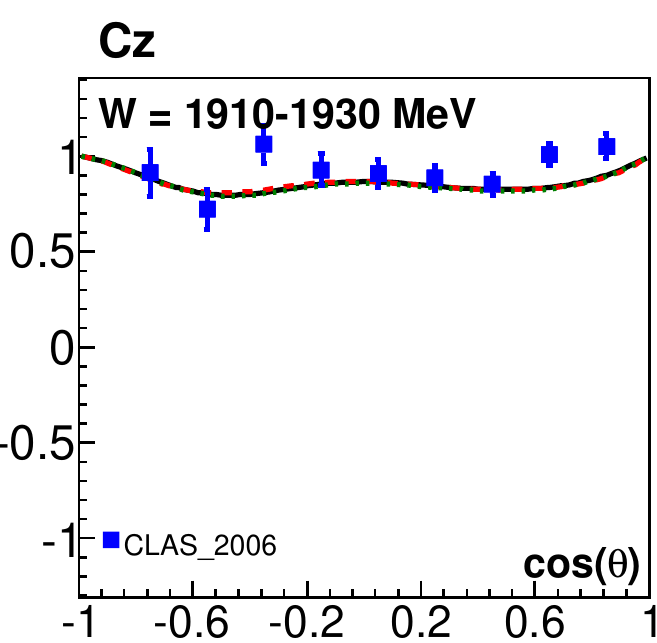}\\
\hspace{-3mm}\includegraphics[width=0.165\textwidth,height=0.17\textwidth]{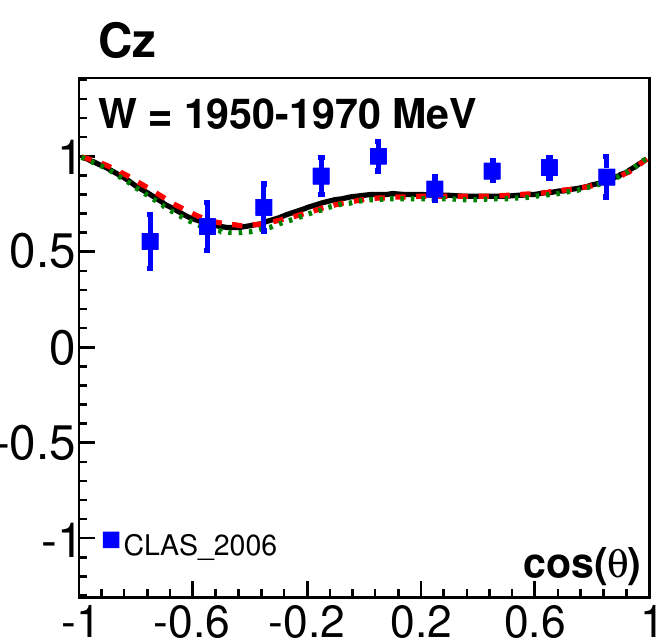}&
\hspace{-4mm}\includegraphics[width=0.165\textwidth,height=0.17\textwidth]{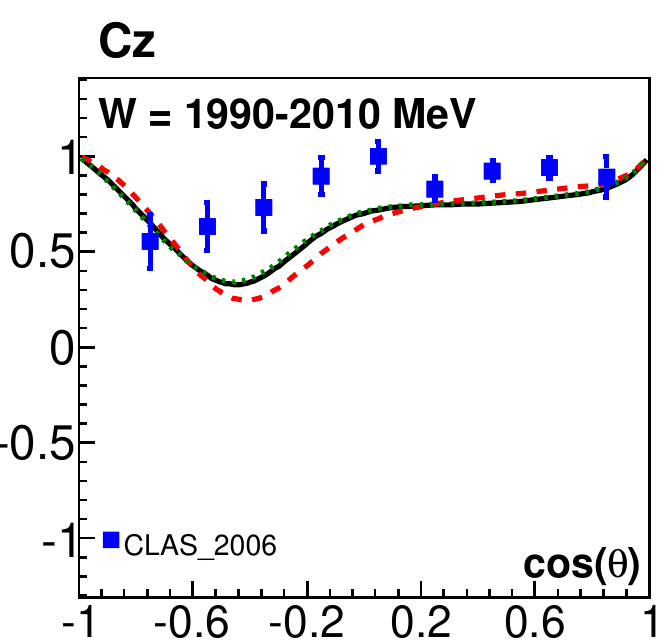}&
\hspace{-4mm}\includegraphics[width=0.165\textwidth,height=0.17\textwidth]{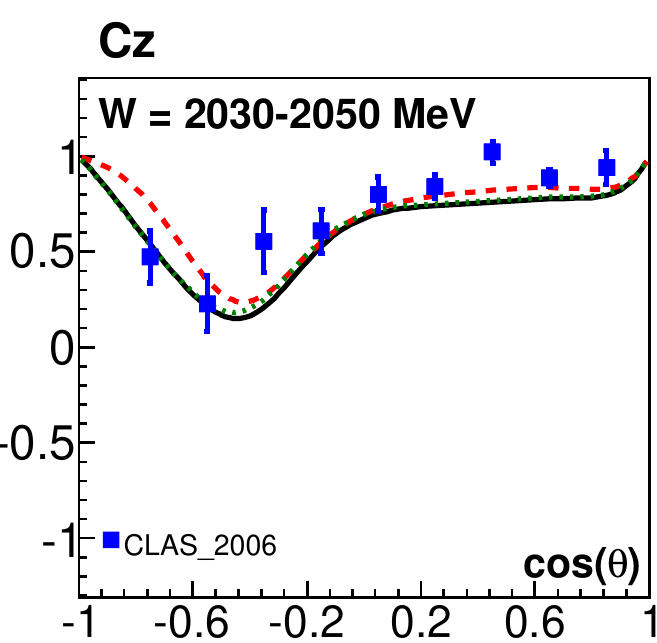}&
\hspace{-4mm}\includegraphics[width=0.165\textwidth,height=0.17\textwidth]{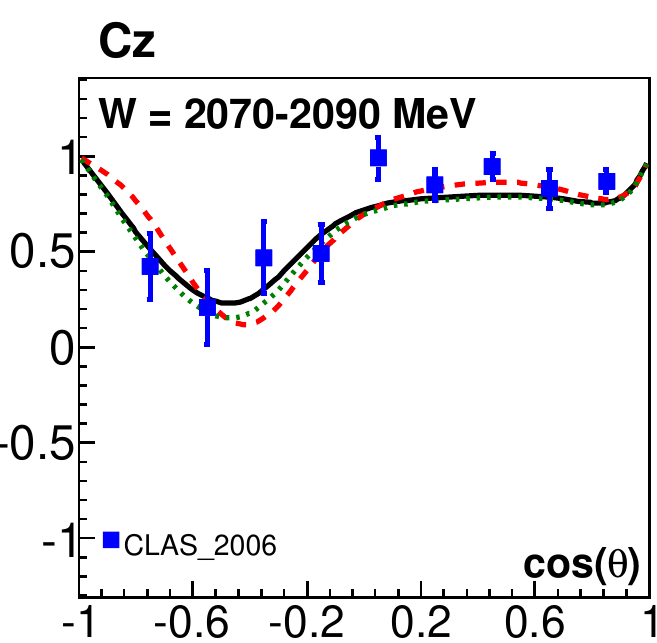}&
\hspace{-4mm}\includegraphics[width=0.165\textwidth,height=0.17\textwidth]{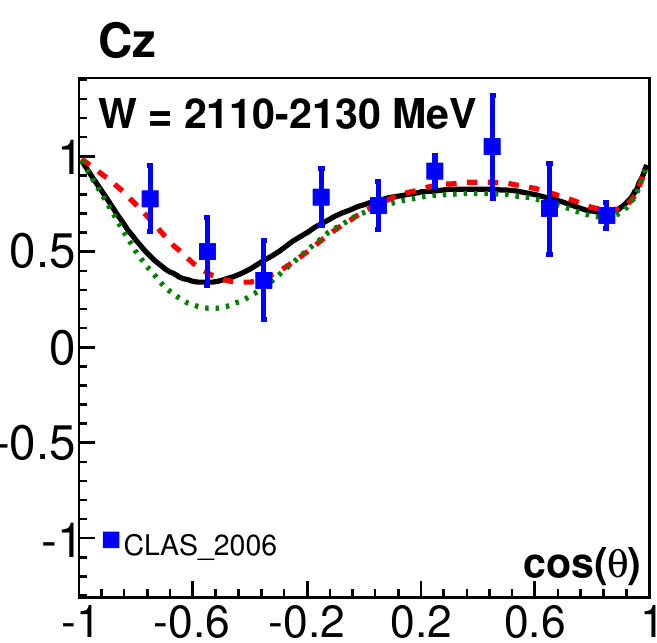}&
\hspace{-4mm}\includegraphics[width=0.165\textwidth,height=0.17\textwidth]{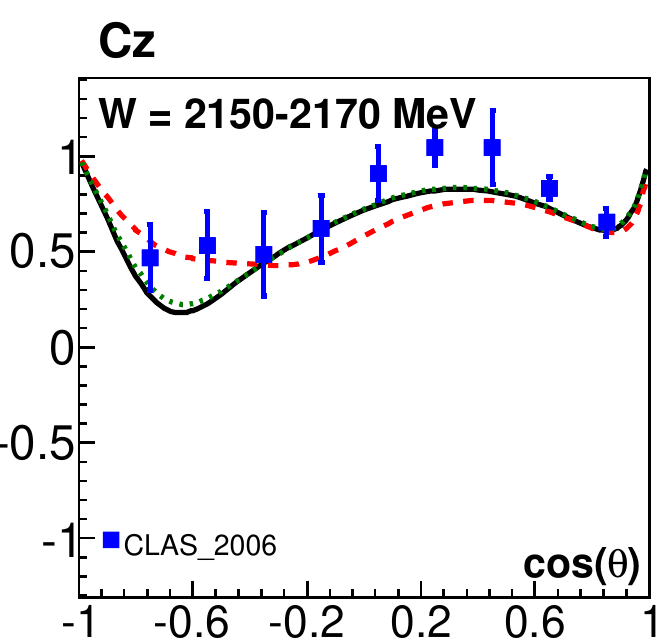}\\
\end{tabular}
\caption{\label{fig:data-g4}(Color online) Fit to the data on $C_x$ and $C_z$
\cite{Bradford:2006ba}  for $\gamma p\to K^+\Lambda$  reaction.
The solid (black) line corresponds the $L+P$ fit, the dashed (red)
line corresponds to fit used to
determine the multipoles of Fig.~\ref{fig:mult}., the dotted (green) line corresponds to BnGa fit.}
\end{figure*}

\subsubsection{Fits to the data}
From the results of the BnGa analysis we expect that in the energy range
considered here the $E_{0+}$, $M_{1-}$, and $E_{1+}$ yield the largest
contributions, followed by $M_{1+}$, $E_{2-}$, and $M_{2-}$. The $E_{2+}$,
$M_{2+}$, $M_{3-}$, $E_{3-}$, $M_{3+}$, $E_{3+}$, $M_{4-}$, $E_{4-}$ all
contribute with increasingly smaller importance, higher multipoles become
negligible. First fits showed that it is not possible, given the statistical
and systematic accuracy of the data, to determine all significant partial waves. Due to strong
correlations between the parameters, the errors became large and the resulting
multipoles showed large point-to-point fluctuations. Hence we decreased the
number of freely fitted multipoles; the higher multipoles were fixed to the
BnGa results. These multipoles are shown in Fig.~\ref{fig:high-l}. Reasonably
small errors were obtained when the four multipoles $E_{0+}$, $M_{1-}$,
$E_{1+}$, and $M_{1+}$ were fitted. The errors increased only slightly when the
multipoles $E_{2-}$, $M_{2-}$, and $E_{2+}$ were fitted in addition but
constrained to the BnGa solution by a penalty function.
 \be
 \chi^2_{pen} = \sum_{\alpha} \frac{(M_\alpha - \hat M_\alpha)^2}{(\delta
\hat M_\alpha)^2} +\sum_{\alpha} \frac{(E_\alpha- \hat E_\alpha)^2}{(\delta
\hat E_\alpha)^2} \label{penalty_1}
\ee
 where $\hat E_\alpha$ and $\hat
M_\alpha$ are the electric and magnetic multipoles from solution with $E_{2-}$,
$M_{2-}$, and $E_{2+}$ fitted freely; $\delta \hat E_\alpha$, $\delta \hat
M_\alpha$ are the multipole errors.

The reaction $\gamma p\to K^+\Lambda$ has been studied extensively by the CLAS collaboration.
The early measurement of the differential cross sections $d\sigma/d\Omega$ \cite{Bradford:2005pt}
was later superseded by a new measurement reporting the differential cross sections and the
recoil polarization \cite{McCracken:2009ra}. The spin transfer from circularly polarized
photons to the final-state $\Lambda$ hyperon, the quantities $C_x$ and $C_z$, were reported in
 \cite{Bradford:2006ba}. The polarization observables $\Sigma, T, O_x, O_z$ have been determined
recently \cite{Paterson:2016vmc}. The data are shown in Figs.~\ref{fig:data-g1}-\ref{fig:data-g3}.
The data are used to determine the photoproduction multipoles in a truncated partial wave
analysis.

\begin{figure*}[pt]
\begin{center}
\begin{tabular}{cccc}
\includegraphics[width=0.245\textwidth,height=0.245\textwidth]{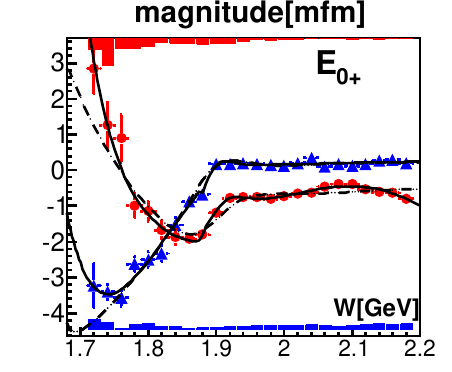}&
\hspace{-2mm}\includegraphics[width=0.245\textwidth,height=0.245\textwidth]{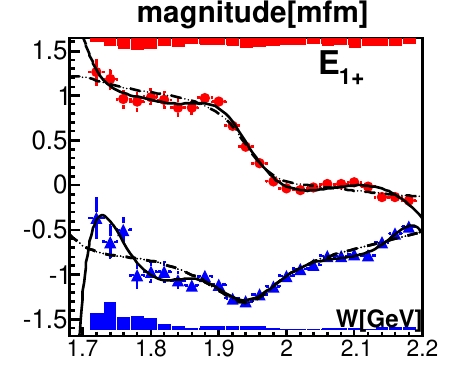}&
\hspace{-5mm}\includegraphics[width=0.25\textwidth,height=0.25\textwidth]{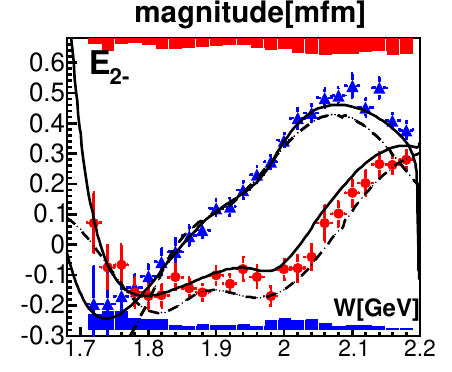}&
\hspace{-6mm}\includegraphics[width=0.25\textwidth,height=0.25\textwidth]{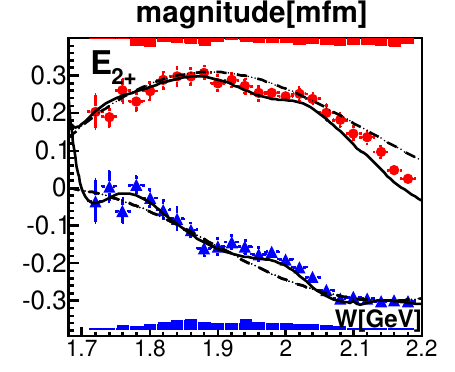}\\
\includegraphics[width=0.245\textwidth,height=0.245\textwidth]{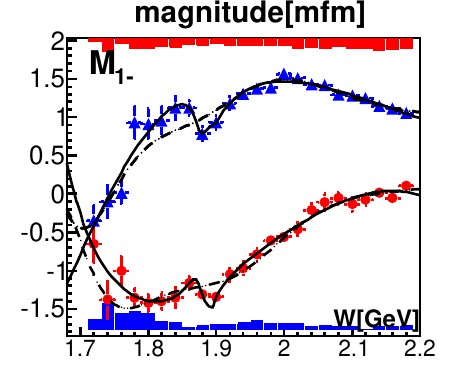}&
\hspace{-2mm}\includegraphics[width=0.245\textwidth,height=0.245\textwidth]{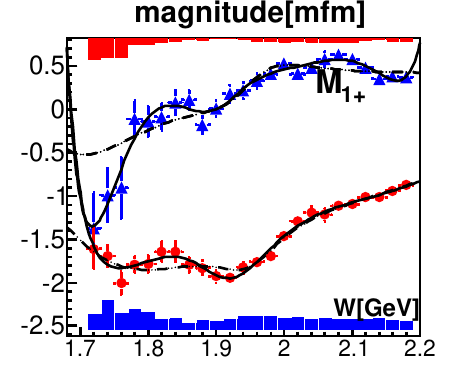}&
\hspace{-5mm}\includegraphics[width=0.25\textwidth,height=0.25\textwidth]{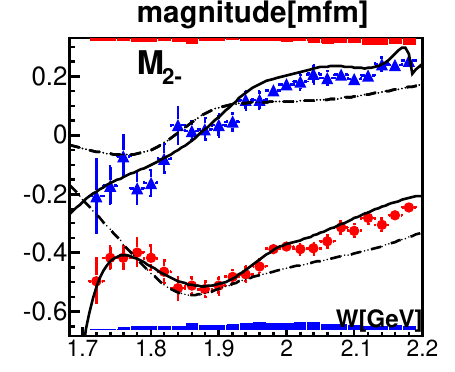}&
\end{tabular}
\end{center}
\caption{\label{fig:mult}(Color online) Real and imaginary parts of the multipoles
derived from fits to the data shown in Figs.~\ref{fig:data-g1}-\ref{fig:data-g3}.
The multipoles $E_{0+}$, $E_{1+}$, $M_{1-}$, and $M_{1+}$ are fitted freely, the
$E_{2-}$, $E_{2+}$, and $M_{2-}$ are forced by a penalty function to stay close to the
BnGa fit. Higher-order multipoles are taken from the BnGa fit. The triangles (blue) mark
the real part of the multipole, circles (red) mark the imaginary part of the multipole.
Solid lines are L+P fits, dashed lines represent the BnGa fit. The
grey (blue/red) areas denote the systematic errors derived from the spread when different
BnGa solutions were used to determine the multipoles. The multipoles are given in mfm
(milli-fermi=attometer).
 \vspace{3mm}  }
\begin{center}
\begin{tabular}{cccc}
\hspace{-3mm}\includegraphics[width=0.25\textwidth,height=0.25\textwidth]{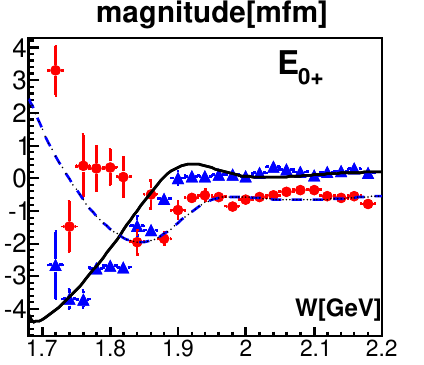}&
\hspace{-4mm}\includegraphics[width=0.25\textwidth,height=0.25\textwidth]{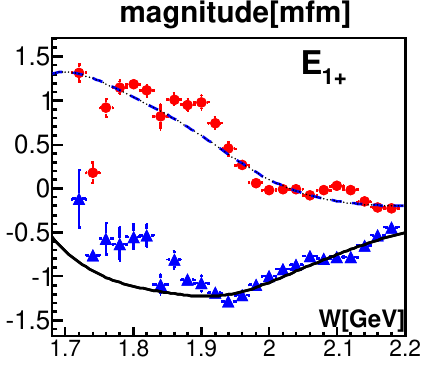}&
\hspace{-3mm}\includegraphics[width=0.25\textwidth,height=0.25\textwidth]{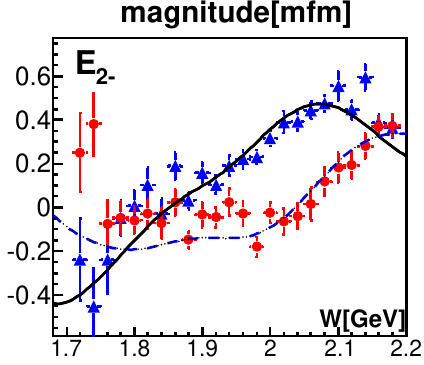}&
\hspace{-2mm}\includegraphics[width=0.25\textwidth,height=0.25\textwidth]{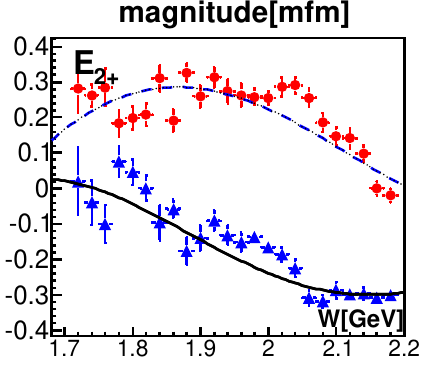}\\
\hspace{-6mm}\includegraphics[width=0.25\textwidth,height=0.25\textwidth]{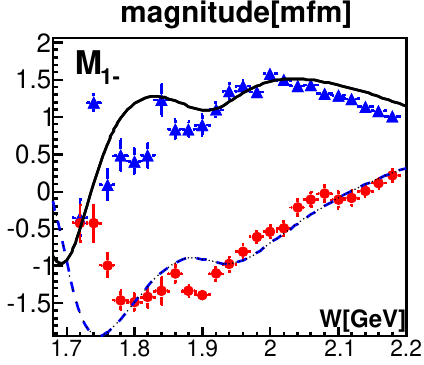}&
\hspace{-4mm}\includegraphics[width=0.25\textwidth,height=0.25\textwidth]{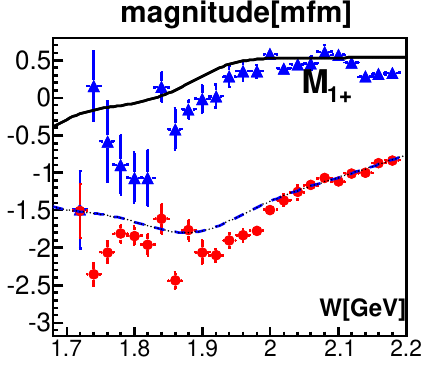}&
\hspace{-3mm}\includegraphics[width=0.25\textwidth,height=0.25\textwidth]{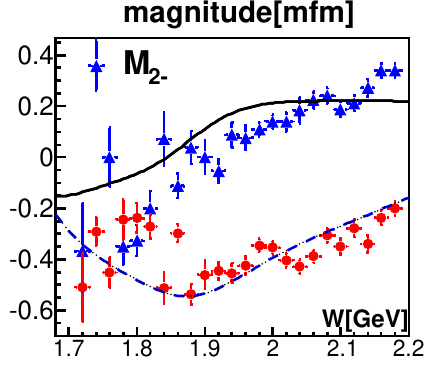}&
\end{tabular}
\end{center}
\caption{\label{fig:test}(Color online) Real and imaginary parts of the $E_{0+}$, $E_{1+}$,
$M_{1-}$, $M_{1+}$,  $E_{2-}$, $E_{2+}$, and $M_{2-}$. These multipoles were fitted freely.
Higher-order multipoles are taken from the BnGa fit. The triangles (blue) mark the real part
of the multipole, circles (red) mark the imaginary part of the multipole.
Solid and dash lines represent the BnGa fit. The multipoles are given in mfm
(milli-fermi=attometer).}
\end{figure*}

The final result for the multipoles are shown in Fig.~\ref{fig:mult}. Strong variations are observed.
The imaginary parts of all multipoles, except $M_{2-}$ and $E_{2+}$, show threshold enhancements due
to $N(1650)1/2^-$ ($E_{0+}$), $N(1710)1/2^+$ ($M_{1-}$), $N(1720)3/2^+$ ($E_{1+}$ and $M_{1+}$),
$N(1700)3/2^-$ ($E_{2-}$). Further structures are clearly seen at about 1900\,MeV in the $E_{0+}$,
$M_{1-}$, $E_{1+}$, $M_{1+}$, $E_{2-}$, $M_{2-}$ multipoles.

These structures emerge reliably when the multipole series is truncated, and only few multipoles
are fitted freely. In Fig.~\ref{fig:test}
we show the results from one of our tests. In this case, the seven largest multipoles,
$E_{0+}$, $M_{1-}$, $E_{1+}$, $M_{1+}$, $E_{2-}$, $M_{2-}$, and $E_{2+}$  were all left free. In several
mass bins, the resulting multipoles show an erratic behavior; the results become unstable.
Likewise, it was important to include the multipoles with large orbital angular momenta. Even though
they are individually all small, neglecting them (by assuming that they are identically zero)
leads to biased results. Furthermore, these multipoles fix the overall phase.

Sandorfi, Hoblit, Kamano, and Lee \cite{Sandorfi:2010uv} have reconstructed the photoproduction
amplitudes for the reaction $\gamma p\to K^+\Lambda$. For the high partial waves, they used the
Born amplitude. Partly, they fitted all waves with $L\leq 3$ freely and determined the phases
as differences to the $E_{0+}$ phase. In other fits, they had the $E_{0+}$ phase free and
fitted all waves with $L\leq 2$. The resulting multipoles showed a wide spread. They concluded that
a very significant increase in solid-angle coverage and statistics is required when all partial waves
up to $L=3$ are to be determined.

\section{BnGa fits to the data}

The BnGa partial wave analysis uses a $K$ matrix formalism to fit data on
pion and photo-induced reactions to extract the leading singularities
of the scattering or production processes. The formalism is described in
detail in a series of publications
\cite{Anisovich:2004zz,Anisovich:2006bc,Anisovich:2007zz,Denisenko:2016ugz}.
Here we briefly outline the dynamical part of the method.

The pion induced reaction $\pi^-p\to K^0\Lambda$ from the initial state
$i=\pi^-p$ to the final state $j= K^0\Lambda$ is described by a
partial wave amplitude $A^{(\beta)}_{ij}$. It is given by a $K$-matrix
which incorporates a summation of resonant and non-resonant terms in
the form
\be
A^{(\beta)}_{ij}=\sqrt{\rho_i} \sum\limits_a
K^{(\beta)}_{ia}\left(I-i\rho K^{(\beta)}\right
)^{-1}_{aj}\sqrt{\rho_j}\,\,.
\label{res_amp}
\ee
The multi-index $\beta$ denotes the quantum numbers of the partial wave,
it is suppressed in the following. The factor $\rho$ represents
the phase space matrix to all allowed intermediate states,
$\rho_{i}$, $\rho_{j}$ are the phase space factors for the initial and the final
state. The $K$ matrix parametrizes resonances
and background contributions:
\begin{equation}
K_{ab}=\sum\limits_\alpha \frac{g^{\alpha}_a
g^{\alpha}_b}{M_\alpha^2-s}+f_{ab}\,.
\end{equation}
Here $g^{\alpha}_{a,b}$ are coupling constants of the pole $\alpha$
to the initial and the final state. The background terms $f_{ab}$
describe non-resonant transitions from the initial to the final
state.

For photoproduction reactions, we use the helicity ($h$)-dependent amplitude for
photoproduction $a^h_{b}$ of the final state $b$ \cite{Chung:1995dx}
\begin{eqnarray}
a^h_{b}&=&P^h_a(I-i\rho K)_{ab}^{-1}\qquad {\rm where}\\
P^h_a&=&\sum\limits_\alpha \frac{A^h_{\alpha}
g^{\alpha}_a}{M_\alpha^2-s}+F_{a}\,.
\end{eqnarray}
$A^h_{\alpha}$ is the photo-coupling of a pole $\alpha$
and $F_{a}$ a non-resonant transition. The helicity amplitudes $A^{1/2}_\alpha$, $A^{3/2}_\alpha$ are defined as residues of the helicity-dependent amplitude at the pole
position and are complex numbers \cite{Workman:2013rca}.

In most partial waves, a constant background term is sufficient to achieve a good fit.
Only the background in the meson-baryon $S$-wave required a more complicated form:
\begin{equation}
f_{ab}=\frac{(a+b\sqrt{s})}{(s-s_0)}.
\end{equation}

Further background contributions are obtained from the reggeized exchange of vector mesons
\cite{Anisovich:2004zz} in the form
\be
A&=&g(t)R(\xi,\nu,t) \quad {\rm where}\\
R(\xi,\nu,t)&=&\frac{1+\xi
exp(-i\pi\alpha(t))}{\sin(\pi\alpha(t))} \left (\frac{\nu}{\nu_0}
\right )^{\alpha(t)} \;.\nonumber
\ee
here, $g(t)=g_0\exp(-bt)$ represents a vertex function and a form factor.
$\alpha(t)$  describes the trajectory, $\nu=\frac 12 (s-u)$, $\nu_0$
is a normalization factor, and $\xi$ the signature of the
trajectory. Pion and and Pomeron exchange both have a positive
signature and therefore \cite{Anisovich:2004zz}:
\be\label{eq9a}
R(+,\nu,t)=\frac{e^{-i\frac{\pi}{2}\alpha(t)}} {\sin
(\frac{\pi}{2}\alpha(t))} \left (\frac{\nu}{\nu_0}\right
)^{\alpha(t)}\;.
\ee

Additional $\Gamma$-functions eliminate the poles at $t<0$:
\be
\sin \left (\frac{\pi}{2}\alpha(t)\right ) \to \sin \left
(\frac{\pi}{2}\alpha(t)\right )  \; \Gamma \left (\frac
{\alpha(t)}{2}\right )\, .
\label{rho_1}
\ee
where the Kaon trajectory is parametrized as
$\alpha(t)=-0.25$ $+ 0.85 t$, with $t$ given in GeV$^2$.

The data on partial wave amplitudes (Fig.~\ref{pipKLambda_fit}) and on the photoproduction
multipoles (Fig.~\ref{fig:mult}) were included in the data base of the BnGa partial wave
analysis. The data are fitted jointly with data on $N\eta$, $\Lambda K$, $\Sigma K$,
$N\pi^0\pi^0$, and $N\pi^0\eta$ from both photo- and pion-induced reactions.
Thus inelasticities in the meson-baryon system are constrained by real data.  A list of the
data used for the fit can be found in
\cite{Anisovich:2011fc,Anisovich:2013vpa,Sokhoyan:2015fra,Gutz:2014wit}
and on our website (pwa.hiskp.uni-bonn.de). In Fig.~\ref{pipKLambda_fit}, the systematic
errors define the error band; in Fig.~\ref{fig:mult}, the systematic error of the real
(imaginary) part of the amplitudes is shown a grey (red/blue) histogram at the bottom (top) line.
The systematic errors are derived by a variation of the model space by adding further
resonances with different spin-parities when the data are fitted.

\section{The Laurent-Pietarinen expansion}
\subsection{Formalism}
The main task of the single channel Laurent-Pietarinen expansion ($SC_{L+P}$)  is extracting pole positions from given partial waves for one reaction.
The driving concept behind the method is to replace
an elaborate theoretical model by a local power-series representation of partial wave
amplitudes~\cite{L+P2013}. The complexity of a partial-wave analysis model
is thus replaced by much simpler model-independent expansion which just exploits analyticity
and unitarity. The L+P approach separates pole and regular part in the form of a  Mittag-Leffler expansion\footnote{Mittag-Leffler expansion \cite{Mittag-Leffler} is the generalization of a Laurent expansion to a more-than-one pole situation. From now on, for simplicity, we will simply refer to this as a Laurent expansion.}, and instead of modeling the regular part using some physical model it uses the conformal-mapping-gener\-ated, rapidly converging power series  with well defined analytic properties called a Pietarinen expansion\footnote{A conformal mapping expansion of this particular type was introduced by Ciulli and Fisher \cite{Ciulli,CiulliFisher}, was described in detail and used in pion-nucleon scattering by Esco Pietarinen \cite{Pietarinen,Pietarinen1}, and named as a Pietarinen expansion by G. H\"{o}hler in \cite{Hohler:1984ux}.} to represent it effectively. So, the method replaces the regular part calculated in a model with the simplest analytic function which has correct analytic properties of the analyzed partial wave (multipole), and fits the given input. In such an approach the model dependence is minimized, and is reduced to the choice of the number and location of L+P branch-points used in the model.
The method is applicable to both, theoretical and experimental input, and represents the first reliable procedure to extract pole positions directly from experimental data, with minimal model bias. The L+P expansion based on the Pietarinen expansion is used in some former papers in the analysis of pion-nucleon scattering data
\cite{Ciulli,CiulliFisher,Pietarinen,Pietarinen1} and in several few-body reactions \cite{L+P2014,L+P2014a,L+P2015}.  The procedure has recently  been generalized also to the multi-channel case \cite{Svarc2016}.

The generalization of the L+P method to a multichannel L+P method, used in this paper, is performed in the following way:
i)~separate Laurent expansions are made for each channel; ii)~pole positions are fixed for all
channels, iii)~residua and Pietarinen coefficients are varied freely; iv)~branch-points are chosen
as for the single-channel model; v)~the single-channel discrepancy function $D_{dp}$ (see  Eq. (5) in ref. \cite{L+P2015}) which quantifies the deviation of the fitted function from the input is generalized
to a multi-channel quantity $D_{dp}^a$ by summing up all single-channel contributions, and vi)~the minimization is performed for all channels in order to obtain the final solution.

The formulae used in the L+P approach are collected in Table~\ref{LplusP}.

L+P is a formalism which can be used for extracting poles from any given set of data, either theoretically generated, or produced directly from experiment. If the data set is theoretically generated, we can never reconstruct the analytical properties of the background put into the model, we can only give the simplest analytic function which on the real axes gives very similar, in practice indistinguishable result from the given model values. Therefore, analyzing partial waves coming directly from experiment is for L+P much more favourable because we do not have such demands. The analytic properties are unknown, so there is no reason why the simplest perfect fit we offer should not be the true result.  As in principle we do not care whether the input is generated by theory or otherwise, in the set of formulas given in Table~\ref{LplusP}. we denote any input fitted with L+P function  $T^{a}(W)$ generically as $T^{a,exp}(W)$.

In this paper we fit partial wave data; the discrete data points coming from a semi-constrained single energy fit of K$\Lambda$ photo-production data, which is obtained in a way that the partial waves with $L>2$  are fixed to Bonn-Gatchina energy dependent partial waves, and lower ones are allowed to be free. We perform a multichannel fit ($MC_{L+P}$) when possible by including single energy data from $\pi N \rightarrow K \Lambda$ process, and we fit both multipoles for the same angular momentum at the same time in the coupled-multipole fit ($CM_{L+P}$).   The regular background part is represented by three Pietarinen expansion series, all
free parameters are fitted. The first Pietarinen expansion with branch-point $x_P$ is restricted
to an unphysical energy range and represents all left-hand cut contributions. The next two Pietarinen
expansions describe the background in the physical range with branch-points $x_Q$ and $x_R$ respecting the analytic properties of the analyzed partial wave. The second branch-point is mostly fixed to the
elastic channel branch-point, the third one is either fixed to the dominant channel threshold,
or left free.  Thus, only rather general physical assumptions about the analytic properties are made
like the number of poles and the number and the position of branch-points, and the simplest analytic
function with a set of poles and branch-points is constructed.

\begin{table*}
\caption{\label{LplusP}Formulae defining the Laurent+Pietarinen (L+P) expansion.\vspace{-5mm}}
\begin{align*}
\label{eq:Laurent-Pietarinen}
T^a(W) &=& \sum _{j=1}^{{N}_{pole}} \frac{x^{a}_{j} + \imath \, \, y^{a}_{j}  }{W_j-W} +
      \sum _{k=0}^{K^a}  c^a_k \, X^a (W)^k  +  \sum _{l=0}^{L^a} d_l^a \, Y^a (W)^l +  \sum
_{m=0}^{M^a} e_m^a \, Z^a (W)^m \nonumber\hspace{50mm} \\
X^a (W )&=&  \frac{\alpha^a-\sqrt{x_P^a-W}}{\alpha^a+\sqrt{x_P^a - W }}; \, \, \, \, \,   Y^a(W ) =  \frac{\beta^a-\sqrt{x_Q^a-W }}{\beta^a+\sqrt{x_Q^a-W }};  \, \, \, \, \,
Z^a(W ) =  \frac{\gamma^a-\sqrt{x_R^a-W}}{\gamma^a+\sqrt{x_R^a-W }}
 \nonumber \hspace{42mm}\\
  D_{dp}^a \ \ \ \ \ & = & \frac{1}{2 \, N_{W}^a - N_{par}^a} \, \, \sum_{i=1}^{N_{W}^a} \left\{ \left[ \frac{{\rm \Re e} \,T^{a}(W^{(i)})-{\rm \Re e} \, T^{a,exp}(W^{(i)})}{ Err_{i,a}^{\rm Re}}  \right]^2 + \right.
            \left. \left[ \frac{{\rm Im} \, T^{a}(W^{(i)})-{\rm \Im m} \, T^{a,exp}(W^{(i)})}{ Err_{i,a}^{\rm \Im m}} \right]^2 \right\}  + {\cal P}^a  \nonumber\hspace{3mm} \\
{\cal P}^{a}\ \ \ \ \  &=& \lambda_c^a \, \sum _{k=1}^{K^a} (c^a_k)^2 \, k^3 +
\lambda_d^a \, \sum _{l=1}^{L^a} (d^a_l)^2 \, l^3 +  \lambda_e^a \,
\sum _{m=1}^{M^a} (e^a_m)^2 \, m^3 \qquad\qquad
 D_{dp}  =  \sum _{a}^{all}D_{dp}^a \nonumber \hspace{40mm}\\
 &&a\, \, .....   \, \,    {\rm  channel \, \,  index} \qquad\qquad
  N_{pole}\; .....\;      {\rm number\; of\; poles} \qquad\qquad
  W_j,W \in \mathbb{C}  \nonumber \hspace{42mm}\\
&& x_i^a, \, y_i^a, \, c_k^a, \, d_l^a, \, e_m^a, \, \alpha^a, \, \beta^a, \, \gamma^a ... \in  \mathbb{R}  \, \,
 \nonumber  \hspace{96mm}\\
 &&  K^a, \, L^a, \, M^a \, ... \, \in  \mathbb{N} \, \, \, {\rm number \, \, of \, \, Pietarinen \, \, coefficients \, \, in \, \, channel \, \, \mathit{a} }.
 \nonumber \hspace{53mm}\\
 && D_{dp}^a \; .....  \; {\rm discrepancy \; function \; in\; channel \; }a \qquad\qquad \hspace{12mm}
N_{W}^a \; .....  \; {\rm number\; of\; energies\; in\; channel \; }a \nonumber \hspace{8mm}\\
 && N_{par}^a \; .....  \; {\rm number\; of\; fitting \; parameters \; in\; channel \; }a
\qquad\qquad {\cal P}^a \, \,  .....   \, \, {\rm Pietarinen  \, \, penalty \, \, function}
\nonumber \hspace{14mm}\\
 && \lambda_c^a, \, \lambda_d^a, \,\lambda_e^a \, \, \,  .....   \, \, {\rm Pietarinen  \, \,  weighting \, \, factors} \nonumber \qquad\qquad\hspace{10mm} x_P^a, \, x_Q^a, \, x_R^a  \in \mathbb{R}
 \, \, \, \,( {\rm  or} \, \, \in \mathbb{C}). \nonumber  \hspace{26mm}
\\
&& Err_{i,a}^{\rm \Re e, \, \Im m} ..... {\rm \, \, minimization \, \, error\, \, of \, \, real \,\, and \, \, imaginary \, \, part \, \, respectively.} \nonumber \hspace{45mm}
\end{align*}\vspace{-3mm}
\end{table*}

In the compilation of our results we show the results of four fits:
a)~the BnGa coupled channel fit to the complete data base including the
energy independent solutions for $\pi^- p\to K^0\Lambda$ and $\gamma p\to K^+\Lambda$
presented here; b)~a single-channel L+P fit to the energy independent solution for
$\pi^- p\to K^0\Lambda$  (${\rm SC_{L+P}^{\, \pi N,K \Lambda}}$) ; c)~a single-channel L+P fit to the energy independent solution for
$\gamma p\to K^+\Lambda$ (${\rm SC_{L+P}^{\, \gamma N,K \Lambda}}$); and d)~ a multi-channel L+P fit to the energy independent solution for
$\pi^- p\to K^0\Lambda$ and $\gamma p\to K^+\Lambda$ (${\rm CC_{L+P}}$).

\subsection{L+P Fits}

\subsubsection{$J^P=1/2^-$-wave}

We have fitted the $J^P=1/2^-$ partial wave from the energy independent amplitude for the reaction
$\pi^- p\to K^0\Lambda$

\begin{figure}[pt]
\begin{center}
\begin{tabular}{ccc}
\hspace{-0.2cm}\includegraphics[width=0.16\textwidth]{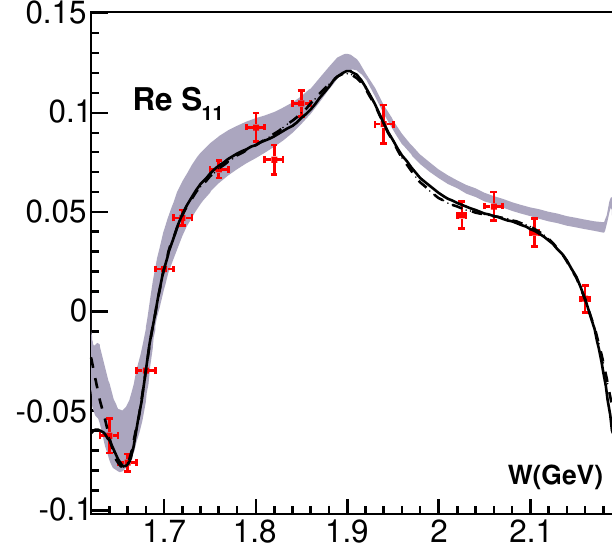}&
\hspace{-0.4cm}\includegraphics[width=0.16\textwidth]{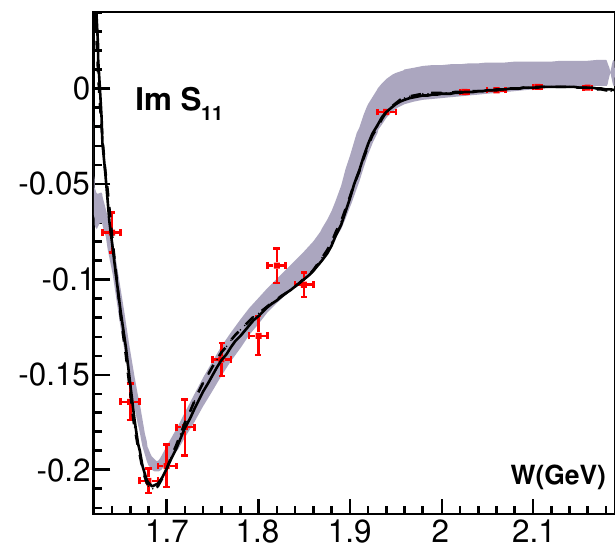}&
\raisebox{-0.82mm}{\hspace{-0.4cm}\includegraphics[width=0.17\textwidth,height=0.162\textwidth]{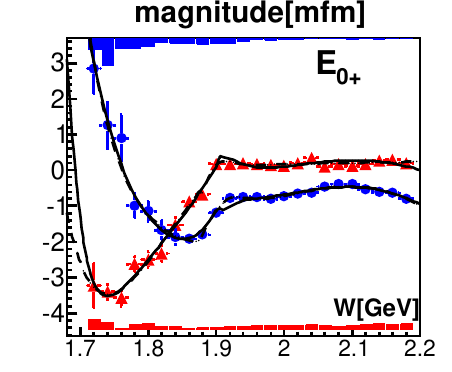}}
\end{tabular}
\end{center}
\caption{\label{fig:E0+}(Color online) Real and imaginary parts of the $J^P=1/2^-$:
S$_{11}$ partial wave amplitude, and the $E_{0+}$ multipole. The grey band represents the
allowed range of BnGa solutions,
the dashed curve the single channel fits ${\rm SC_{L+P}^{\, \pi N,K \Lambda}}$ and
${\rm SC_{L+P}^{\, \gamma N,K \Lambda}}$, and the solid curves
the  L+P fit to both data sets. }
\end{figure}
\noindent
in a  ${\rm SC_{L+P}^{\, \pi N,K \Lambda}}$ fit. A $\chi^2 =2.45$ was
obtained for the 28 data points with 23 parameters.
We needed two poles, one at 1667\,MeV and second one at 1910\,MeV. Due to the low-statistics of the data,
the results from the single-channel fit show large errors.

The 48 data points on the $E_{0+}$ multipole from $\gamma p\to K^+\Lambda$ required only one pole close to
1900\,MeV. The strong peak at low mass of the imaginary part of the $E_{0+}$ multipole is reproduced
by the function $Y^{a}(W)$ with a branching point at the $K^+\Lambda$ threshold. Note that the lowest
mass bin for the $E_{0+}$ multipole starts at 1700\,MeV, significantly above the $N(1650)1/2^-$ mass.
The data were described with a
$\chi^2=0.48 $ and 19 parameters in a ${\rm SC_{L+P}^{\, \gamma N,K \Lambda}}$ fit. Compared to
the pion-induced reaction, the errors on the higher-mass resonance (at 1900\,MeV) are considerably reduced.

The common fit to both data sets (with 76 data points) used two poles, the fit resulted in a $\chi^2=0.86$
for 37 parameters. The results are shown in Table~\ref{tab:results} and Figs.~\ref{fig:mult}
and \ref{fig:E0+}.

The real part of the pole positions of the $N(1650)1/2^-$ resonance are nicely consistent
when the three values are compared, the imaginary part is likely too narrow in the L+P fit.
The magnitudes of the inelastic pole residue are consistent at the $2\sigma$ level when the
BnGa and CC L+P fits are compared. The phases, however, seem to be inconsistent.

\begin{figure}[pt]
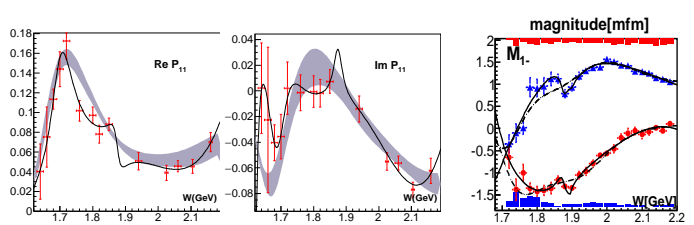

\begin{center}
\vspace*{1mm}
\begin{tabular}{ccc}
\hspace{-0.3cm}\includegraphics[width=0.16\textwidth]{p11_real.pdf}&
\hspace{-0.4cm}\includegraphics[width=0.16\textwidth]{p11_imag.pdf}&
\raisebox{-0.82mm}{\hspace{-0.155cm}\includegraphics[width=0.17\textwidth,height=0.158\textwidth]{M_1-.pdf}}
\end{tabular}
\end{center}
\caption{\label{fig:M1-}(Color online) Real and imaginary parts of the $J^P=1/2^+$:
P$_{11}$ partial wave amplitude, and the $M_{1-}$ multipole. The grey band represents the
allowed range of BnGa solutions,
the dashed curve the single channel fits ${\rm SC_{L+P}^{\, \pi N,K \Lambda}}$ and
${\rm SC_{L+P}^{\, \gamma N,K \Lambda}}$, and the solid curves
the L+P fit to both data sets.  }
\end{figure}

The $N(1895)1/2^-$ pole positions are
well defined with acceptable errors and consistent when the four analyses are compared, only
the single-channel L+P fit to photoproduction data returns a slightly too narrow width.
All four analyses yield compatible magnitudes of the inelastic pole residues, the phases
disagree at the $2\sigma$ level. The magnitudes and the phases of the $E_{0+}$ multipole
determined by the BnGa fit agree well with the values of the L+P fits
within the rather large uncertainties.
Note that the errors in the CC L+P and BnGa fits have different
origins: The L+P errors are of a statistical nature, the BnGa errors are derived from the spread of results
of a variety of different fits. Both approaches establish the
need for $N(1650)1/2^-$ and unquestionably require $N(1895)1/2^-$.

\subsubsection{$J^P=1/2^+$-wave}

We have fitted the $J^P=1/2^+$-wave using the P$_{11}$ energy independent amplitude for the $\pi^- p\to K^0\Lambda$
reaction and the $M_{1-}$ multipole from $\gamma p\to K^+\Lambda$. The first data set $\pi^- p\to K^0\Lambda$
required two poles. The first pole was located near 1700\,MeV, the second one was found near 2100\,MeV even though
with large error bars: the admitted range covers masses from $\sim 1790$ to $\sim 2375$\,MeV. The
photoproduction data required only one pole close to 1900\,MeV. The CC L+P fit to both data sets was performed with two poles.

\begin{table*}[pt]
\renewcommand{\tabcolsep}{2.pt}
\caption{\label{tab:results}Pole parameters for the $J^P=1/2^-$ and  $1/2^+$ waves from the BnGa multichannel partial-wave analysis (BnGa), from the single channel L+P fits to fits to $\pi N\to K\Lambda$ and  $\gamma p\to K^+\Lambda$ (SC), and from the combined L+P fit to $\gamma p\to K^+\Lambda$ and $\pi^-p\to K^0\Lambda$ (CC). Masses and widths are given in MeV.
The acronym $|a{_1}(\pi N\to K\Lambda)|$ stands for the normalized residua
$2|Res_{_1}(\pi N\to K\Lambda)|/\Gamma_{\rm pole}$, $|A{_1}(E_{0+})|$ for the residua
$|Res_{_1}(E_{0+}(\gamma N\to K\Lambda))|$; they are given in GeV$^{-1/2}$.
The phases $\Theta$ are given in degrees $(360^\circ \equiv 2\pi)$; UNDET for stands for
``state undetermined''.
${\rm SC_{L+P}^{\, \pi N,K \Lambda}}$  for single-channel L+P fit to $\pi N \to K \Lambda$, ${\rm SC_{L+P}^{\, \gamma N,K \Lambda}}$  for single-channel L+P fit to $\gamma N \to K \Lambda$, ${\rm CC_{L+P}^{\, \pi N,K \Lambda}}$  for coupled-channel L+P fit to $\pi N \to K \Lambda$ and $\gamma N \to K \Lambda$, etc . Under PDG, the PDG estimates are given
or our own estimates from PDG entries. \vspace{2mm}}
\begin{center}
\def\arraystretch{1.4}
\footnotesize
\begin{tabular}{|c|ccccc|ccccc|}
  \hline
                   &  PDG             &    BnGa   & ${\rm SC_{L+P}^{\, \pi N,K \Lambda}}$  & ${\rm SC_{L+P}^{\, \gamma N,K \Lambda}}$   &   ${\rm CC_{L+P}}$
                    &  PDG             &    BnGa   & ${\rm SC_{L+P}^{\, \pi N,K \Lambda}}$  & ${\rm SC_{L+P}^{\, \gamma N,K \Lambda}}$   &   ${\rm CC_{L+P}}$
                                    \tabularnewline   \hline
&\multicolumn{5}{c|}{\boldmath$N(1650)1/2^-$\unboldmath}&\multicolumn{5}{c|}{\boldmath$N(1710)1/2^+$\unboldmath}\\\hline
 M$_1$                       &   1640-1670            &1658\er10&1667\er43&  & 1662\er49
                            &  1670-1770           &1690\er15&1723\er16&  & 1697\er23     \tabularnewline
$\Gamma_1$                  &    100-170              &102\er8 & 75\er16        &      & 59\er16
                            &   80-380            &155\er25  & 37\er14     &     & 84\er34         \tabularnewline
$|a{_1}(\pi N\to K\Lambda)|$&  0.23\er0.09        &0.26\er0.10&   $<0.35$      &     & 0.05\er0.05
                            &   0.17\er0.06        &0.16\er0.05& $<0.09$    &   & $<0.20$            \tabularnewline
$\Theta_1(\pi N)$           &\script (110\er20)\oo  &\script(110\er20)\oo& \script(-123\er237)\oo &  & \script(-95\er33)\oo
                            &\script (85\er9)\oo   &\script(-160\er25)\oo& \script(-45\er62)\oo &      & \script(-120\er83)\oo        \tabularnewline
$10^3 \cdot|A{_1}(E_{0+}/M_{1-})|$    &                      &32\er5      &   &   & \script UNDET
                            &                      &32\er16      &         &      & \script UNDET   \tabularnewline
$\Theta_1(E_{0+}/M_{1-})$   &                      & \script(0\er12)\oo&       &     & \script UNDET
                            &                      &\script(-40\er30)\oo&  &   & \script UNDET
                                                                                                 \tabularnewline
 \hline
&\multicolumn{5}{c|}{\boldmath$N(1895)1/2^-$\unboldmath}&\multicolumn{5}{c|}{\boldmath$N(1880)1/2^+$\unboldmath}\\\hline
M$_2$                       & 1905\er 20           &1895\er15&  1910\er64    &   1901\er18 & 1906\er17
                            & 1870\er40            &1860\er40& 2081\er293 &   1876\er11      & 1878\er11           \tabularnewline
$\Gamma_2$                  &  100\er40           &132\er30 & 119\er24    &   68\er 18 & 100\er 11
                            &  220\er 50           &230\er50 & 48\er183    &31\er9    & 33\er9                    \tabularnewline
$|a{_1}(\pi N\to K\Lambda)|$&  0.05\er0.02         &0.09\er0.03& 0.06\er0.03        &       & 0.03\er0.01
                            &  0.02\er0.01         &0.05\er0.02&  \script UNDET    &    & 0.15\er0.05               \tabularnewline
$\Theta_2(\pi N\to K\Lambda)$&\script (-90\er30)\oo &\script (8\er30)\oo&\script (-77\er107)\oo &   & \script (87\er27)\oo
                            &\script(32\er10)\oo  &\script (27\er30)\oo&  \script UNDET &  & \script (-82\er9)\oo       \tabularnewline
$10^3 \cdot|A{_2}(E_{0+}/M_{1-})|$     &                      & 22\er17     &         & 30\er21 & 51\er25
                            &                      & 18\er12     &         & 8\er6  &  8\er5              \tabularnewline
$\Theta_2(E_{0+}/M_{1-})$    &                      &\script(-25\er30)\oo&         &\script(-80\er 47)\oo & \script(-73\er 30)\oo
                            &                      &\script(90\er70)\oo         &         & \script(60\er40)\oo      & \script (59\er40)\oo                 \tabularnewline
\hline
\end{tabular}
\vspace{2mm}
\end{center}
\caption{\label{tab:results1}Pole parameters for the $J^P=3/2^+$, $3/2^-$, and $5/2^-$  waves from the BnGa multichannel partial-wave analysis (BnGa). The $E_{1+}$ and $M_{1+}$ ($E_{2-}$ and $M_{2-}$) multipoles
exciting the $J^P=3/2^+$ ($3/2^-$) partial waves are fitted in a coupled multipole CM L+P fit, the
$E_{2+}$ in a single channel SC L+P fit. Masses and widths are given in MeV.
The acronym  $|A{_1}(E_{1+}/E_{2-}/E_{2+})|$ stands for the residues
$|Res_{1}(E_{0+}(\gamma N\to K\Lambda))|$, $|Res_{_1}(E_{2-}(\gamma N\to K\Lambda))|$, etc.,
which are given in GeV$^{-1/2}$ units;
the phases $\Theta$ are given in degrees $(360^\circ \equiv 2\pi)$.
${\rm CM_{L+P}^{\, \pi N,K \Lambda}}$  denotes the coupled-multipole L+P fit to $\gamma N \to K \Lambda$,
${\rm SC_{L+P}^{\, \pi N,K \Lambda}}$ the single-channel L+P fit to $\gamma N \to K \Lambda$.
Under PDG, the PDG estimates are given or our own estimates from PDG entries.   \vspace{2mm}}
\begin{center}
\def\arraystretch{1.4}
\footnotesize
\begin{tabular}{|c|ccc|ccc|ccc|}

\hline
                             &  PDG                 &  BnGa   &    ${\rm CM_{L+P}}$
                            &  PDG                 &  BnGa   &   ${\rm CM_{L+P}}$
                            &  PDG                 &  BnGa   &   $\rm SC_{L+P}^{\, \gamma N,K \Lambda}$        \tabularnewline
                                                        \hline
&\multicolumn{3}{c|}{\boldmath$N(1900)3/2^+$\unboldmath}&\multicolumn{3}{c|}{\boldmath$N(1875)3/2^-$\unboldmath}
&\multicolumn{3}{c|}{\boldmath$N(2060)5/2^-$\unboldmath}\\\hline
 M$_1$                      &    1900-1940     &1945\er35                    & 1912\er30
                            &    1800-1950     &1870\er25                    &  1977\er41
                            &    2030-2130     & 2030\er25      &  2019\er51          \tabularnewline
$\Gamma_1$                  &    130-300       &$135^{+70}_{-30}$            & 166\er30
                            &    150-250       &210\er25                     & 120\er50
                            &    300-450       &   $350_{ -30}^{  +80}$        &  141\er67               \tabularnewline
$10^3 \cdot|A{_1}(E_{1+}/E_{2-}/E_{2+})|$&               &45\er12                &
                            &                  &11\er10                &
                            &                  &   8\er6                          &                \tabularnewline
$\Theta_1(E_{1+}/E_{2-}/E_{2+})$     &                  &\script(-100\er20)\oo        &
                            &                  &\script(40\er50)\oo          &
                            &                  &  \script(-100\er80)\oo                           &                    \tabularnewline
$10^3 \cdot|A{_1}(M_{1+}/M_{2-})/M_{2+}|$    &                  &80\er30                &
                            &                  & 9\er8               &
                            &                  &  60\er18                           &                   \tabularnewline
$\Theta_1(M_{1+}/M_{2-}/M_{2+})$   &                  &\script(95\er30)\oo          &
                            &                  &\script(-30\er100)\oo        &
                            &                  &\script(-170\er10)\oo                             &                    \tabularnewline
\hline
\end{tabular}
\end{center}
\end{table*}

The results are shown in Table~\ref{tab:results} and Figs.~\ref{fig:mult} and~\ref{fig:M1-}. The 28 data
points for $\pi^- p\to K^0\Lambda$ were fitted with 23 parameters and a $\chi^2=0.67$. The 48 data points
on the $M_{1-}$ multipole were described with a $\chi^2=0.366$ and 19 parameters. The common fit to both data
sets resulted in a $\chi^2=0.505$ for 41 parameter. Both approaches, the BnGa and CC L+P fit,
establish the need for $N(1710)1/2^+$, and unquestionably require $N(1880)1/2^+$.

The $N(1710)1/2^+$ mass is consistent in the CC L+P and the BnGa fits, its width tends to be
smaller in the CC L+P fit, see Tables~\ref{tab:results} but the difference is $1.7\sigma$ only.
The magnitudes of the inelastic residue for this resonance have large error bars in the L+P fits
and cover zero, we give upper limits only. The limits are compatible with the BnGa result. In spite of
the large errors in the magnitudes, the phases are consistent.

The masses of the $N(1880)1/2^+$ resonance from the BnGa and CC L+P fits are compatible but not the widths.
The inelastic residues disagree slightly. Both, the single-channel SC L+P and the coupled-channel CC L+P
fit, agree that the $N(1880)1/2^+$ width should be smaller than $\sim 40$\,MeV while BnGa finds
a normal hadronic width.
However, we have performed a CC L+P fit imposing a mass of 150\,MeV. When the result of the CC L+P fit
is compared to the observables with 674 data points (Figs.~\ref{fig:data-g1}
to \ref{fig:data-g4}), the fit deteriorates only minimally, the $\chi^2$ increases by 4.5 units.
We conclude that the $N(1880)1/2^+$ resonance is definitely required in this
nearly model-independent analysis and that it has a normal hadronic width.
The magnitudes of the inelastic residues and of the $M_{1-}$ multipole
agree reasonably well, the phases of the inelastic residues are again inconsistent while
the $M_{1-}$ multipole phases agree well within their uncertainties.

\subsubsection{$J^P=3/2^+$-wave}
The $J^P=3/2^+$-wave was not derived from the pion induced reaction $\pi^-p\to K^0\Lambda$, so
the two photoproduction multipoles
$E_{2-}$ and $M_{2-}$ were fitted simultaneously in the coupled-multipoles L+P mode (${\rm CM_{L+P}}$). The CM L+P fit
required only one pole close to 1900\,MeV, no $N(1720)3/2^+$ was needed. Due to the presence of
important thresholds ($\Sigma K$, $N(1520)3/2^-\pi$, $N(1535)1/2^-\pi$), the $N(1720)3/2^+$ resonance
has a rather complicated pole structure, and we refrain from discussing this resonance here.
The fit to the 96 data points in the two data sets is shown in Fig.~\ref{fig:mult}. The fit
returned a $\chi^2=0.42$ for 35 parameters.
The results are shown in Table~\ref{tab:results1}. The poles from the L+P and BnGa
fits are fully consistent. We conclude that $N(1900)3/2^+$ is definitely confirmed in this
nearly model-independent analysis.

\subsubsection{$J^P=3/2^-$}
Due to limited statistics, the $J^P=3/2^-$-wave could not be derived from the pion induced
reaction $\pi^-p\to K^0\Lambda$. Thus, only the two photoproduction multipoles $E_{1+}$
and $M_{1+}$ were fitted in the coupled-multipoles CM L+P mode (${\rm CM_{L+P}}$). The L+P
fit to the 96 data points in the two data sets returned a $\chi^2=0.55$ for 36 parameters,
the fit is shown in Fig.~\ref{fig:mult}. The fit required only one pole close to 1900\,MeV,
no $N(1700)3/2^-$ was needed. A low-mass pole at about 1700\,MeV is required in the BnGa
fit but due to the complicated pole structure in this mass region, we again refrain from
discussing its properties here. The results of the L+P and the BnGa fits are shown in
Table~\ref{tab:results1}. The poles from the L+P and BnGa
fits are found to be inconsistent. In the BnGa model, a mass of 1870\er 25\,MeV is found,
and there is a second pole -- not discussed here -- at 2150\,MeV. The L+P fit does not
find evidence for a two-pole structure and places the mass of the one pole at 1977\er41\,MeV.

\subsubsection{$J^P=5/2^-$-waves}
The $J^P=5/2^-$-wave was not derived from the pion induced reaction $\pi^-p\to K^0\Lambda$,
and in this case only the $E_{2-}$ multipole could be determined from the data.
The single channel L+P mode ($\rm SC_{L+P}^{\, \gamma N,K \Lambda}$ ) was hence used to
fit the data. The fit required one pole at about 2000\,MeV.
The fit to the 48 data points in the two data sets returned a $\chi^2=0.60$ for 25 parameters.
The results are shown in Table~\ref{tab:results1} and Fig.~\ref{fig:mult}. The pole positions
from the L+P and BnGa fits are fully consistent. We conclude that $N(2060)5/2^-$ is confirmed.

\section{Comparison to other groups}

\begin{figure}[pb]
\begin{center}
\begin{tabular}{cc}
\hspace{-0.3cm}\includegraphics[width=0.21\textwidth]{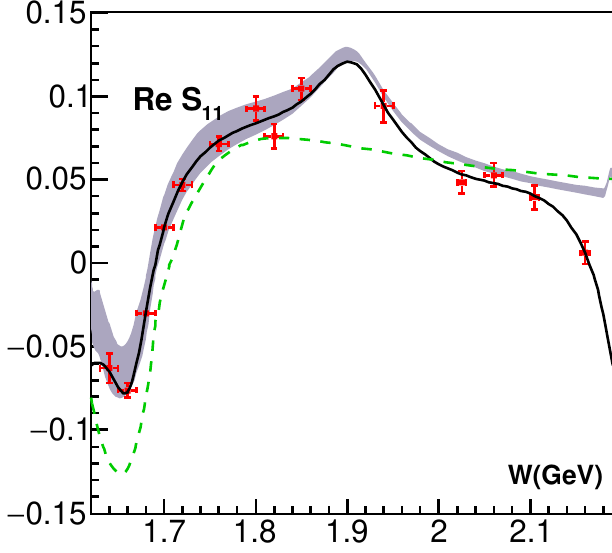}&
\hspace{-0.33cm}\includegraphics[width=0.21\textwidth]{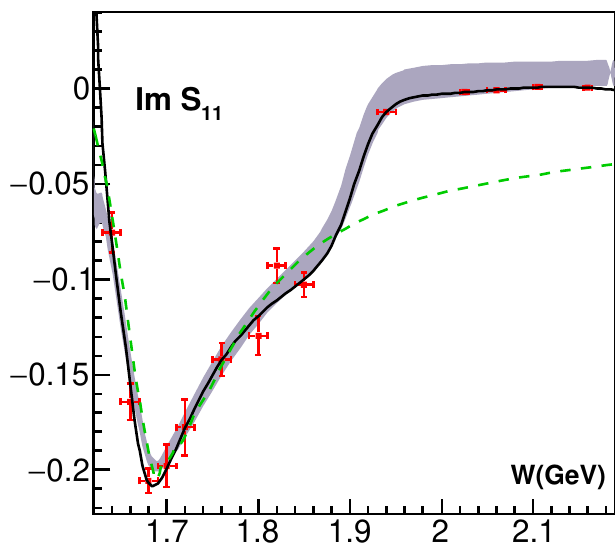}\\
\hspace{-0.3cm}\includegraphics[width=0.21\textwidth]{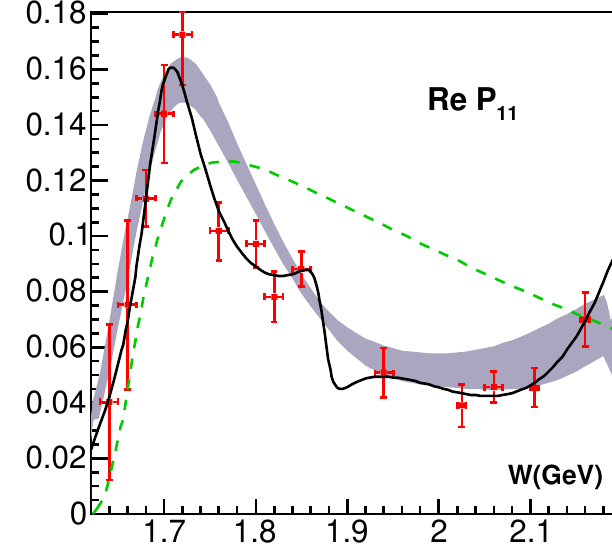}&
\hspace{-0.33cm}\includegraphics[width=0.21\textwidth]{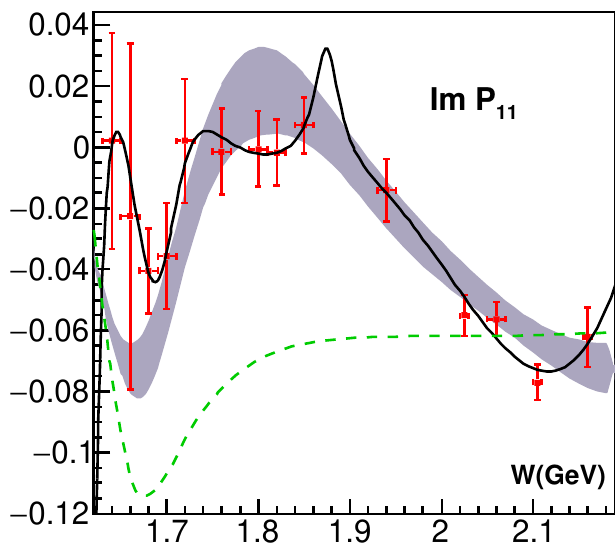}
\end{tabular}
\end{center}
\caption{\label{fig:comp1}
Real and imaginary part of the (dimensionless) $S_{11}$ and $P_{11}$ waves~\cite{Anisovich:2014yza}.
The energy-dependent solution BnGa2011-02 is shown as error band. The solid curve
represents a L+P fit. The dashed (green) curve is given by the solution J\"uBo2015-B of the
J\"uBo group~\cite{Ronchen:2015vfa}. The BnGa and J\"uBo groups use a different sign convention.
The J\"uBo amplitudes are shown with an inverted sign.
 }
\end{figure}
Figure~\ref{fig:comp1} shows the real and imaginary parts of low-$L$ partial-wave
amplitudes from Refs.~\cite{Anisovich:2014yza} and ~\cite{Ronchen:2015vfa}.
The amplitudes are similar in magnitude but differ in their shape. The J\"uBo fit
does not contain $N(1895)1/2^-$, the third resonance in the $J^P=1/2^-$ wave that
is confirmed here and in a recent
analysis of $\gamma p\to \eta , \eta'p$~\cite{Kashevarov:2017kqb}. Both the analysis in
Ref.~\cite{Ronchen:2015vfa} and this work, introduce $N(1710)1/2^+$ - a resonance not
needed in Ref.~\cite{Arndt:2006bf} - but here we find evidence for an additional
resonance in this partial wave, $N(1880)1/2^+$. Thus the differences in the partial-wave amplitudes
are to be expected.

There is a large number of papers devoted to partial wave analyses
of the reaction $\gamma p\to K^+\Lambda$. We discuss here only recently published
papers which include at least one measurement of a double polarization variable.

\begin{figure*}[pt]
\begin{center}
\begin{tabular}{cccc}
\includegraphics[width=0.245\textwidth,height=0.245\textwidth]{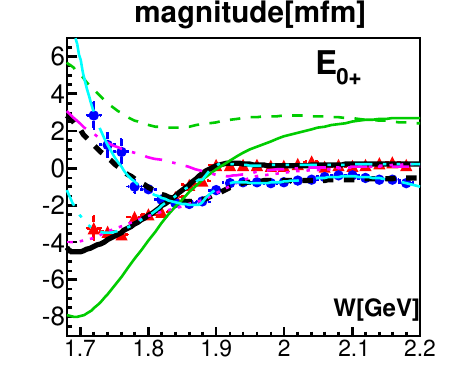}&
\hspace{-2mm}\includegraphics[width=0.245\textwidth,height=0.245\textwidth]{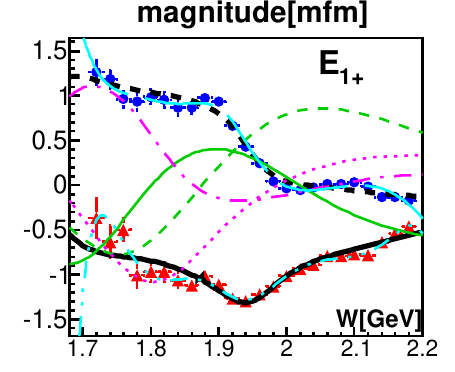}&
\hspace{-2mm}\includegraphics[width=0.25\textwidth,height=0.25\textwidth]{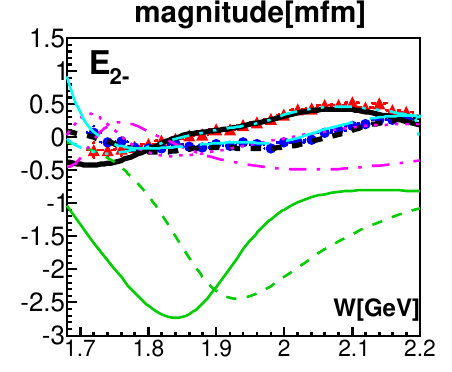}&
\hspace{-2mm}\includegraphics[width=0.25\textwidth,height=0.25\textwidth]{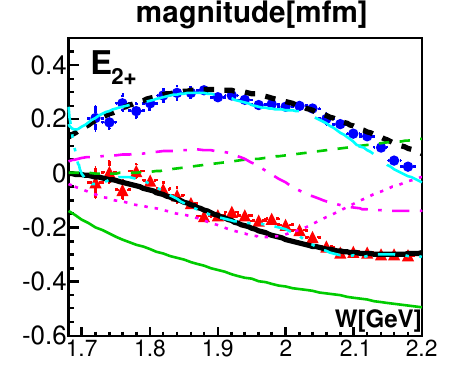}\\
\includegraphics[width=0.245\textwidth,height=0.245\textwidth]{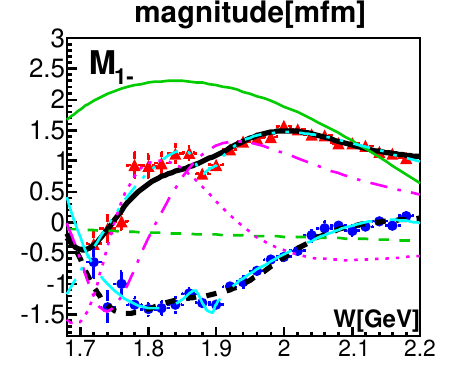}&
\hspace{-2mm}\includegraphics[width=0.245\textwidth,height=0.245\textwidth]{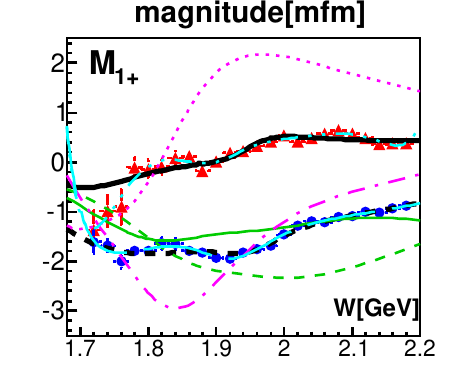}&
\hspace{-2mm}\includegraphics[width=0.25\textwidth,height=0.25\textwidth]{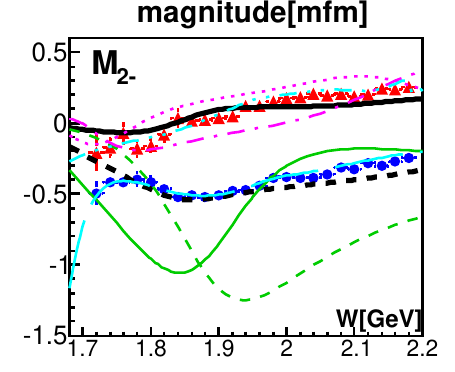}&
\end{tabular}
\end{center}
\caption{\label{fig:comp2}(Color online) Comparison of the real and imaginary parts of the
$E_{0+}$, $E_{1+}$, $M_{1-}$, $M_{1+}$, $E_{2-}$, $E_{2+}$, and $M_{2-}$ multipoles.
The result of the energy-independent analysis are shown with error bars; the BnGa
fit~\cite{Anisovich:2011fc} to the real part is represented by (black) thick curves,
to the imaginary part by thick dashed curves. The L+P fit, shown by (cyan) long-dashed
and long-dashed-dotted curves, often coincides with the BnGa fit. The fit of
Ref.~\cite{Skoupil:2016ast} is shown by thin (green) solid or dashed curves, and the fit of
Refs.~\cite{Mart:2015jof,Mart:2017mwj} by thin (magenta) dotted or dashed dotted curves, again
for the real or imaginary parts, respectively. Refs.~\cite{Mart:2015jof,Mart:2017mwj} use
a different sign convention. These amplitudes are shown with an inverted sign.
 \vspace{-2mm}  }
\end{figure*}
The Gent group proposes a methodology based on Bayesian inference to determine those
resonances which contribute to $\gamma p\to K^+\Lambda$~\cite{DeCruz:2011xi,DeCruz:2012bv}.
They try different groups of 11 resonances and find that the fit with
$N(1535)1/2^-$, $N(1650)1/2^-$, $N(1680)5/2^+$, $N(1720)3/2^+$, $N(1875)3/2^-$,
$N(1880)1/2^+$, $N(1900)3/2^+$, and $N(2000)5/2^+$ has the highest evidence.

In a similar model, Skoupil and Byd\v{z}ovsk\'y \cite{Skoupil:2016ast} use
alternatively 15 or 16 resonances. They confirm
the findings of the Gent group but report evidence that $N(1880)1/2^+$ should
be replaced by $N(1860)5/2^+$.

A number of groups have analyzed pion or photo-induced reactions with a Kaon and a
$\Lambda$ hyperon in the final state. Wu, Xie, and Chen~\cite{Wu:2014yca} studied
the reaction $\pi^-p\to K^0\Lambda$ up to $W = 1.76$\,GeV in an isobar model; the isobars include
hyperon exchanges in the $u$-channel and $K^*$ exchange in the $t$-channel. The leading
$s$-channel contributions were found to be due $N(1535)1/2^-$, $N(1650)1/2^-$ and $N(1720)3/2^+$ formation.
Xiao, Ouyang, Wang, and Zhong~\cite{Xiao:2016dlf} studied the mass range below 1.8\,GeV
and emphasize the leading role of $N(1535)1/2^-$ and $N(1650)1/2^-$. The J\"ulich-Bonn (J\"uBo)
group~\cite{Ronchen:2015vfa} described
the data on $\pi^-p\to K^0\Lambda$ simultaneously with other pion-induced reactions
in an analytic, unitary, coupled-channel approach. SU(3) flavor symmetry was used to
relate both the $t$- and the $u$-channel exchanges. The authors fit the available data
(see Fig.~\ref{pipKLambda_data}); all resonances found in the GWU
analysis~\cite{Arndt:2006bf} were introduced in the fit and four further ones.

Mart, Clympton and Arifi \cite{Mart:2015jof,Mart:2017mwj} take into account the set of
resonances used in the BnGa analysis~\cite{Anisovich:2011fc}. They find that spin-5/2
resonances play an important role and have to be taken into account. In their best fit, the
authors use 17 $N^*$ resonances. The three
resonances $N(1650)1/2^-$, $N(1720)3/2^+$, and $N(1900)3/2^+$ provide the most
important contributions.

In Fig.~\ref{fig:comp2}, the photoproduction multipoles from the BnGa analysis and those of
Skoupil and Byd\v{z}ovsk\'y~\cite{Skoupil:2016ast} and of Mart, Clympton and Arifi~\cite{Mart:2015jof}
are compared. There is no much similarity even though partly the same resonances are used. But possibly,
this is not too surprising. In a comparison of the best studied process, $\gamma p \to \pi N$,
significant differences were observed in the multipoles obtained by the BnGa, J\"uBo, and GWU groups
\cite{Beck:2016hcy} even though all three groups were capable of describing the data reasonably well.
However, new data enforced a considerable reduction of the spread of the three results.
In any case, the comparison demonstrates that further work is needed before the
$\gamma p\to K^+\Lambda$ reaction can be considered as well understood.

\noindent
\section{Summary}

For a long time it has been anticipated that photoproduction experiments will provide measurements that are
sufficient in number and statistical accuracy to construct the four complex amplitudes governing
the photoproduction of an octet baryon and a pseudoscalar meson. A determination of these four
amplitudes requires the measurement with sufficient accuracy of at least eight carefully
selected observables \cite{Chiang:1996em}, and
one phase still remains undetermined. Alternatively, the multipoles driving the excitation of specific
partial waves can be deduced from the data in a truncated partial wave analysis.

In this paper, we have performed such a truncated partial wave analysis of the reaction
$\gamma p\to K^+\Lambda$. The CLAS experiments studied this reaction and reported data on
the differential cross section $d\sigma/d\Omega$, on the polarization observables $P$, $T$ and
$\Sigma$, and on the spin correlation parameters $O_x$, $O_z$, $C_x$, $C_z$. The data cover the
resonance region from 1.71 to 2.13\,GeV, mostly in 20\,MeV wide bins. Thus at the moment, these data
offer the best chance to perform a truncated partial wave analysis.

In a first step, we determined
the number of multipoles that can be deduced from the data. When the number of free multipoles is increased
in the energy-independent analysis, the errors in the determination of the multipoles increases, and one
has to balance precision on the one hand and the number of multipoles on the other hand. It turned
out that only the four largest multipoles, $E_{0+}$, $M_{1-}$, $E_{1+}$, $M_{1-}$, can be determined
without constraints when a good precision of the multipoles is required. In addition, three further
multipoles, $E_{2-}$, $M_{2-}$, $E_{1+}$, could be derived from the data when a penalty function
forced the fit not to deviate too much from an energy dependent solution.
In addition to the photoproduction multipoles, we also used partial wave amplitudes for the
reaction $\pi^- p\to K^0\Lambda$ which had been determined earlier.

The energy-dependent solution was found within the BnGa approach. In this approach, a large
number of data on pion and photo-induced reaction is fitted in a coupled channel analysis.
The data base includes $N\pi$, $N\eta$, $\Lambda K$, $\Sigma K$, $N\pi\pi$, and $N\pi\eta$
final states and, in an iterative procedure, the partial wave amplitudes and photoproduction
multipoles derived here. The higher photoproduction multipoles that could not be determined
in the fits to the CLAS data were kept fixed to multipoles from the BnGa analysis.

All multipoles considered here, $E_{0+}$, $M_{1-}$, $E_{1+}$, $M_{1+}$, $E_{2-}$, $M_{2-}$,
$E_{1+}$, are fitted within a Laurent-Pietarinen expansion. This expansion exploits the
analytic structure of the S-matrix. In the vicinity of a resonance position (and reasonably
close to the real axis), the photoproduction amplitude is determined by poles and the
opening of thresholds. When this analytic structure is imposed, fits to the photoproduction
multipoles and partial wave amplitudes require no further dynamical input, the fits do not
impose any model bias. The Laurent-Pietarinen fits were performed to the photoproduction
multipoles, to the partial wave amplitudes from the $\pi^-\to K^0\Lambda$ reaction, and to both
in a coupled channel fit. The results are then compared to those from the BnGa fit.

The two resonances $N(1895)1/2^-$ and $N(1900)3/2^+$ are firmly established. The results
on their masses, widths, and other properties agree well. Also the $N(1880)1/2^+$ resonance
is definitely required but there remains the question of the width: within the
Laurent-Pietarinen expansion, its width is 40\,MeV or less while its width within the BnGa
approach is about 150\,MeV. The statistical significance of the narrow width is however
very small.

The two resonances $N(1875)3/2^-$ and $N(2060)5/2^-$ are derived from photoproduction
multipoles which are constrained to follow the BnGa solution. In the $J^P=3/2^-$ partial
wave, BnGa finds two poles; in the Laurent-Pietarinen fit, only one pole is observed
at a mass in between the two BnGa poles. The BnGa and Laurent-Pietarinen results on $N(2060)5/2^-$
are nicely consistent.

Summarizing, we can claim that several resonances found in the BnGa energy-dependent
multichannel analysis are confirmed by fits based on a Laurent-Pietarinen expansion
with a minimal model dependence.\\

{\small
This work is supported by the \textit{Deutsche Forschungsgemeinschaft} (SFB/TR110), \textit{Deutsche Forschungsgemeinschaft} (SFB 1014)
the \textit{US Department of Energy under contract DE-AC05-06OR23177, the U.K.~Science and Technology Facilities Council grant ST/L005719/1, and the
\textit{Russian Science Foundation} (RSF 16-12-10267).}
}

\end{document}